\documentclass[twocolumn]{aastex631}

\usepackage{amsmath}
\usepackage{amssymb}
\usepackage{graphicx}
\usepackage{hyperref}
\usepackage{xcolor}
\usepackage{needspace}
\usepackage{etoolbox}
\pretocmd{\section}{\needspace{5\baselineskip}}{}{}
\pretocmd{\subsection}{\needspace{4\baselineskip}}{}{}
\AtBeginDocument{\patchcmd{\thebibliography}{\small}{\footnotesize}{}{}}

\AtBeginDocument{\renewcommand{\paragraph}[1]{\par\smallskip\noindent\textit{#1}\enspace\ignorespaces}}

\newcommand{\deepdisc}{\textsc{DeepDISC}}
\newcommand{\scarlet}{\textsc{Scarlet}}
\newcommand{\sextractor}{\textsc{SExtractor}}
\newcommand{\detectron}{\textsc{Detectron2}}
\newcommand{\swinb}{Swin-B}
\newcommand{\mvit}{MViTv2}
\newcommand{\dpone}{DP1}
\newcommand{\dptwo}{DP2}
\newcommand{\rsp}{Rubin Science Pipelines}

\newcommand{\ecdfs}{ECDFS}
\newcommand{\edfs}{EDFS}

\newcommand{\CgalA}{84.2}
\newcommand{\CstarA}{49.1}
\newcommand{\CgalCWA}{81.5}
\newcommand{\CstarCWA}{64.0}
\newcommand{\CgalFA}{84.5}
\newcommand{\CstarFA}{66.6}

\newcommand{\CgalH}{84.9}
\newcommand{\CstarH}{74.3}
\newcommand{\CgalCWH}{84.6}
\newcommand{\CstarCWH}{76.6}
\newcommand{\CgalFH}{87.0}
\newcommand{\CstarFH}{78.2}

\newcommand{\Cgalbase}{40.2}
\newcommand{\Cstarbase}{13}
\newcommand{\CgalIN}{83.9}
\newcommand{\CstarIN}{50.5}
\newcommand{\Cgaldelta}{0.3}
\newcommand{\Cstardelta}{1.4}

\newcommand{\HSTComp}{31.0}
\newcommand{\HSTPur}{94.6}
\newcommand{\HSTUB}{14.9}
\newcommand{\HSTUBlo}{12.5}
\newcommand{\HSTUBhi}{17.8}
\newcommand{\HSTDDComplim}{24.8}
\newcommand{\HSTGalRec}{94.8}
\newcommand{\HSTStarRec}{71.4}
\newcommand{\HSTRubinCompHeldout}{44.0}
\newcommand{\HSTRubinPurHeldout}{77.6}

\newcommand{\HSTRubinComp}{43.8}
\newcommand{\HSTRubinPur}{75.4}
\newcommand{\HSTRubinUB}{18.0}
\newcommand{\HSTNclean}{34{,}612}
\newcommand{\HSTNcandels}{27{,}438}%
\newcommand{\HSTNdpone}{15{,}839}
\newcommand{\HSTComplim}{25.2}

\newcommand{\HSTRubinUBlo}{16.3}
\newcommand{\HSTRubinUBhi}{19.8}
\newcommand{\HSTRubinUBext}{17.7}
\newcommand{\HSTRubinUBspatial}{22.3}
\newcommand{\HSTRubinGalRec}{85.3}
\newcommand{\HSTRubinStarRec}{61.5}
\newcommand{\HSTNshared}{15{,}928}
\newcommand{\HSTNmatchpur}{11{,}945}
\newcommand{\HSTNmatchcomp}{12{,}030}
\newcommand{\HSTSetPur}{95.8}
\newcommand{\HSTSetComp}{33.4}
\newcommand{\HSTSetUB}{17.3}
\newcommand{\HSTSetGalRec}{89.1}
\newcommand{\HSTSetStarRec}{67.5}

\newcommand{\Npatches}{101}
\newcommand{\Ntilestrain}{994}
\newcommand{\Ntilesval}{152}
\newcommand{\Ntilestest}{312}
\newcommand{\Ntilestot}{1{,}458}
\newcommand{\NtilestrainPre}{1{,}025}
\newcommand{\NtilesvalPre}{153}
\newcommand{\NtilestestPre}{323}
\newcommand{\Ntilestotpre}{1{,}501}
\newcommand{\Ntilesdrop}{43}
\newcommand{\Ntilesdropdr}{4}
\newcommand{\Ntilesdropcov}{41}

\newcommand{\Nanntrain}{301{,}455}
\newcommand{\Nannmaxtile}{556}
\newcommand{\Ngaltrain}{163{,}687}
\newcommand{\Nstartrain}{28{,}346}
\newcommand{\Nobjtrain}{192{,}033}
\newcommand{\Nobjval}{27{,}739}
\newcommand{\Nobjtest}{68{,}505}
\newcommand{\Ngalval}{23{,}856}
\newcommand{\Nstarval}{3{,}883}
\newcommand{\Ngaltest}{58{,}710}
\newcommand{\Nstartest}{9{,}795}
\newcommand{\Ngalall}{246{,}253}
\newcommand{\Nstarall}{42{,}024}
\newcommand{\Nobjall}{288{,}277}
\newcommand{\Ntilestrainraw}{1{,}244}
\newcommand{\Ntilesvalraw}{208}
\newcommand{\Ntilestestraw}{376}
\newcommand{\Ntilesrawtot}{1{,}828}
\newcommand{\Ntilesqcdrop}{370}
\newcommand{\Ntilesfillmask}{325}
\newcommand{\Nanndroplast}{8{,}147}
\newcommand{\Nobjdroplast}{5{,}318}
\newcommand{\Nannfinaltot}{454{,}443}
\newcommand{\Nanncrop}{346{,}798}
\newcommand{\NuOnly}{13}
\newcommand{\NuAny}{28}
\newcommand{\NyAny}{16}

\newcommand{\Nbrightstar}{447}
\newcommand{\Nremoved}{61}
\newcommand{\NstarTilingRaw}{51{,}216}

\newcommand{\NstarScarlet}{51{,}652}
\newcommand{\NstarStageOneRem}{1{,}198}
\newcommand{\StageOneStarRemFrac}{2.3}

\newcommand{\UBfrac}{2.1}
\newcommand{\Shredfrac}{97.5}
\newcommand{\Resfrac}{0.1}
\newcommand{\EqUnresfrac}{0.3}
\newcommand{\UBfracTwo}{10.2}
\newcommand{\ShredfracTwo}{76.6}
\newcommand{\ResfracTwo}{11.3}
\newcommand{\EqUnresfracTwo}{1.9}
\newcommand{\Ncomplex}{12{,}453}
\newcommand{\NcomplexTwo}{16{,}179}
\newcommand{\OverallComp}{83.4}
\newcommand{\OverallPur}{79.5}
\newcommand{\NgtMatch}{109{,}816}
\newcommand{\NpredMatch}{115{,}180}
\newcommand{\NmatchedMatch}{91{,}549}
\newcommand{\DDComplimGal}{26.5}
\newcommand{\DDComplimStar}{26.3}

\newcommand{\DDnormAcc}{97.1}
\newcommand{\DDnormGal}{97.8}
\newcommand{\DDnormStar}{94.3}

\newcommand{\DDswAcc}{95.9}
\newcommand{\DDswGal}{96.1}
\newcommand{\DDswStar}{95.2}

\newcommand{\RFacc}{98.3}
\newcommand{\RFgalRec}{99.5}
\newcommand{\RFstarRec}{95.2}
\newcommand{\RFgalPrec}{98.2}
\newcommand{\RFstarPrec}{98.6}
\newcommand{\RFgalFone}{98.8}
\newcommand{\RFstarFone}{96.9}

\newcommand{\NholdoutHO}{1{,}104}
\newcommand{\NcommonHO}{1{,}026}

\newcommand{\DDnormAccHO}{97.4}
\newcommand{\DDnormGalHO}{97.5}
\newcommand{\DDnormStarHO}{97.0}
\newcommand{\DDnormGalPrecHO}{99.1}
\newcommand{\DDnormStarPrecHO}{91.8}
\newcommand{\DDnormGalFoneHO}{98.3}
\newcommand{\DDnormStarFoneHO}{94.3}
\newcommand{\DDswAccHO}{96.2}
\newcommand{\DDswGalHO}{95.7}
\newcommand{\DDswStarHO}{97.8}
\newcommand{\DDswGalPrecHO}{99.3}
\newcommand{\DDswStarPrecHO}{86.9}
\newcommand{\DDswGalFoneHO}{97.5}
\newcommand{\DDswStarFoneHO}{92.0}
\newcommand{\DDswfAccHO}{96.1}
\newcommand{\DDswfGalHO}{95.7}
\newcommand{\DDswfStarHO}{97.4}
\newcommand{\DDswfGalPrecHO}{99.2}
\newcommand{\DDswfStarPrecHO}{86.8}
\newcommand{\DDswfGalFoneHO}{97.4}
\newcommand{\DDswfStarFoneHO}{91.8}
\newcommand{\RFAccHO}{98.5}
\newcommand{\RFGalHO}{99.2}
\newcommand{\RFStarHO}{96.1}
\newcommand{\RFGalPrecHO}{98.9}
\newcommand{\RFStarPrecHO}{97.4}
\newcommand{\RFGalFoneHO}{99.1}
\newcommand{\RFStarFoneHO}{96.7}

\newcommand{\RFagree}{94.18}

\newcommand{\Nrfcommon}{45{,}916}
\newcommand{\Nrfagree}{43{,}243}
\newcommand{\Nrfagreegal}{42{,}877}
\newcommand{\Nrfagreestar}{366}
\newcommand{\Nrfdisagree}{2{,}673}
\newcommand{\NrfdisagreeDDstar}{2{,}441}
\newcommand{\NrfdisagreeDDgal}{232}
\newcommand{\Nrfddgal}{43{,}109}
\newcommand{\Nrfddstar}{2{,}807}
\newcommand{\Nrfrfgal}{45{,}318}
\newcommand{\Nrfrfstar}{598}
\newcommand{\RFagreeHeldout}{93.6}

\shorttitle{DeepDISC-LSST-DP1}
\shortauthors{Yadav et al.}

\begin{document}

\correspondingauthor{Prince Yadav, Xin Liu}
\email{princey2@illinois.edu, xinliuxl@illinois.edu}

\title{\deepdisc{} on LSST Data Preview 1: \\ Scene-Level Detection, Deblending, and Star/Galaxy Classification \\ in the ECDFS and EDFS Fields}

\author[0009-0003-2770-5537]{Prince Yadav}
\affiliation{Department of Physics, University of Illinois at Urbana-Champaign, 1110 West Green Street, Urbana, IL 61801, USA}

\author[0009-0001-2278-8199]{Yaswant Sai Ejjagiri}
\affiliation{Department of Astronomy, University of Illinois Urbana-Champaign, 1002 West Green Street, Urbana, IL 61801, USA}
\affiliation{National Center for Supercomputing Applications, University of Illinois Urbana-Champaign, Urbana, IL 61801, USA}

\author[0000-0003-0049-5210]{Xin Liu}
\affiliation{Department of Astronomy, University of Illinois Urbana-Champaign, 1002 West Green Street, Urbana, IL 61801, USA}
\affiliation{National Center for Supercomputing Applications, University of Illinois Urbana-Champaign, Urbana, IL 61801, USA}
\affiliation{Center for Artificial Intelligence Innovation, University of Illinois Urbana-Champaign, 1205 West Clark Street, Urbana, IL 61801, USA}

\author[0009-0005-7923-054X]{Grant Merz}
\affiliation{Department of Astronomy, University of Illinois Urbana-Champaign, 1002 West Green Street, Urbana, IL 61801, USA}

\author{the LSST Dark Energy Science Collaboration}

\begin{abstract}
We present the first application of \deepdisc{}, a scene-level deep-learning framework, to real LSST data, using the Data Preview~1 (\dpone{}) coadded images of the Extended Chandra Deep Field-South and the Euclid Deep Field-South fields. In a single pass, \deepdisc{} jointly detects, segments, and classifies sources as stars or galaxies, rather than treating these tasks as separate steps as in traditional pipelines. We warm-start from a model pretrained on LSST simulations and fine-tune it on \dpone{} six-band images, using the \dpone{} catalog's extendedness flag (point-like versus extended) as provisional class labels. The classification part of the network is then fine-tuned alone on a curated sample of stars and galaxies with spectroscopic, Gaia, and photometric identifications, so the final output is a star/galaxy classification. Fine-tuning on real \dpone{} observations substantially improves source recovery relative to the un-fine-tuned model, raising galaxy and star completeness from \Cgalbase\% and \Cstarbase\% to \CgalA\% and \CstarA\% under strict class-aware matching; under class-agnostic matching the fine-tuned model reaches \CgalH\% and \CstarH\%. On a common sample of sources that neither classifier saw in training, \deepdisc{} classifies to within about one percentage point of a published photometry-based Random Forest for the same field (\DDnormAccHO\% vs.\ \RFAccHO\%); at fainter magnitudes \deepdisc{} labels more sources as stars than the Random Forest, and independent labels favor the Random Forest in most of these disagreements. Cross-matching to deeper HST/CANDELS imaging on a held-out footprint gives an unrecognized-blend fraction for \deepdisc{} of \HSTUB\% (95\% CI \HSTUBlo--\HSTUBhi\%), consistent with or modestly below the \HSTRubinUB\% we measure for the production \dpone{} catalog. This work establishes a baseline for \deepdisc{} on real LSST data and a foundation for self-supervised pretraining and for LSST Data Preview~2.

\end{abstract}


\section{Introduction}
\label{sec:intro}

The Vera C.\ Rubin Observatory Legacy Survey of Space and Time (LSST; \citealt{Ivezic2019}) will deliver an unprecedented volume of optical imaging data over the next decade. Over $\sim 18{,}000$~deg$^2$ and six LSSTCam photometric bands ($u$, $g$, $r$, $i$, $z$, and $y$; hereafter $ugrizy$), the repeat visits will be stacked into deep coadded images reaching a nominal $5\sigma$ depth of $r \sim 27.5$, on which LSST is expected to detect $\sim 20$ billion galaxies and $\sim 17$ billion stars. The traditional approach to analyzing such surveys, in which detection, deblending, photometry, classification, and photometric redshift estimation are performed sequentially by independent pipelines, faces three structural problems at LSST scale and depth: (i) crowding and blending make detection and deblending fundamentally co-dependent; (ii) bright stars produce diffraction spikes, scattered light, and saturation artifacts that contaminate large neighborhoods; and (iii) the per-object cutout abstraction throws away contextual information that is essential for distinguishing genuine sources from artifacts and from neighbors.

Blending is now the dominant systematic in deep imaging surveys. In Hyper Suprime-Cam (HSC) Wide, Deep, and UltraDeep layers, respectively 58\%, 66\%, and 74\% of detections lie in regions with multiple peaks above $5\sigma$ \citep{Bosch2018}; LSST 10-year depths are comparable to HSC UltraDeep layers \citep{Sanchez2021}. Unrecognized blends (UBs), where multiple true sources are detected as one, are particularly damaging: they bias photometry by $\gtrsim 0.75$~mag for $\sim 12\%$ of artificially blended CANDELS galaxies \citep{Boucaud2020}, increase shear-measurement noise by $\sim 14\%$ at LSST depth \citep{Dawson2016}, and propagate into cosmological-parameter shifts at the level of current Stage-III \citep{MacCrann2022} and forecasted Stage-IV \citep{Nourbakhsh2022} analyses.

LSST Data Preview~1 (\dpone{}; \citealt{LSSTDP1}), although shallower than the eventual 10-yr survey by $\sim 1.5$ magnitudes, is the first dataset in which detection and classification methods can be tested against the actual Rubin instrument and pipeline, and it includes the Extended Chandra Deep Field-South (\ecdfs{}) and the Euclid Deep Field-South (\edfs{}), the deepest extragalactic fields in the release. Two \dpone{} benchmarks frame this work. The official \dpone{} source-injection tests \citep[their \S5.6]{LSSTDP1} show that at the 80\% completeness threshold, ${\sim}20\%$ of injected sources (primarily galaxies) are misclassified as stars based on the \rsp{} \texttt{refExtendedness} parameter (the pipeline's binary morphological star/galaxy separator, derived from the difference between PSF and model magnitudes, with $0$ for point-like and $1$ for extended sources), and the fraction of correctly classified injected stars drops to ${\sim}50\%$ at $i = 23.8$~mag (corresponding to 90\% detection completeness). Complementing this, \citet{Adari2025} cross-matched \dpone{} to HST and found that 18\% of objects in \ecdfs{} are unrecognized blends, consistent with simulation-based predictions from \citet{Troxel2023,Dawson2016,Nourbakhsh2022}.

This work uses \deepdisc{} \citep{Merz2023}, a scene-level deep-learning framework that addresses these problems by detecting, segmenting, and classifying every source in an image in a single pass rather than treating these tasks as separate steps. Built on the Mask R-CNN family of detectors \citep{He2017,Burke2019}, \deepdisc{} has been demonstrated on real HSC imaging \citep{Merz2023}, on simulated LSST data with photometric-redshift estimation \citep{Merz2025}, on JWST NIRCam imaging \citep{Merz2026}, on Euclid Q1 imaging \citep{Jiang2026}, and on simulated joint Rubin+Roman scenes, where scene-level inference reduced the unrecognized-blend fraction relative to the LSST production pipeline \citep{EjjagiriMerz2026}. Until now, however, \deepdisc{} had not been applied to real LSST imaging.

This paper presents the first deployment of \deepdisc{} on \dpone{} data. We focus on the \ecdfs{} and \edfs{} fields because of their extragalactic science focus, low stellar density ($\sim 4{:}1$ galaxy-to-star ratio), and rich ancillary multi-wavelength imaging that supports future cross-survey extensions. In this paper, classification means a binary star/galaxy decision: neither of our label sources defines a quasar class, so quasars are not separated out here; we plan to add one in later work (\S\ref{sec:future}).

The paper is organized as follows. Section~\ref{sec:data} describes the \dpone{} \ecdfs{} and \edfs{} data and the tile preparation. Section~\ref{sec:gt} details the pseudo-ground-truth construction, the over-deblending diagnostic, and the multi-step filtering procedure. Section~\ref{sec:methods} summarizes the \deepdisc{} architecture, training configuration, and matching protocols. Section~\ref{sec:results} presents detection, classification, and matching performance. Section~\ref{sec:discussion} discusses limitations, comparison to traditional Rubin pipeline outputs and to the \citet{Adari2025} HST cross-match, and the implications for upcoming \dpone{} extensions and \dptwo{}. Section~\ref{sec:conclusion} concludes.

\section{Data}
\label{sec:data}

\subsection{LSST Data Preview 1 and the target fields}
\label{sec:dp1}

\dpone{}, available to Rubin data rights holders, is based on commissioning observations with the LSSTComCam camera between October 24, 2024 and January 11, 2025. It comprises $\sim 15$~deg$^2$ of imaging across seven discrete, non-contiguous fields, processed through the \rsp{} into deep coadded images and Object catalogs. We focus on the two deepest extragalactic fields in \dpone{}, namely \ecdfs{}  and \edfs{}, summarized in Table~\ref{tab:fields}.

\ecdfs{} is one of the most heavily observed extragalactic fields in \dpone{} (854 visits in total) and reaches the deepest $r$-band coadd depth in the release. It is also the field covered by the HST/CANDELS catalog used in \S\ref{sec:hst_validation}, which lets us measure completeness, purity, and the unrecognized-blend fraction against reference that does not depend on the Rubin pipeline.

\edfs{} was selected to overlap with the corresponding Euclid deep field; its band coverage is more limited, but it remains a high-priority extragalactic field. Both fields have low stellar density (galaxy:star $\sim 4{:}1$ in the \dpone{} Object catalog; $\sim 6{:}1$ in our final training annotation set), supporting deep coadditions and weak-lensing science cases. Our footprint comprises five tracts (2234, 2394, 4848, 4849, 5063) spanning the two fields; the outlying tracts 5064 and 2393, which sit on the very edge of the \dpone{} all-band footprint and contain only 166 and 25 full-band sources respectively, are excluded.
\begin{table}
\centering
\caption{LSST \dpone{} extragalactic deep fields used in this work.}
\label{tab:fields}
\renewcommand{\arraystretch}{1.50}
\setlength{\tabcolsep}{3pt}

\begin{tabular}{ccccc}
\hline\hline
 & RA & Dec & 5$\sigma$ $r$ & \# of Visits  \\
Field  & (deg) & (deg) & (mag) &  ($ugrizy$) \\
\hline
\ecdfs{} & 53.16 & $-28.10$ & 26.0 & 43, 230, 236, 162, 153, 30 \\
\edfs{}  & 59.10 & $-48.73$ & 25.5 & 20, 61, 87, 42, 42, 20\\
\hline
\end{tabular}
\textit{Note:} Coordinates are field centers; analyzed regions span $\sim 1$~deg radius around each. Depths from \citet{LSSTDP1}.
\end{table}

Rubin images are delivered on a sky tessellation of overlapping \emph{tracts}, each subdivided into \emph{patches}. Each \dpone{} patch is $3400{\times}3400$ pixels, which we cut into smaller square $760{\times}760$-pixel \emph{tiles}, the units on which \deepdisc{} is trained and evaluated (\S\ref{sec:tiles}); tiles are named \texttt{T}(tract)\_\texttt{P}(patch)\_\texttt{C}(cutout origin in pixels) throughout the paper. Each labelled object on a tile (its bounding box, pixel mask, and class) is an \emph{annotation}, the standard term for a single training label in object detection. We retrieve the six-band coadded images and Object catalogs for the \dpone{} tracts covering both fields, discard any patch that contains NaN values in any band or lacks $\geq 70\%$ valid pixel coverage in every band, and keep each remaining patch as a six-band image together with its WCS.

\begin{figure*}
\centering
\includegraphics[width=0.9\textwidth]{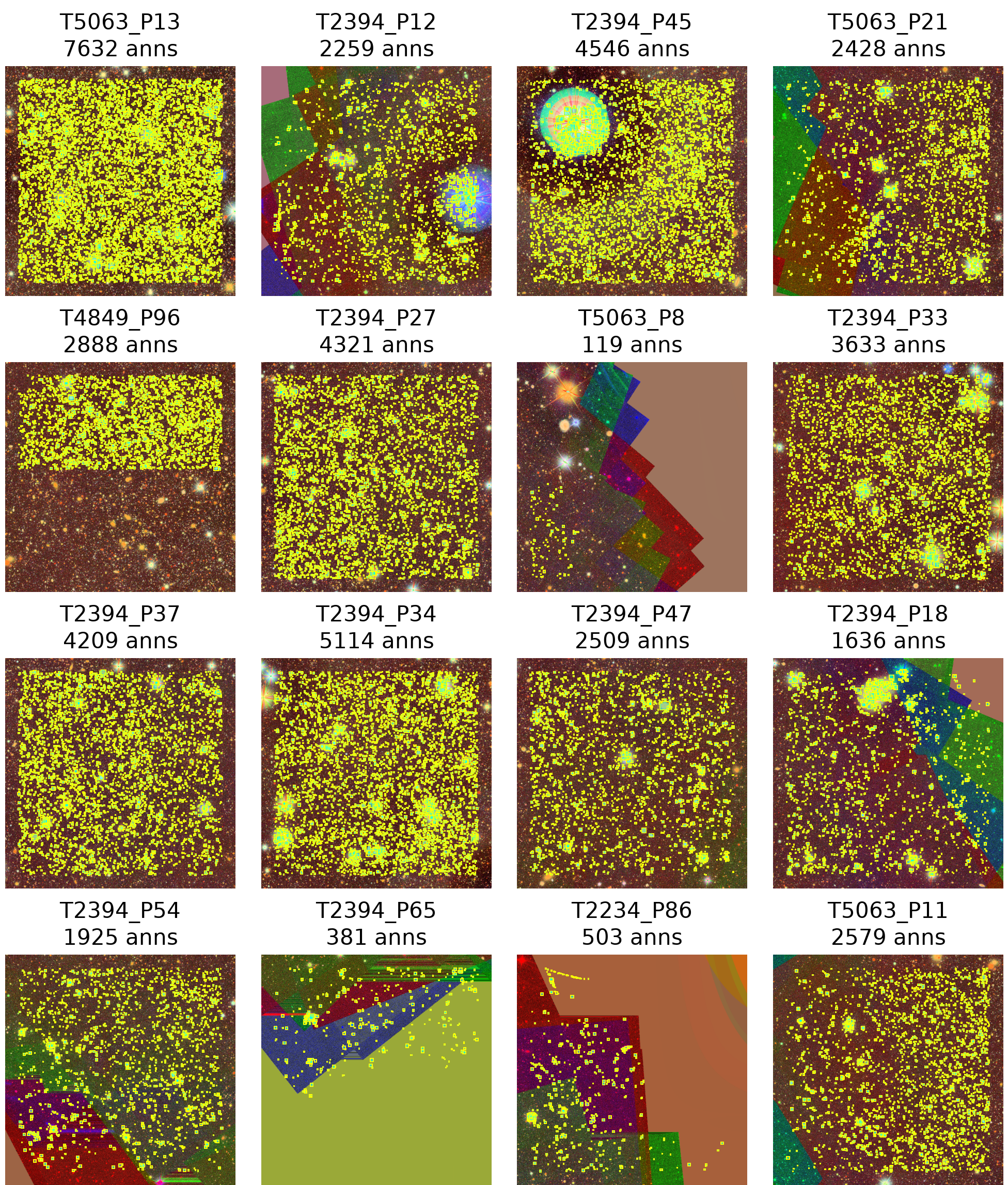}
\caption{A montage of representative \dpone{} patches (each labelled by tract/patch and annotation count;  annotations overlaid in yellow) illustrating the range of data-quality situations that motivate our curation pipeline (\S\ref{sec:gt}). Some patches are densely and fully annotated (e.g.\ \texttt{T5063\_P13} with $7{,}632$ annotations); some are only partially annotated, where our per-patch pseudo-ground truth is incomplete (e.g.\ \texttt{T2394\_P18} with $1{,}636$ annotations); some contain band-dropout regions (the false-colour areas, where one or more bands are missing and our pseudo-ground truth has no labels) (e.g.\ \texttt{T5063\_P8} with $119$ annotations); and several are contaminated by bright stars with saturated cores and diffraction spikes (e.g.\ \texttt{T2394\_P45} with $4{,}546$ annotations). If used unfiltered, such regions inject inconsistent or missing pseudo-ground-truth labels that destabilise training; these are precisely the cases handled by the band-dropout/fill rejection, bright-star masking, and annotation-density and image-quality gates of \S\ref{sec:gt}.}
\label{fig:annotation_overview}
\end{figure*}

\subsection{Normalization and tile preparation}
\label{sec:tiles}

We apply a per-band normalization arcsinh stretch ($\mathrm{asinh}(\mathrm{flux}/b)$ with $b = 0.1$~nJy on the native coadd pixel values) followed by per-tract/per-patch $z$-score normalization: we verified on a 50-image sample that all six normalized bands have mean $\approx 0$ and standard deviation $\approx 1$. This $b$ is far below the per-pixel sky noise (2.6--46~nJy across $ugrizy$). Retraining the headline model with $b$ set to the sky noise, holding the rest of the pipeline unchanged, changes overall completeness by $\sim 0.25\%$ but changes star completeness by 0.9--3.0 percentage points across configurations. 
We keep $b = 0.1$~nJy, the setting under which every model in this paper was trained. The weak overall sensitivity agrees with \citet{Merz2023}, who found the \deepdisc{} architecture to be robust to contrast scaling on instance-segmentation tasks (unlike ResNet architectures, whose detection accuracy varied substantially between $z$-scale and Lupton \citep{Lupton2004} contrast scalings).

Before tiling we apply a 200-pixel border crop to each normalized patch, reducing the $3400{\times}3400$ frame to $3000{\times}3000$ and updating the WCS accordingly. Neighbouring patches overlap, and each source in the overlap is listed in the Object catalog of only one of them. A source that lies in the outer 200 pixels of our patch but is listed under the neighbouring patch therefore has no catalog entry, and hence no annotation, in our patch: it is a real object with no label. The 200-pixel border crop removes this outer strip so that sources in patch overlaps do not appear unlabelled in training images; across the combined run it retains $346{,}798$ of $348{,}661$ annotations ($99.47\%$). Tiling marks the boundary between patch-level and tile-level processing (Table~\ref{tab:curation}): every step so far acts on whole patches, and every filter from here on acts on individual tiles within a split. We slice each cropped $(6, 3000, 3000)$ patch into overlapping $(6, 760, 760)$ tiles with a 560-pixel stride (200-pixel overlap). Each annotation consists of a rectangular bounding box and a pixel-mask outline for one source (their construction is described in \S\ref{sec:scarlet_gt}); annotations that cross a tile boundary are clipped to the tile, and any that are left with zero area are dropped. We additionally apply a tile-level annotation-spread filter that discards tiles whose annotation centroids do not span more than 50\% of the tile extent in both axes, removing sparsely-populated edge tiles. The same tiling configuration is used for training, validation, and test splits; the model is fully convolutional in the spatial dimensions and supports variable-size inference, as also exploited in the HSC analysis of \citet{Merz2023}. The patch-level training/validation/test split (70/11/20 of \Npatches{} patches, random seed 42) followed by tiling and the spread filter yields \Ntilestrainraw{} training, \Ntilesvalraw{} validation, and \Ntilestestraw{} test tiles; the tile-level filtering of \S\ref{sec:filter} then reduces these to the final image budget summarized in Table~\ref{tab:tiles}. Figure~\ref{fig:cutout_tiling} (Appendix~\ref{app:tiles}) illustrates this tiling for an example patch, including which cutouts are kept versus dropped.

Tile-level data quality varies substantially across these fields (Figure~\ref{fig:annotation_overview}): annotation coverage is incomplete in places, some regions lack one or more bands entirely, and bright stars contaminate their neighbourhoods. Training on unfiltered tiles would inject inconsistent or missing labels, so we apply a multi-stage curation procedure, described in full in \S\ref{sec:gt}. After curation the analysed dataset comprises \Nobjall{} objects (\Ngalall{} galaxies, \Nstarall{} stars) on \Ntilestot{} tiles (Table~\ref{tab:tiles}).

\begin{table}
\raggedright
\caption{Final image-tile budget for the \ecdfs{}+\edfs{} run after tile-level filtering. Tile and object counts are both measured on the same final tile set, after fill-value masking, bright-star masking, the minimum-annotation density filter, and the final two-gate image-quality pass (\S\ref{sec:tiles}); the immediate counts before that final pass are \NtilestrainPre{}/\NtilesvalPre{}/\NtilestestPre{} (\Ntilestotpre{} total), and the counts before any tile-level filtering are \Ntilestrainraw{}/\Ntilesvalraw{}/\Ntilestestraw{}. Galaxy/star counts are the extended/point-like classes of the \dpone{} catalog's \texttt{refExtendedness} flag carried by the annotations (\S\ref{sec:scarlet_gt}); the labels are provisional and are replaced for classification in Stage 2 (\S\ref{sec:staged_training}).}
\label{tab:tiles}
\renewcommand{\arraystretch}{1.25}
\setlength{\tabcolsep}{14pt}

\normalsize

\begin{tabular}{cccc}
\hline\hline
Split & Tiles & Galaxies & Stars \\
\hline
Training   & \Ntilestrain{} & \Ngaltrain{} & \Nstartrain{} \\
Validation & \Ntilesval{}   & \Ngalval{}   & \Nstarval{} \\
Test       & \Ntilestest{}  & \Ngaltest{}  & \Nstartest{} \\
\hline
\end{tabular}
\textit{Note.} Counts are unique objects, deduplicated by \texttt{objectId}, on the final \Ntilestot{}-tile set, i.e.\ after the \Ntilesdrop{} tiles removed by the image-quality pass (predominantly partial band coverage) have been excluded. Galaxy:star ratio $\sim 6{:}1$ in the training annotation set.
\end{table}

\section{Pseudo-Ground-Truth Construction}
\label{sec:gt}

Table~\ref{tab:curation} (Appendix~\ref{app:curation}) lists every curation step of \S\ref{sec:data} and this section in the order applied, labelled as patch-level or tile-level, with its parameters and the tile and object counts at each stage.

\subsection{Scarlet-based pseudo-ground truth}
\label{sec:scarlet_gt}

We use \scarlet{} \citep{Melchior2018} as our pseudo-ground-truth source for deblending. Throughout, \emph{pseudo-ground truth} denotes the labels we construct from \rsp{} products, rather than from external validation, to train the detector (Stage 1 of the two-stage scheme of \S\ref{sec:staged_training}): per-object masks and bounding boxes from \scarlet{}, each carrying the \dpone{} Object catalog's \texttt{refExtendedness} class, which separates extended from point-like sources and which we call galaxy and star for brevity. Where the context is unambiguous we shorten pseudo-ground truth to \emph{ground truth}. \scarlet{} performs joint multi-band constrained matrix factorization with priors on monotonicity, symmetry, and PSF-matched morphology, and is the production deblender within the \rsp{}. For each patch, we run \scarlet{} through the \rsp{} \texttt{meas\_extensions\_scarlet} package to retrieve source footprints, segmentation masks, and peak positions, and convert these into bounding-box and instance-mask annotations in the COCO format \citep{Lin2014} used by our training pipeline. We adopt a strict lower threshold of $2\sigma$ for mask generation. The choice of $2\sigma$ as the lower threshold was validated by an Otsu multi-threshold analysis in both pixel-intensity and per-object signal-to-noise-ratio (SNR) space, which selects $\sim 2\sigma$ as the natural background/source separator (consistent across all six bands). Each annotation inherits its provisional class from the \texttt{refExtendedness} value of the corresponding source in the \dpone{} Object catalog (extended, labelled galaxy; point-like, labelled star). A representative pair of these pseudo-ground-truth tiles, with the \scarlet{} annotations overlaid exactly as fed into \deepdisc{} Stage-1 (detection) training, is shown in Appendix~\ref{app:gt_samples} (Figure~\ref{fig:gt_samples}).

Any biases in the upstream deblender (under-deblending, over-deblending, or classification errors) are therefore inherited by the model, as \citet{Merz2023} found for HSC. Section~\ref{sec:overdeblend} characterizes a specific over-deblending failure mode and Section~\ref{sec:matchedset} introduces matched-set metrics to mitigate the impact of label imperfection on reported performance.

\subsection{Over-deblending diagnostic}
\label{sec:overdeblend}

\begin{figure}[htbp]
\centering
\includegraphics[width=0.45\textwidth]{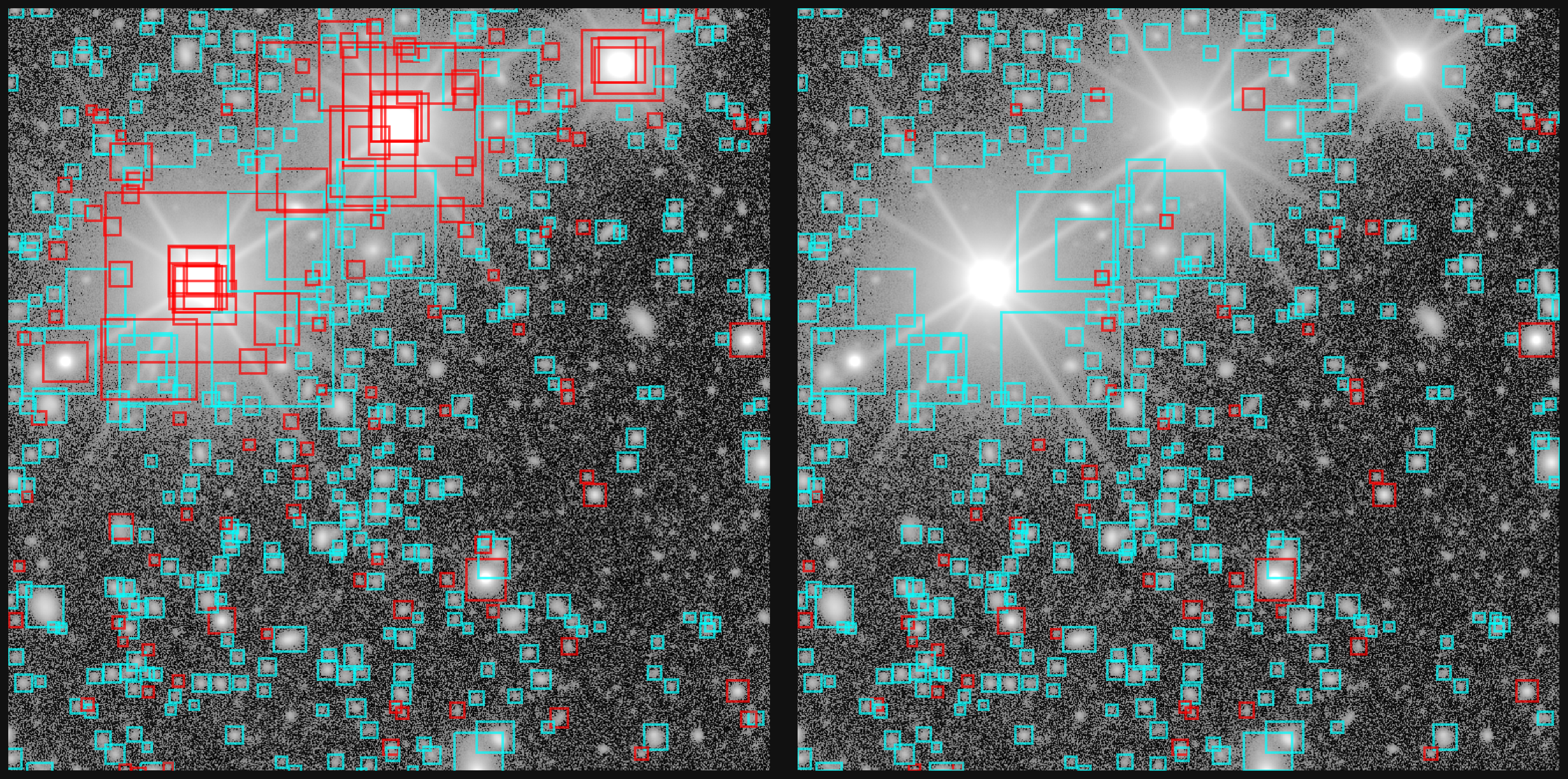}
\caption{Before (left) and after (right) two-step annotation filtering on tile \texttt{T5063\_P14\_C\_0\_1680}, centered on three saturated bright stars. Boxes show \scarlet{} annotations (cyan: galaxy; red: star). The raw \scarlet{} catalog (left, 450 annotations) contains dense clusters of spurious \texttt{star} boxes shredding each bright-star core; the two-step cleanup removes the spurious star boxes (red) while leaving galaxy boxes (cyan) untouched, yielding the training pseudo-ground truth (right, 372 annotations). The cleanup is parametrized by the inner and outer exclusion zones described in \S\ref{sec:filter}, so only the regions around the bright stars are affected and all other annotations are left entirely untouched, as the full tile shows.}
\label{fig:filter_before_after}
\end{figure}

During visual inspection of the training set, we identified a recurrent failure mode in which \scarlet{} pseudo-ground truth contains many overlapping components on individual bright or saturated sources. A representative example is a single bright star associated with $\sim 15$ overlapping bounding boxes labeled \texttt{star}, all clustered within the saturated core; Figure~\ref{fig:filter_before_after} shows such a case, where the cleanup described below reduces a tile centred on three bright stars from 450 annotations to 372. We refer to this artifact as ``shredding'' following the deblending literature \citep[e.g.,][]{Bosch2018}. Analogous \sextractor{}+\scarlet{} shredding around bright stars and extended galaxies was reported by \citet{Merz2023} on HSC, where it degraded detection performance for the largest objects and motivated a magnitude-difference cut at the labeling stage.

If left unfiltered, these annotations train \deepdisc{} to predict many overlapping spurious detections at every bright star, severely degrading classification metrics on the test set. We therefore apply a two-step filtering procedure described in the next subsection. Even after this cleanup, at test time \deepdisc{} still returns more components than \scarlet{} for most complex groups; that residual \scarlet{}-anchored over-deblending is characterized in \S\ref{sec:blend_results}.

\subsection{Two-step annotation filtering}
\label{sec:filter}

Our filtering pipeline is intentionally asymmetric in class: galaxy annotations are never directly modified in the following two steps, because the shredding signature is specific to stars. Over the combined Deep-Fields \scarlet{} catalog the stellar SNR distribution carries a pronounced high tail (the median is $4$, but the 90th percentile is $163$ and the 99th is $1409$), which is what fragments of a saturated core produce. Stars are therefore subject to two steps of cleanup that target known shredding modes around bright/saturated sources.

\paragraph{First filter (patch level): percentile cut on the raw \scarlet{} catalog.}
We compute per-class distributions of bounding-box width, height, area, and signal-to-noise ratio (SNR) over the entire \scarlet{}-annotated catalog. For stars only, we then remove any annotation that exceeds the 98th percentile of the width, height, or area distributions, or the 99th percentile of the SNR distribution. The intuition is that the shredded components of a saturated star are simultaneously high-SNR and large in area; legitimate stars rarely exceed the joint cut. Over the combined Deep-Fields \scarlet{} catalog these percentiles fall at width/height $45$~pixels, area $1980$~pixel$^2$, and SNR $1409$. Galaxies are untouched by this filter.

\paragraph{Second filter (tile level): two-zone bright-star mask.}
After tiling (\S\ref{sec:tiles}), we apply a per-tile bright-star sweep targeted at PSF halos and diffraction-spike contamination. For each surviving star annotation with area $> 1600$~pixel$^2$ we define an effective radius $r_{\rm eff} = \sqrt{\mathrm{area}/\pi}$ and two concentric exclusion zones:
\begin{itemize}
\item \emph{Inner zone} ($\leq 0.6\,r_{\rm eff}$): all other star annotations within this radius are removed unconditionally;
\item \emph{Outer zone} ($\leq 2.0\,r_{\rm eff}$): star annotations are removed only if their footprint overlaps the bright star's area by more than 1\% (intersection-over-self criterion).
\end{itemize}
Galaxy annotations are never removed by either zone.

\paragraph{Other tile-level filters.}
Because the training/validation/test split is made at patch level before tiling (\S\ref{sec:tiles}), the tile-level filters act on each split separately; we quote totals here and give the per-split counts in Table~\ref{tab:tiles}. Some pixels carry a no-data placeholder value rather than a measurement; we identify that value in each band as the most common pixel value in the tile and drop any object label falling inside such a region. We further require each retained tile to contain at least 50 valid annotations after cleanup. Tiling and the spread filter yield \Ntilesrawtot{} tiles, on which the following filters act. Fill-value and artifact masking removes \Ntilesfillmask{} of these, and the minimum-annotation filter a further two. A final image-quality pass then applies two gates to every remaining tile: rejection of low-dynamic-range tiles (mean Lupton standard deviation below $0.003$) and rejection of tiles with partial band coverage (any band with standard deviation below $10\%$ of the per-tile median). This pass removes \Ntilesdrop{} tiles (\Ntilesdropdr{} low-dynamic-range, \Ntilesdropcov{} partial-coverage; two tiles fail both gates), leaving \Ntilestot{} tiles (Table~\ref{tab:tiles}). Tile-level filtering therefore discards \Ntilesqcdrop{} of the \Ntilesrawtot{} candidate tiles ($20\%$) in total; the full funnel and the losses at each step are given in Appendix~\ref{app:curation} (Table~\ref{tab:curation}). The gate is driven by the shallowest bands: \NuOnly{} tiles are rejected on $u$ alone and would be retained if $u$ were exempted (per-band breakdown in Table~\ref{tab:curation}); we keep all six bands on an equal footing so that every training tile carries usable signal in all six. The counts above are in three non-interchangeable units: patch-level annotations (\Nanncrop{}), tile-level appearances (\Nannfinaltot{}, since sources recur across overlapping tiles), and unique objects (\Nobjall{}; Table~\ref{tab:tiles}). Appendix~\ref{app:tiles} shows examples of retained and removed tiles (Figure~\ref{fig:final_kept_removed}) and the three principal rejection categories (Figure~\ref{fig:qc_reasons}).

\paragraph{Quantitative impact.}
The first filter removes \NstarStageOneRem{} of the \NstarScarlet{} \scarlet{} star annotations and the second a further \Nremoved{} at \Nbrightstar{} bright-star centres (Table~\ref{tab:curation}). Although the absolute removal is small, these are precisely the annotations that would otherwise teach the network to emit spurious detections at every bright star (\S\ref{sec:overdeblend}). We retain the filtered annotations in a separate auxiliary catalog so that downstream qualitative evaluation of \deepdisc{} predictions in those regions remains possible.

We note that the percentile cuts are computed over the combined multi-tract Deep-Fields catalog and the zone radii are empirical; before scaling to additional fields or to \dptwo{} we plan to validate threshold choices against an independently-curated subset of the \rsp{} \scarlet{} outputs and to consider replacing the post-hoc filter with model-driven alternatives: \texttt{scarlet2} \citep{Sampson2024}, for example, replaces \scarlet{}'s hand-designed monotonicity and symmetry constraints with a learned source-morphology prior, which should suppress shredding at source rather than after the fact.

\paragraph{Band-dropout regions.}
At the survey edge some tiles contain regions where one or more $ugrizy$ bands are absent. \scarlet{} produces no annotations there because the upstream catalog and deblender do not cover them, not because of a band-count requirement (\scarlet{} imposes none, and $0.15\%$ of our annotations are built from a single band). Left in, such regions would add detections with no pseudo-ground-truth counterpart and bias the reported purity low. They are removed by the fill-value masking above, which drops tiles dominated by no-data pixels (53 of the \Ntilestestraw{} raw test tiles, $\sim 14\%$) and discards labels inside the residual no-data pixels, which cover $<0.1\%$ of the retained tiles. All metrics in \S\ref{sec:results} are therefore computed on the retained tiles.

\section{Methods}
\label{sec:methods}

\subsection{Architecture}
\label{sec:architecture}

We use the \deepdisc{} framework \citep{Merz2023,Merz2025} implemented in the open-source \detectron{} library \citep{wu2019detectron2} with a Cascade Mask R-CNN meta-architecture, i.e.\ Mask R-CNN \citep{He2017} whose box-refinement head is applied three times in sequence at progressively stricter overlap thresholds \citep{Cai2018}. The Generalized R-CNN structure consists of three components: (i) a per-image feature-extraction backbone, (ii) a Region Proposal Network (RPN) that generates candidate object regions from anchored boxes on the backbone feature maps, and (iii) region-of-interest (ROI) heads that refine those proposals and emit per-object predictions \citep{Merz2023}. A \emph{head} is a small sub-network applied to each candidate object: the classification head returns its class, the box head a refined bounding box, and the mask head a pixel mask. The backbone in this work is a \swinb{} vision transformer \citep{Liu2021Swin} (the ``Base''-size Swin model, whose attention is computed within shifted local windows so that it scales to large images) with a feature pyramid network \citep{Lin2017} for multi-scale feature extraction. \swinb{} is used in its standard configuration \citep{Liu2021Swin} and has $86.8$~M parameters; with the FPN, RPN, and ROI heads, the full detector has $139$~M parameters, $48$~M of them in the cascade box and mask heads. For each proposal, \texttt{RoIAlign} \citep{He2017} samples the feature map at a regular grid of points inside the box, using bilinear interpolation so that no rounding to the pixel grid occurs, and returns a $7{\times}7{\times}256$ feature patch ($14{\times}14{\times}256$ for the mask head); every proposal, whatever its size, therefore reaches the heads in the same format, and the operation has no learnable parameters. These patches are passed to a cascade of three classification + bounding-box regression heads at IoU thresholds $\{0.5, 0.6, 0.7\}$ \citep{Cai2018} (intersection-over-union being the overlap area of a predicted and a true bounding box divided by their combined area), each stage re-pooling from the boxes refined by the previous one, plus a mask head for instance segmentation. Figure~\ref{fig:stage1_train} illustrates this architecture and the end-to-end Stage-1 data flow, from the six-band tile through the \swinb{} backbone / FPN pyramid, the RPN, and \texttt{RoIAlign} to the per-object classification, bounding-box, and mask outputs. For a $760{\times}760$-pixel tile at $0.2\arcsec$ per pixel, the $4{\times}4$-pixel patch embedding yields a token grid of about $190{\times}190$, the only point at which the image itself is read; the four \swinb{} stages then halve the grid three times (strides 4, 8, 16, and 32 pixels, with attention restricted to shifted $7{\times}7$-token windows), and the FPN combines their outputs into five maps, P2--P6, at strides 4--64 pixels, i.e.\ $0.8\arcsec$--$12.8\arcsec$ per cell.

Anchors are the starting boxes from which the RPN works, and they are not learned: at every cell of every pyramid level, three boxes with aspect ratios $1{:}2$, $1{:}1$, and $2{:}1$ are laid down, with base sizes of 8, 16, 32, 64, and 128 pixels on P2--P6 (the \deepdisc{} setting, in place of \detectron{}'s 32--512-pixel defaults, which would be larger than most \dpone{} sources). The RPN, one small network shared across the levels, gives every anchor an objectness score and a box correction; in training, an anchor is a positive example if it overlaps a truth box with an IoU of at least $0.7$, or is the best-overlapping anchor for that box, and a negative example below $0.3$. All five levels are used, but not equally. Since 97\% of our annotation boxes are 8--32 pixels on a side (median 14 pixels), the 8-, 16-, and 32-pixel anchors of P2--P4 cover nearly all sources, and the 64- and 128-pixel anchors of P5--P6 are left to the few bright stars and large galaxies, up to $\sim 200$ pixels. The ROI heads then pool their features from P2 for essentially every object, because 99.99\% of the boxes lie below the 112-pixel size at which \detectron{} assigns a box to the next level; the coarser levels reach the detections mainly through the context that the FPN folds into P2.

We adopt the region-based Mask R-CNN family on which \deepdisc{} is built; three properties of the data make it a natural fit. Deblending requires a pixel mask for every object, not only a bounding box, which Mask R-CNN provides natively \citep{He2017}. A \dpone{} tile holds $\sim 300$ sources, most of them 8--32 pixels across, so detection has to operate on a fine feature map with small anchors, which the FPN and the anchor settings described above provide. The cascade's staged box refinement at IoU $0.5$, $0.6$, and $0.7$ \citep{Cai2018} is designed to tighten localization, which matters when neighbours are a few pixels apart. The same family has been validated on astronomical images by \citet{Burke2019}, \citet{Merz2023,Merz2025}, and \citet{Jiang2026}, on whose framework and checkpoints this work builds directly. The choice of \swinb{} follows the backbone comparison of \citet{Merz2023}, who evaluated six ResNet-family and two transformer-family configurations (\swinb{} and \mvit{}; \citealt{li2022mvitv2}) on real HSC Deep+UltraDeep imaging and found the transformers ahead on nearly every metric: galaxy average precision (AP, the standard detection score: the purity of the detections averaged over completeness, at a range of IoU match thresholds) rose from $\sim 46$ to $\sim 53$ and star AP from $\sim 26$ to $\sim 36$, with a median bounding-box IoU of $0.94$ on matched objects. \citet{Jiang2026} adopted the same backbone for \deepdisc{} on Euclid Q1. \mvit{} was marginally ahead at fixed compute; the decisive factor here is the availability of an in-domain \swinb{} checkpoint (\S\ref{sec:multiband}), and the controlled \mvit{} ablation is deferred to \S\ref{sec:future}.

We omit the photometric-redshift ROI head used in \citet{Merz2025}, which is the subject of a companion paper (Krishnan et al. in prep). The total loss is the sum of RPN classification + bounding box, ROI cascade classification + bounding box (per stage), and mask losses, following the standard \detectron{} formulation (see \citealt{Merz2025} eq.~1, with the redshift term omitted).

\subsection{Multi-band input}
\label{sec:multiband}

\dpone{} provides six-band coadd images. Following \citet{Merz2025}, we adapt the \swinb{} patch embedding from 3 to 6 input channels with a 6-channel patch-embedding convolution, inherited directly from the six-band LSST checkpoint below (for the ImageNet-21k comparison of Appendix~\ref{app:init}, the 3-channel weights are duplicated across the two band triplets); downstream layers are unchanged because they operate on token embeddings independent of channel count. The arcsinh stretch + $z$-score normalization (\S\ref{sec:tiles}) is applied at runtime; we do not apply explicit Lupton or asinh-RGB scaling at the model input, since \citet{Merz2023} found the \swinb{} backbone to be robust to scaling choices on instance-segmentation tasks. All models reported in the main text are initialized from an in-domain checkpoint, a \swinb{} Cascade Mask R-CNN pretrained for 50 epochs on $\sim 30{,}000$ simulated LSST DC2 scenes following \citet{Merz2025}, which we refer to throughout as the \emph{LSST checkpoint}. The checkpoint warm-starts the backbone, FPN, RPN, and mask head; the three cascade classification and box-regression heads are re-initialized and trained from scratch in Stage 1 (\S\ref{sec:training}), and Stage 2 starts from the Stage-1 result. Appendix~\ref{app:init} compares this with an ImageNet-21k initialization.

\subsection{Hybrid label sources and staged training}
\label{sec:staged_training}

Our supervisory signal combines two label sources, each used where it is most reliable: the \dpone{} catalog's \texttt{refExtendedness} supervises the Stage-1 star/galaxy classes, and the \citet{Gatto2026} curated sample supersedes it in Stage 2. This mirrors the strategy of \citet{Jiang2026} on Euclid Q1 (Euclid MER segmentation maps for detection and deblending, DESI DR1 spectroscopy for classification).
\begin{itemize}
\item \emph{Stage-1 labels (detection, segmentation, and provisional star/galaxy classes)}: \scarlet{}-derived per-object pixel masks (\S\ref{sec:scarlet_gt}), each carrying the \texttt{refExtendedness} class of the original \dpone{} Object catalog. The masks are pixel-accurate by construction and are used by the production \rsp{}, but the \texttt{refExtendedness} classes degrade beyond $i \approx 24$~mag \citep{Bosch2018,LSSTDP1}.

\item \emph{Stage-2 classification labels}: the curated star/galaxy sample of \citet{Gatto2026} in \ecdfs{}, 15{,}611 sources (11{,}250 galaxies, 4{,}361 stars) drawn from spectroscopic surveys, Gaia DR3, and photometric stellar identifications. These labels are independent of the upstream deblender and are not subject to the \texttt{refExtendedness} ceiling at $i \gtrsim 24$~mag.
\end{itemize}

Exact Rubin \texttt{objectId} matching assigns \citet{Gatto2026} labels to 8{,}370 \scarlet{} sources on the Stage-1 training tiles (6{,}151 galaxies, 2{,}219 stars) and 3{,}934 on the held-out test tiles (3{,}132 galaxies, 802 stars), together 78.8\% of the 15{,}611-source catalog (magnitude distribution in Figure~\ref{fig:label_mag_dist}). \citet{Gatto2026} sources that fall on a tile but have no \texttt{objectId} match are added as synthetic annotations (961 on the training tiles, 271 on the test tiles), so that every detectable \citet{Gatto2026} source in a split is represented: 9{,}331 sources on the training tiles and 4{,}205 on the test tiles.

To prevent evaluation leakage, the Stage-2 split follows the Stage-1 tile geography rather than a random source-level partition: sources on the training tiles form the Stage-2 training set, sources on the held-out test tiles form the evaluation set, and a boundary guard on the 78 sources detected in both regions ensures the two sets share no \citet{Gatto2026} source.

\paragraph{Provenance of the stellar labels.}
The \citet{Gatto2026} stellar labels are heterogeneous: Gaia DR3 identifications dominate at $r \lesssim 22$~mag, while the faint end, where \texttt{refExtendedness} fails ($i \gtrsim 24$~mag), rests on photometric classifications from SIMPLE \citep{Damen2011}, COMBO-17 \citep{Wolf2004}, and DESI Legacy \citep{Duncan2022} with a stellar purity of $\sim$85\% \citep{Gatto2026}. Our 2{,}635-star Stage-2 training sample is accordingly $\sim$16\% spectroscopic, $\sim$47\% Gaia DR3, and $\sim$37\% photometric (median $r \approx 22.6$ for the photometric stars), with a magnitude distribution comparable to \citet[their Figure~1]{Gatto2026} apart from a slight faint-end deficit from our restriction to \scarlet{}-detected sources (Figure~\ref{fig:label_mag_dist}).

\begin{figure*}[htbp]
\centering
\includegraphics[width=\textwidth]{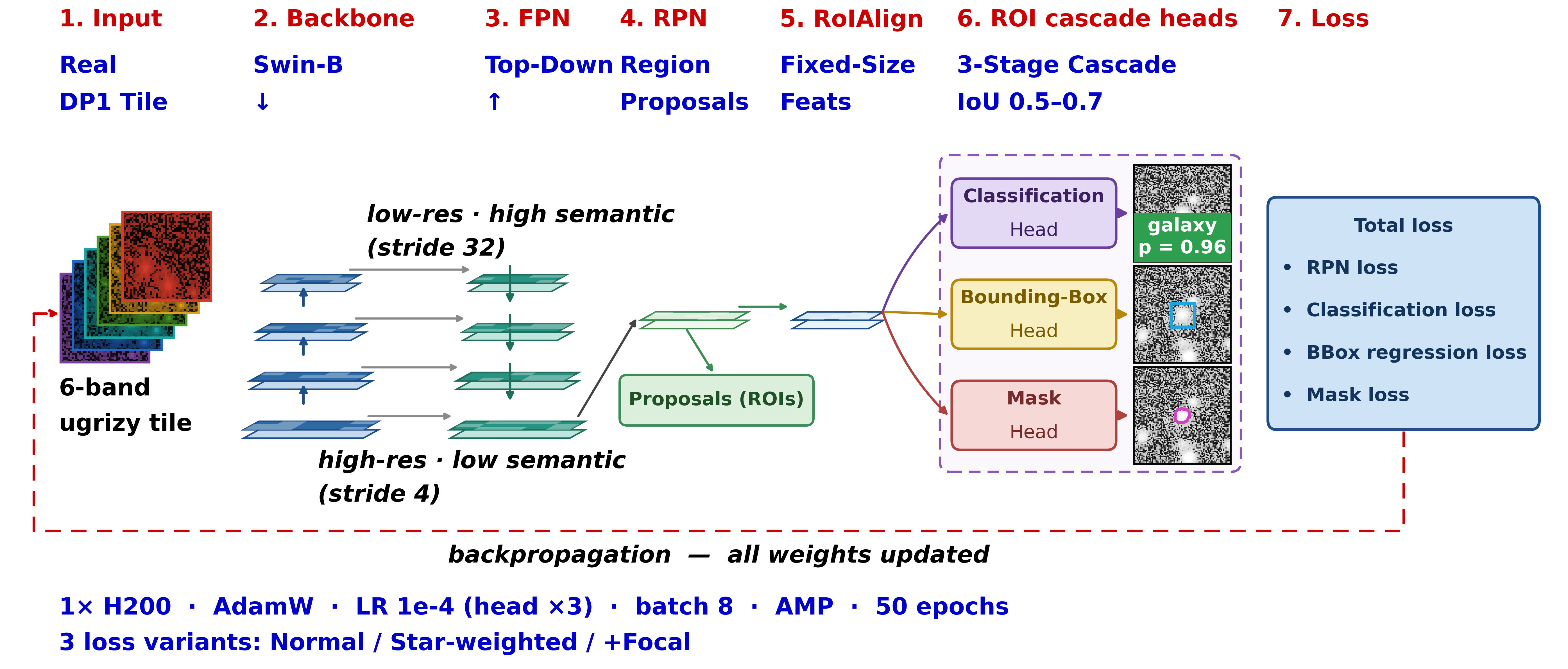}
\caption{\textbf{Stage-1 training: joint detection, deblending, and star/galaxy
classification.} The seven numbered stages trace a single forward pass, illustrated
on a real \dpone{} \ecdfs{} tile. \emph{(1)}~The six-band deep-coadd cutout enters
as a six-channel image (shown as the tinted band stack). \emph{(2)}~The \swinb{}
backbone progressively downsamples it into a feature hierarchy (spatial stride
$4\!\rightarrow\!32$; low-resolution but high-semantic at the top), and \emph{(3)}~a
feature pyramid network (FPN) rebuilds a high-resolution, semantically-rich pyramid
through a top-down pathway with lateral connections. \emph{(4)}~The region proposal
network (RPN) proposes candidate objects (ROIs), and \emph{(5)}~\texttt{RoIAlign}
pools a fixed-size feature patch for each ROI. \emph{(6)}~A three-stage ROI cascade
(IoU thresholds $0.5/0.6/0.7$) feeds the classification, bounding-box, and mask
heads; the panels show these outputs for one genuinely detected source: the
predicted class probability ($\mathrm{galaxy}$, $p{=}0.96$), the bounding box, and
the instance-segmentation mask. \emph{(7)}~The total loss sums the two RPN terms (objectness binary cross-entropy
and smooth-$L_1$ anchor regression), the per-cascade-stage terms (softmax
cross-entropy for classification and smooth-$L_1$ box-delta regression), and the
per-pixel binary-cross-entropy mask term, and is back-propagated through \emph{all}
weights. The network is warm-started from the LSST checkpoint
(\S\ref{sec:multiband}; backbone, FPN, RPN, and mask-head weights loaded, cascade box heads
reinitialized) and fine-tuned end-to-end on \scarlet{}-derived box and mask labels
and catalog-\texttt{refExtendedness} class labels (provisional; \S\ref{sec:scarlet_gt}). The three class-imbalance
variants share this identical architecture and schedule and differ \emph{only} in
the ROI classification term: plain softmax cross-entropy (``normal''), a
star-reweighted cross-entropy (star weight $\sqrt{N_{\rm gal}/N_{\rm star}}$), and a
star-weighted softmax focal loss ($\gamma{=}2$); see \S\ref{sec:training}.}
\label{fig:stage1_train}
\end{figure*}

We adopt a two-stage training protocol that adapts the staged approach of \citet{Jiang2026}, who train detection and classification jointly and freeze the backbone only when adding photometric-redshift heads; here the second stage is instead the classification step itself:
\begin{enumerate}
\item \emph{Stage 1 (detection)}: starting from the LSST checkpoint (\S\ref{sec:multiband}), we train the full network with all weights free, using the \scarlet{}-derived bounding-box and instance-mask labels together with the catalog \texttt{refExtendedness} classes (provisional; superseded in Stage 2).  This stage establishes the detection and segmentation features, and corresponds to the ``no class weight'' configuration reported throughout. The wall-clock time on NVIDIA H200 GPU is $<$ 2 hrs for Stage 1 training.
\item \emph{Stage 2 (classification head fine-tune)}: we freeze the backbone, FPN, RPN, and detection-cascade heads from Stage 1, and using the \citet{Gatto2026} class labels of the sources located on the Stage-1 \emph{training} tiles (the geographic split above), we finetune only the classification head. \scarlet{}-only annotations are masked from the classification loss in this stage. This stage is computationally cheap compared to Stage 1 because only the classification head is updated: the wall-clock time on NVIDIA H200 GPU is $<$ 15 min for Stage 2 classification head finetuning
\end{enumerate}
The two-stage protocol is summarized schematically in Figure~\ref{fig:stage1_train} (detection) and, in the identical layout, Figure~\ref{fig:stage2_train} (classification-head fine-tuning; Appendix~\ref{app:stage2_schematic}).

This protocol preserves the detection and instance-segmentation performance demonstrated in Stage 1, while replacing the catalog \texttt{refExtendedness} class labels (noisy at faint magnitudes) with the curated \citet{Gatto2026} labels wherever they are available. We report Stage-1 detection metrics (Tables~\ref{tab:performance}, \ref{tab:performance_hung}) and Stage-2 classification metrics (Table~\ref{tab:gatto_compare}) separately, so that each set of metrics is reported against the cleanest available ground truth. As an ablation, we also report Stage-1 metrics with these \texttt{refExtendedness} labels as a sanity check, against which the Stage-2 numbers should be compared.

\subsection{Training configuration}
\label{sec:training}

We train Stage 1 (Figure~\ref{fig:stage1_train}) for 50 epochs with the AdamW optimizer at a base learning rate of $10^{-4}$ (with a $3{\times}$ multiplier on the ROI-head learning rate), batch size 8, and AMP mixed-precision, warm-started from the LSST checkpoint (\S\ref{sec:multiband}). We apply a linear warmup followed by step decays at 50\%, 75\%, and 90\% of training. Gradient clipping with norm 1 is applied to stabilize early training of the freshly-initialized cascade box heads. We employ standard spatial augmentations (random flips, $90^\circ$ rotations) but no Gaussian blur or noise injection, since the training data are already PSF-convolved \citep{Merz2023}. The training set comprises \Nanntrain{} \scarlet{}-derived annotations across the \Ntilestrain{} training tiles, 303 per tile on average and at most \Nannmaxtile{}. The RPN's proposal budget is set from this density in two steps. First, on each pyramid level it keeps the $3{\times}\Nannmaxtile{} = 1{,}668$ anchors with the highest objectness scores; on P2 this is the top $1.5\%$ of about $10^5$ anchors, whereas the coarsest level has fewer anchors than the cap, so the large sources proposed there are never cut. Non-maximum suppression (NMS) then removes any box that overlaps a higher-scoring one at IoU above $0.7$. Second, the best $2{\times}\Nannmaxtile{} = 1{,}112$ proposals from all levels together are passed to the ROI heads, which is two to four candidate boxes for every true source, even in the densest tile.

We additionally run two Stage-1 class-imbalance variants to test whether the $\sim 6{:}1$ galaxy-to-star ratio in our training annotation set (\Nobjtrain{} objects: \Ngaltrain{} galaxies, \Nstartrain{} stars; Table~\ref{tab:tiles}) is suppressing star recall:
\begin{itemize}
\item \emph{Star-weighted} run: the ROI classification loss is reweighted by a star class-weight factor of $\sqrt{N_{\rm gal}/N_{\rm star}}$.
\item \emph{Star-weighted + focal} run: the $\sqrt{N_{\rm gal}/N_{\rm star}}$ star weight is combined with a softmax focal loss ($\gamma = 2.0$) on the ROI classification head, replacing the default cross-entropy.
\end{itemize}
All three Stage-1 runs (no class weight, star-weighted, star-weighted~$+$~focal) use the same architecture, optimizer, and schedule; only the loss layer differs, and no repeat-factor sampling is used. Training is performed on a single NVIDIA H200 GPU with activation checkpointing enabled to reduce peak memory consumption and finishes under 2 hours clock time. Training-convergence diagnostics for the headline run (loss curves and the expected Cascade R-CNN stage ordering) are collected in Appendix~\ref{app:loss_curves} (Figure~\ref{fig:losses}).

For Stage 2 (Figure~\ref{fig:stage2_train}; Appendix~\ref{app:stage2_schematic}), we finetune the classification head with the same optimizer and a gentle learning rate of $10^{-6}$ for an additional 30 epochs, with the backbone, FPN, RPN, and detection cascade explicitly frozen via PyTorch \texttt{requires\_grad=False} to prevent any drift in the detection features. Stage 2 uses cross-entropy loss with class re-weighting computed at the annotation level (i.e.\ without deduplicating sources that recur across overlapping tiles) on the Gatto-labeled subset of our training tiles (the 8{,}370 exact matches plus 961 synthetic recoveries of \S\ref{sec:staged_training}, counted once per (duplicated) annotation rather than once per unique source; explicitly, $\sqrt{11{,}193 / 4{,}224} = 1.6278$, with class weights galaxy=1.0, star=1.6278, background=1.0). As with Stage~1, we train four Stage-2 head variants (plain cross-entropy with no class weight, focal, star-weighted, and star-weighted$+$focal) and report the same three loss configurations used for Stage~1 (no class weight, star-weighted, star-weighted$+$focal; Tables~\ref{tab:performance}--\ref{tab:performance_hung}) in Table~\ref{tab:gatto_compare} and Figure~\ref{fig:cls_quality}, the latter alongside the \texttt{refExtendedness}-derived label baseline of the training catalog itself. The focal-only head is trained but not reported: on the held-out evaluation set of \S\ref{sec:cls_results} its matched set and its per-object predictions are identical to those of the star-weighted$+$focal head, so it would duplicate that column.

\subsection{Inference and matching}
\label{sec:inference}

At inference time, \deepdisc{} produces a bounding box, instance mask, classification score, and class assignment for each detected object. Predictions with classification score below a threshold $\tau$ are discarded; we sweep $\tau \in [0.01, 0.95]$ to construct purity/completeness curves. Throughout, completeness is the fraction of true objects that are recovered and purity the fraction of detections that correspond to a true object. We use these names for detection; for the star/galaxy classification of \S\ref{sec:cls_results} we use the equivalent terms recall and precision, so that detection and classification metrics are not confused. $F_1$ denotes the harmonic mean of the pair in either case. The operating point used for all reported test-set inference was fixed on the validation set by a coarse-then-fine sweep over the per-image detection cap and NMS threshold, selecting $\mathrm{TopK}=485$, $\mathrm{NMS}=0.7$, and score threshold $\tau = 0.22$ (best combined $F_1$ averaged over the two classes, $F_1^{\rm comb}=0.643$; galaxy/star detection $F_1=0.807/0.479$; the full sweep is shown in Appendix~\ref{app:sweep}, Figure~\ref{fig:sweep}). TopK is the maximum number of detections a tile may return after the ROI stage; the RPN budgets scale with it, $3{\times}$TopK per level before NMS and $2{\times}$TopK in total after it. NMS $=0.7$ is the overlap above which two boxes are merged, applied once to the proposals and once to the final detections. A cap of 485 is 1.4 times the average number of annotations in a test tile (352) but below the densest tiles (up to 558): after the score cut, 87 of the 312 test tiles still return exactly 485 detections. Raising the cap admits spurious detections faster than true sources, which is why the validation sweep settles on this value; at $\mathrm{TopK}=1000$ the combined $F_1$ drops from $0.64$ to $0.56$ (Figure~\ref{fig:sweep}). We compute two matching protocols, each addressing a different question.

\paragraph{(a) 5-condition class-aware matching.} A predicted object is counted as a match for a ground-truth object only if all five of the following hold: (i) sky-coordinate proximity within $0.5\arcsec$; (ii) same predicted and ground-truth class (the test-tile truth classes being the catalog \texttt{refExtendedness} labels; \S\ref{sec:scarlet_gt}); (iii) bounding-box IoU $\geq 0.30$ for galaxies and $\geq 0.05$ for stars (the looser IoU floor for stars reflects their compact, near-point-source morphology, where small positional offsets dominate the IoU denominator); (iv) the ground-truth object is not yet assigned to another prediction; (v) prediction score above $\tau$. This is our headline reporting metric and the protocol used to compute the completeness numbers in \S\ref{sec:headline}.

\paragraph{(b) Two-step FOF + Hungarian matching.} For blend-aware metrics we adopt the two-step protocol of \citet{EjjagiriMerz2026}, whose FOF grouping step follows the catalog-merging approach of \citet{Mao2021}. The first step forms friends-of-friends (FOF) groups by linking truth and predicted objects within a fixed angular linking length $L$ ($L = 1\arcsec$ as primary, $L = 2\arcsec$ as a sensitivity check). Each FOF group is characterized by the number of truth objects $n_t$ and detections $n_d$. The second step runs the Hungarian algorithm within each FOF group using a strict $0.5\arcsec$ inner match radius and class-agnostic adjacency; classes are used post-match only. Complex groups ($n_t > 1$ or $n_d > 1$) are then classified from their truth and detection counts into the categories listed in Table~\ref{tab:fofcats}. From these we report completeness $C$, purity $P$, resolved-blend, unrecognized-blend, and shred fractions. The FOF+Hungarian decomposition is more informative than a single AP or completeness number because a method can be highly complete on a per-truth basis while still failing on blends; the UB fraction surfaces this. The same framework was applied by \citet{Adari2025} to the LSSTComCam \dpone{} \ecdfs{} catalog cross-matched to HST, providing a direct empirical reference value (UB $\approx 18\%$).

\begin{table}
\raggedright
\caption{Two-step FOF+Hungarian group categories used in this work. $n_t$ = truth objects in group, $n_d$ = detections in group; ``paired'' means assigned by the Hungarian step within the $0.5\arcsec$ radius. The count convention follows \citet{Adari2025}.}
\label{tab:fofcats}
\renewcommand{\arraystretch}{1.4}
\setlength{\tabcolsep}{1.5pt}

\footnotesize
\begin{tabular}{ll}
\hline\hline
Category & Condition \\
\hline
Isolated match     & $n_t = 1,\ n_d = 1$ \\
Missed             & $n_d = 0,\ n_t \geq 1$ \\
Spurious           & $n_t = 0,\ n_d \geq 1$ \\
Resolved blend     & $n_t = n_d > 1$, all paired within $0.5\arcsec$ \\
Equal-count unres. & $n_t = n_d > 1$, at least one object unpaired \\
Unrec.\ blend      & $n_t > n_d$ \\
Shred              & $n_d > n_t$ \\
\hline
\end{tabular}%

\end{table}

\subsection{Matched-set metrics for label-bias mitigation}
\label{sec:matchedset}

\scarlet{} pseudo-ground truth is neither complete nor pure (\S\ref{sec:gt}). Standard precision/recall and AP metrics implicitly assume both, and so penalize the model for (i) detecting genuine sources that are absent from our pseudo-ground truth, and (ii) failing to reproduce shredded annotations. To separate classification performance from labeling artifacts, we follow \citet{Merz2023} and additionally compute classification metrics on the \emph{matched set} only: the subset of detections with a one-to-one association to a ground-truth object. Within this set we compute per-class precision, recall, and the harmonic-mean $F_1$ score. \citet{Merz2023} validated this approach against HST-based labels in COSMOS and found that it largely removes the AP-score depression caused by upstream label-generation artifacts.

\subsection{HST/CANDELS true-purity validation}
\label{sec:hst_validation}

The matched-set metrics in \S\ref{sec:matchedset} mitigate but do not eliminate the dependence of headline numbers on \scarlet{}-derived pseudo-ground truth: a detection that has no \scarlet{} counterpart is excluded from the matched-set computation, so we cannot tell from \scarlet{} alone whether such a detection is a true source or an artifact. To obtain a true completeness, true purity, and absolute UB-fraction measurement, we cross-match our \dpone{} \ecdfs{} detection catalog against the HST/CANDELS GOODS-S F160W-selected catalog \citep{Guo2013} from the CANDELS imaging \citep{Grogin2011,Koekemoer2011}, which is significantly deeper than \dpone{} (CANDELS GOODS-S reaches $H_{\rm F160W} \approx 28$, roughly two magnitudes below the \dpone{} \ecdfs{} $r \approx 26$ detection floor) and is independent of the upstream \scarlet{}/\rsp{} pipeline. The protocol parallels the HST-COSMOS validation that \citet{Merz2023} performed for HSC and the JWST/NEXUS validation that \citet{Jiang2026} performed for Euclid Q1 ($\sim 90\%$ true purity).

We restrict the comparison to the spatial intersection of the CANDELS GOODS-S footprint with the \dpone{} \ecdfs{} patch footprint, eroded $5\arcsec$ inward so that blend-neighbour searches never clip at an edge. Of the \HSTNclean{} clean (\texttt{FLAGS}${}=0$) CANDELS WFC3 F160W sources, \HSTNcandels{} enter the cross-match; on the \dpone{} side, \HSTNshared{} detections lie on the shared footprint and \HSTNdpone{} enter the $0.5\arcsec$ match. Cross-matching uses a $0.5\arcsec$ radius, consistent with \citet{Adari2025} and \citet{Jiang2026}. We define:
\begin{itemize}
\item \emph{True completeness}: fraction of CANDELS sources within the joint footprint that are recovered by \deepdisc{} (one or more \deepdisc{} detections within the matching radius).
\item \emph{True purity}: fraction of \deepdisc{} detections within the joint footprint that have a CANDELS counterpart.
\item \emph{HST-anchored UB fraction}: fraction of CANDELS source pairs separated by less than the \dpone{} effective resolution that are detected as a single source by \deepdisc{}; directly comparable to the \citet{Adari2025} HST-anchored measurement of $\approx 18\%$ for the \rsp{} Object catalog on the same field.
\end{itemize}
Sources that fall on the masked CANDELS footprint (e.g., near saturated stars in the HST imaging) are excluded from both numerator and denominator. Completeness is also resolved as a function of $F160W$ magnitude; the resulting $50\%$ completeness limits and all HST-anchored numbers are reported in \S\ref{sec:hst_results}.

\section{Results}
\label{sec:results}

\subsection{Headline detection performance}
\label{sec:headline}

\begin{figure*}[t]
\centering
\includegraphics[width=0.95\textwidth]{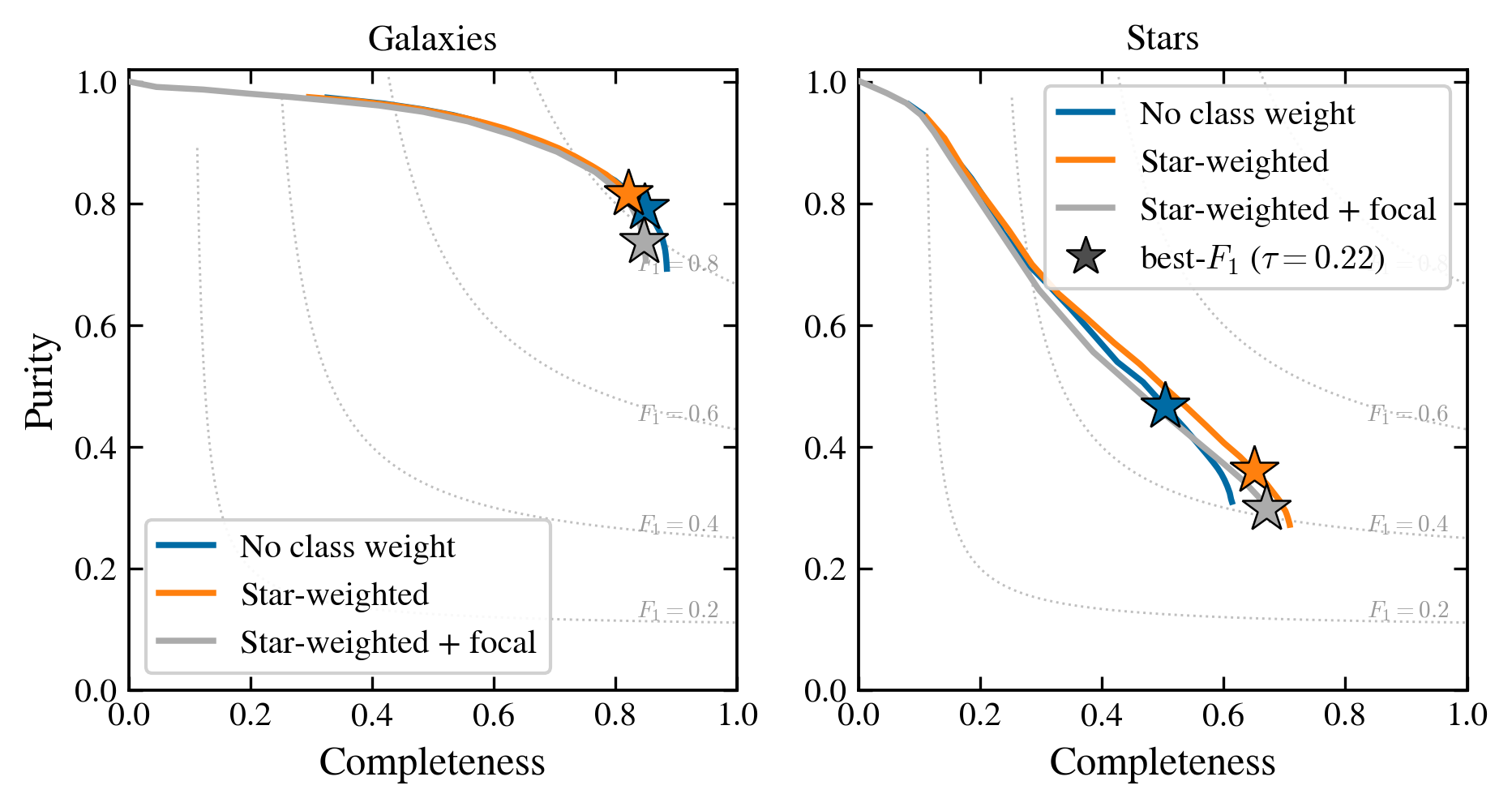}
\caption{Purity--completeness curves over the score-threshold sweep $\tau \in [0.01, 0.95]$ for galaxies (left) and stars (right) on the held-out \dpone{} \ecdfs{}+\edfs{} test set, shown for the three Stage-1 loss variants (no class weight, star-weighted, star-weighted$+$focal). The star marker on each curve is the best-$F_1$ operating point ($\tau = 0.22$, $\mathrm{TopK}{=}485$, $\mathrm{NMS}{=}0.7$); dotted arcs are lines of constant $F_1$. Galaxy recovery (left) is high and nearly variant-independent; the loss variants trade a few points of star completeness/purity (right), consistent with the headline numbers in Tables~\ref{tab:performance}--\ref{tab:performance_hung}.}
\label{fig:detection_quality}
\end{figure*}

\begin{figure*}[t]
\centering
\includegraphics[width=0.95\textwidth]{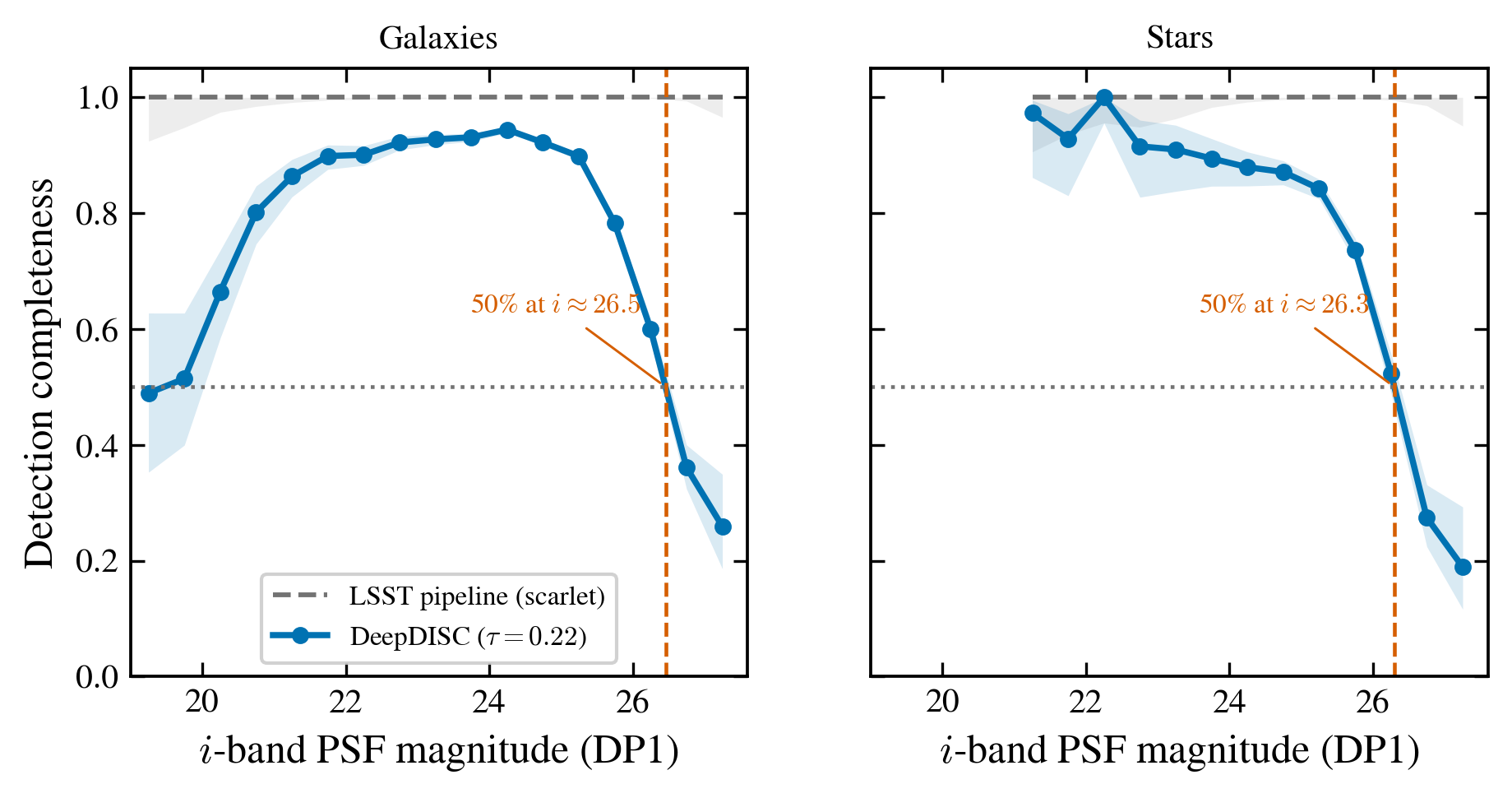}
\caption{Detection completeness at the fixed operating point ($\tau = 0.22$, $\mathrm{TopK}{=}485$, $\mathrm{NMS}{=}0.7$) as a function of \dpone{} $i$-band PSF magnitude, for galaxies (left) and stars (right), for the headline (no class weight) model on the \ecdfs{} test tiles (the only field with local DP1 photometry; \edfs{} is consistent in aggregate). Matching is the class-blind $0.5\arcsec$ sky cross-match behind the overall \OverallComp\% completeness (Table~\ref{tab:blends}); shaded bands are Wilson 95\% confidence intervals, and bins require $\geq 30$ truth appearances with valid DP1 $i$-band photometry ($72\%$ of \ecdfs{} appearances). The grey dashed curve is the \scarlet{} SNR$\geq$2 detection list evaluated against the same curated truth; it is near-unity \emph{by construction}, since the truth derives from the \scarlet{} catalog (\S\ref{sec:scarlet_gt}), so the \deepdisc{} curve reads as completeness relative to the \dpone{} catalog rather than absolute depth (the HST-anchored absolute limit is $F160W \approx \HSTDDComplim$; Figure~\ref{fig:hst_ub}). \deepdisc{} recovers $\gtrsim 90\%$ of galaxies over $21.5 \lesssim i \lesssim 25.5$ and crosses 50\% at $i \approx \DDComplimGal$ (galaxies) and $i \approx \DDComplimStar$ (stars). The bright-end galaxy dip is dominated by centre offsets on extended sources under the fixed $0.5\arcsec$ criterion rather than missed detections: re-matching $i < 20.5$ galaxies at $1.5\arcsec$ recovers $95\%$ completeness.}
\label{fig:completeness_mag}
\end{figure*}

Tables~\ref{tab:performance} and \ref{tab:performance_hung} summarize Stage-1 detection completeness for our training configurations, evaluated under the two matching protocols defined in \S\ref{sec:inference}: a strict 5-condition class-aware match (Table~\ref{tab:performance}; \S\ref{sec:inference}a) and a sky-coordinate FOF + Hungarian match in which class is used only post-hoc for evaluation (Table~\ref{tab:performance_hung}; \S\ref{sec:inference}b). All metrics are computed on the retained tiles of \S\ref{sec:filter}, which exclude band-dropout regions. The full purity--completeness trade-off across the score-threshold sweep, for all three loss variants, is shown in Figure~\ref{fig:detection_quality}. Figure~\ref{fig:completeness_mag} decomposes the recovery at the adopted operating point by magnitude, comparing \deepdisc{} against the \scarlet{} detection list evaluated on the same curated truth.

\paragraph{In-domain pretraining: a warm start, not a prerequisite.}
The same \swinb{} architecture, evaluated without \dpone{} fine-tuning, achieves only $\sim$\Cgalbase\% galaxy and $\sim$\Cstarbase\% star completeness on the test set, consistent with severe domain shift from simulated DC2 to real \dpone{}. \dpone{} fine-tuning (no class weighting, 5-condition match) raises this to $\sim$\CgalA\% / $\sim$\CstarA\%, a $\sim 2\times$ improvement for galaxies and a $\sim 4\times$ improvement for stars. Initialization has little influence on this outcome: an ImageNet-21k checkpoint fine-tuned identically reaches near-parity with the in-domain checkpoint (Appendix~\ref{app:init}).

\paragraph{Matching-protocol dependence is large for stars.}
The stellar completeness numbers are highly sensitive to the matching protocol. Under the strict 5-condition match, the headline (no class weight) configuration recovers only $\sim$\CstarA\% of stars, vs.\ $\sim$\CstarH\% under the FOF+Hungarian match (Table~\ref{tab:performance_hung}); the gap arises because the 5-condition protocol enforces class agreement and a per-class IoU floor at the matching step, whereas the Hungarian protocol allows cross-class matches and uses class only for post-match evaluation. The galaxy numbers are far more stable across protocols ($\sim$\CgalA\% vs.\ $\sim$\CgalH\%). The gap therefore consists mainly of stars that are detected but not labelled star at Stage 1: the FOF+Hungarian numbers measure detection alone, whereas the 5-condition numbers count a star only when it is both detected and classified as a star. Both protocols are reported here for transparency; the FOF+Hungarian numbers are the more directly comparable to traditional source-catalog completeness measurements (e.g., the LSST production \rsp{} pipeline; \citealt{EjjagiriMerz2026,Adari2025}).

\paragraph{Class imbalance trades galaxy for star recall.}
Star-weighted training (\S\ref{sec:training}) shifts the trade-off in the expected direction: under the 5-condition match, galaxy completeness drops from $\sim$\CgalA\% to $\sim$\CgalCWA\% while star completeness rises from $\sim$\CstarA\% to $\sim$\CstarCWA\%; under Hungarian matching, the corresponding shifts are $\sim$\CgalH\% $\to$ $\sim$\CgalCWH\% (galaxies) and $\sim$\CstarH\% $\to$ $\sim$\CstarCWH\% (stars). The Hungarian-matching numbers show that the apparent ``rebalancing benefit'' is smaller than the 5-condition numbers alone suggest: the unweighted model already recovers $\sim$\CstarH\% of stars under the protocol most directly comparable to traditional catalog matching. Star weighting compensates for the $\sim 6{:}1$ galaxy:star ratio of our \scarlet{} training annotations ($\sim 4{:}1$ in the broader \dpone{} Object catalog), and the star-weighted$+$focal configuration has the highest detection completeness for both classes under both protocols (Tables~\ref{tab:performance}--\ref{tab:performance_hung}). We nonetheless keep the unweighted model as the headline: in the Stage-2 classification step, star weighting lowers star precision and overall accuracy (Table~\ref{tab:gatto_compare}), and the weight itself is tuned to the class ratio of this particular training set, which will differ for other footprints.

\begin{table}[t]
\raggedright
\caption{Stage-1 detection completeness on \dpone{} \ecdfs{}+\edfs{} on the retained tiles (\S\ref{sec:filter}), under the strict class-aware match (\S\ref{sec:inference}a), on the final \Ntilestest{}-tile test set at $\tau{=}0.22$. All values percent.}
\label{tab:performance}
\renewcommand{\arraystretch}{1.5}
\setlength{\tabcolsep}{4pt}

\begin{tabular}{lcc}
\hline\hline
Configuration                                  & Galaxies & Stars \\
\hline
LSST checkpoint baseline (no \dpone{} FT)                & \Cgalbase    & \Cstarbase    \\
\dpone{} FT, no class weight (headline)        & \CgalA       & \CstarA       \\
\dpone{} FT, star-weighted                     & \CgalCWA     & \CstarCWA     \\
\dpone{} FT, star-weighted + focal             & \CgalFA      & \CstarFA      \\
\hline
\end{tabular}%
\end{table}

\begin{table}[b]
\raggedright
\caption{Stage-1 detection completeness on \dpone{} \ecdfs{}+\edfs{} on the retained tiles (\S\ref{sec:filter}), under the FOF + Hungarian match (\S\ref{sec:inference}b), on the final \Ntilestest{}-tile test set at $\tau{=}0.22$. Class is used only for post-match evaluation; cross-class matches are allowed at the assignment step. All values percent. The star-weighted$+$focal configuration gives the best overall recovery, raising star completeness from \CstarH\% (headline) to \CstarFH\%.}
\label{tab:performance_hung}
\renewcommand{\arraystretch}{2}
\setlength{\tabcolsep}{6pt}

\begin{tabular}{lcc}
\hline\hline
Configuration                                  & Galaxies & Stars \\
\hline
\dpone{} FT, no class weight (headline)        & \CgalH       & \CstarH       \\
\dpone{} FT, star-weighted                     & \CgalCWH     & \CstarCWH     \\
\dpone{} FT, star-weighted + focal             & \CgalFH      & \CstarFH      \\
\hline
\end{tabular}%

\end{table}

\subsection{Blend-aware metrics from two-step matching}
\label{sec:blend_results}

\begin{figure}
\centering
\includegraphics[width=\columnwidth]{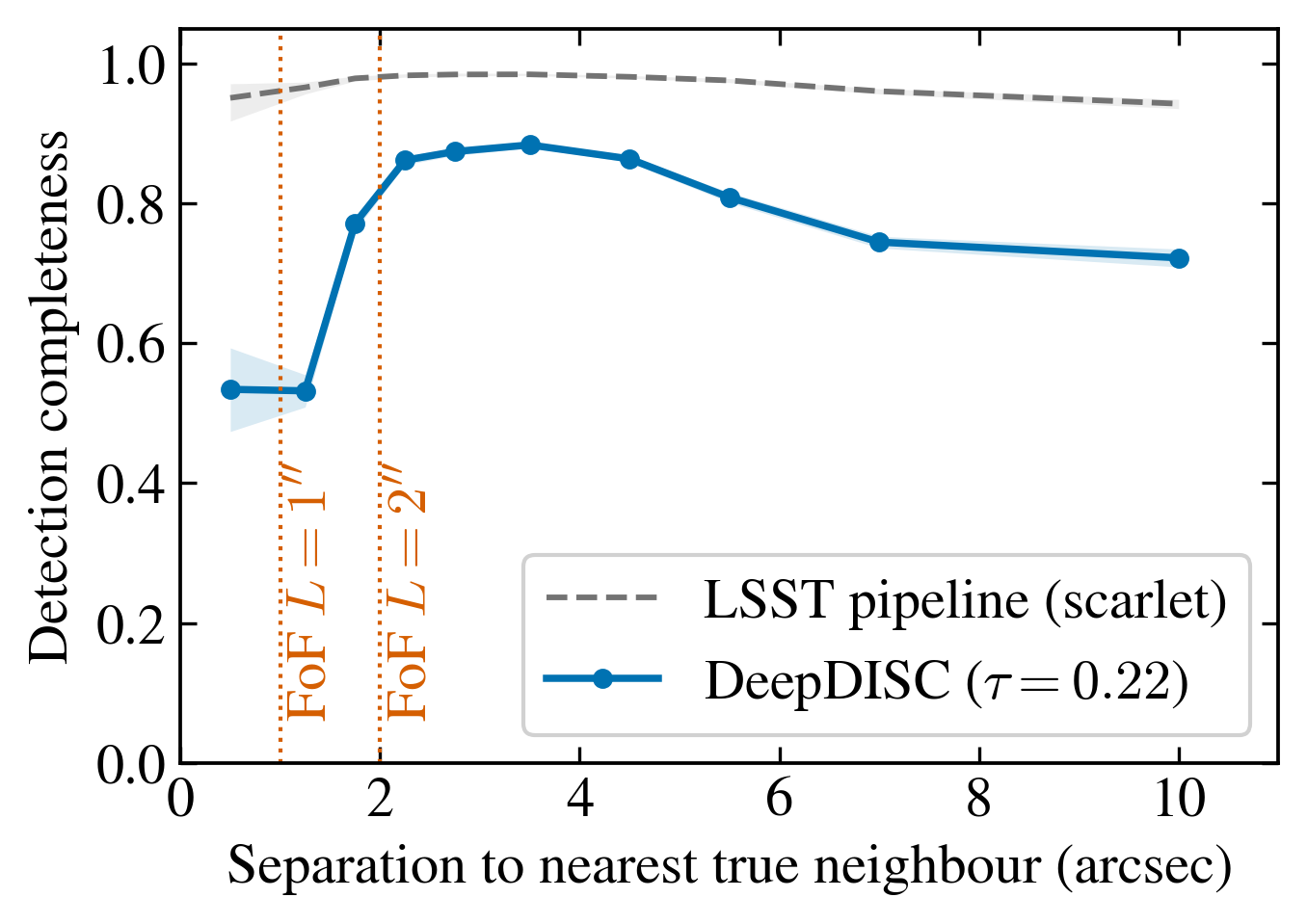}
\caption{Class-agnostic detection completeness at the fixed operating point versus separation to the nearest true neighbour, over the full \ecdfs{}+\edfs{} test set (Wilson 95\% bands; separations computed within each $760\times760$ tile). Dotted verticals mark the two FOF linking lengths of Table~\ref{tab:blends}. \deepdisc{} recovers $\sim 88\%$ of sources whose nearest neighbour lies at $2\arcsec$--$4\arcsec$ but only $\sim 53\%$ below $1.5\arcsec$ ($\approx 2\times$ the Rubin $0.7\arcsec$ PSF), while the \scarlet{} reference stays $\gtrsim 95\%$ by construction: close pairs are the dominant completeness cost of \deepdisc{} detection. The mild decline for the most isolated sources ($\gtrsim 6\arcsec$) persists at fixed magnitude.}
\label{fig:completeness_nnsep}
\end{figure}

Applying the FOF+Hungarian protocol of \S\ref{sec:inference}b to the headline model initialized from the LSST checkpoint, we obtain the blend-decomposed metrics in Table~\ref{tab:blends}, computed at the fixed operating point ($\tau = 0.22$, $\mathrm{TopK}=485$, $\mathrm{NMS}=0.7$) on the \Ntilestest{}-tile test set (\NgtMatch{} objects, \NpredMatch{} predictions, \NmatchedMatch{} matched). The UB fraction ($\sim$\UBfrac\% at $L=1\arcsec$, $\sim$\UBfracTwo\% at $L=2\arcsec$) can be placed alongside two reference points: (i) the LSST production-pipeline UB fraction on the simulated DC2 footprint, $\sim 11\%$ \citep{EjjagiriMerz2026}; and (ii) the empirical HST-cross-match UB fraction on the same real \ecdfs{} field, $18\%$ \citep{Adari2025}. These are not like-for-like comparisons: a \scarlet{}-anchored UB fraction cannot register blends that \scarlet{} itself merged, so it understates the UB fraction relative to an external truth, and the HST-anchored comparison of \S\ref{sec:hst_results} is the appropriate one. The dominant blend failure mode is over-deblending (shredding): $\sim$\Shredfrac\% of complex FOF groups at $L=1\arcsec$ (and $\sim$\ShredfracTwo\% at $L=2\arcsec$) contain more predictions than truth objects. We adopt $L=1\arcsec$ as the primary linking length (matching the \dpone{} astrometric scale) and report $L=2\arcsec$ as a sensitivity check. Two points are important when interpreting this number. First, ``shredding'' here is measured \emph{against the \scarlet{} pseudo-ground truth}, and the \scarlet{} catalog is itself an aggressive deblender; a complex FOF group is flagged as shredded whenever \deepdisc{} returns even one more component than \scarlet{} assigned, so the metric counts both genuine fragmentation and cases where \deepdisc{} deblends a source that \scarlet{} had already split. The shred fraction is therefore an upper bound on true over-deblending relative to a fixed reference, not an absolute fragmentation rate. Second, against CANDELS both \deepdisc{} and the DP1 production catalog show shred rates of $\sim 0\%$ (\S\ref{sec:hst_results}; Table~\ref{tab:hst_validation}): relative to a higher-resolution external truth, the Rubin $0.7\arcsec$ PSF \emph{merges} blends rather than splitting them. Within the CANDELS footprint, therefore, the extra components relative to \scarlet{} do not correspond to single HST sources being split, and the two references bracket the deblending behavior from opposite sides (over-segmentation relative to \scarlet{} versus under-segmentation relative to HST). Figure~\ref{fig:completeness_nnsep} shows the same operating point as a function of crowding: \deepdisc{} completeness falls from $\sim 88\%$ for well-separated sources to $\sim 53\%$ below $1.5\arcsec$, localizing the completeness cost of blending that the FOF taxonomy of Table~\ref{tab:blends} aggregates.
\begin{table*}
\raggedright
\caption{Two-step FOF+Hungarian blend metrics for the headline (no class weight) \deepdisc{} model on the held-out \dpone{} \ecdfs{}+\edfs{} test set, at both linking lengths. Overall (class-agnostic) completeness/purity are \OverallComp\%/\OverallPur\% (matched \NmatchedMatch{} of \NgtMatch{} objects vs.\ \NpredMatch{} predictions at score $\geq 0.22$, $\mathrm{TopK}=485$, on \Ntilestest{} tiles). DC2 simulation and DP1 ECDFS reference values from \citet{EjjagiriMerz2026,Adari2025} for context. The shred fraction is measured against the \scarlet{} reference and is interpreted, alongside the contrasting $\sim 0\%$ HST-anchored shred rate, in \S\ref{sec:blend_results}.}
\label{tab:blends}
\renewcommand{\arraystretch}{2}
\setlength{\tabcolsep}{26pt}

\begin{tabular}{lcccc}
\hline\hline
Metric & This work & This work & DC2 (DM) & DP1 ECDFS \\
       & ($L{=}1\arcsec$) & ($L{=}2\arcsec$) & (sim.) & (HST) \\
\hline
Resolved blend & \Resfrac\%    & \ResfracTwo\%   & 1\%    & \nodata     \\
Unrec.\ blend  & \UBfrac\%     & \UBfracTwo\%    & 11.2\% & 18\%    \\
Shred          & \Shredfrac\%  & \ShredfracTwo\% & 0.15\% & $\sim 0\%$ \\
Equal-count unres. & \EqUnresfrac\% & \EqUnresfracTwo\% & \nodata & \nodata \\
\hline
Complex groups & \Ncomplex{} & \NcomplexTwo{} & \nodata & \nodata \\
\hline
\end{tabular}%

\end{table*}

\begin{figure*}[!ht]
\centering
\includegraphics[width=0.95\textwidth]{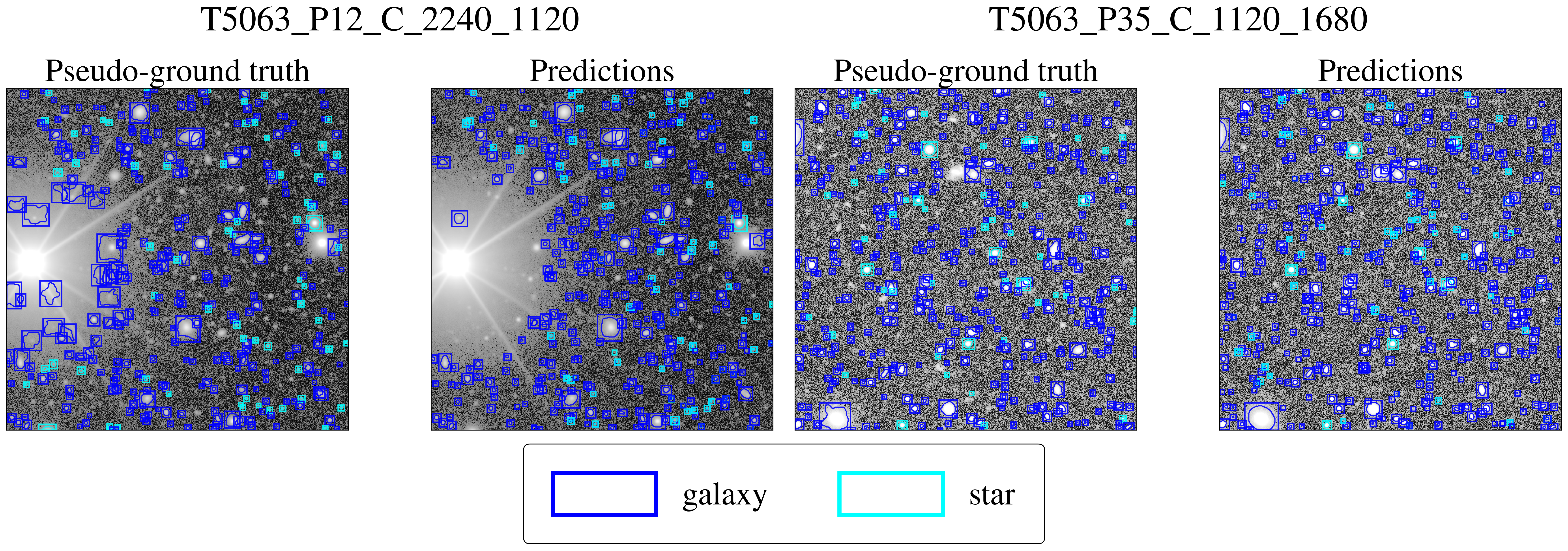}
\caption{Comparison between our curated pseudo-ground-truth test tiles (labelled \emph{Pseudo-ground truth}) and \deepdisc{} predictions at a score threshold of $\geq 0.22$ (labelled \emph{Predictions}) for two representative held-out test tiles from Tract~5063. Bounding boxes and mask-boundary outlines are colour-coded by class (blue for galaxies, cyan for stars; shared legend at bottom). As discussed in \S\ref{sec:filter}, our pseudo-ground-truth curation removes most of the overlapping star boxes produced by shredding near bright stars. In the first example (\texttt{T5063\_P12\_C\_2240\_1120}) a few shredded annotations remain near the bright star; these are galaxy annotations, since the bright-star curation acts only on star annotations, so only blue annotations survive. \deepdisc{} does not reproduce these residual \scarlet{} annotations around the bright star in this tile. The second example illustrates the predicted classes, bounding boxes, and segmentation masks in a crowded field. Left to right: \texttt{T5063\_P12\_C\_2240\_1120}, with 338 ground-truth objects (273 galaxies, 65 stars) against 447 predicted objects (363 galaxies, 84 stars); and \texttt{T5063\_P35\_C\_1120\_1680}, with 424 ground-truth objects (368 galaxies, 56 stars) against 480 predicted objects (411 galaxies, 69 stars).}
\label{fig:gt_vs_pred_examples}
\end{figure*}

Figure~\ref{fig:gt_vs_pred_examples} shows two representative held-out test tiles with the curated pseudo-ground truth and \deepdisc{} predictions plotted side by side, as a qualitative complement to the detection and blend-decomposition metrics above.

\subsection{Star/galaxy classification on the matched set}
\label{sec:cls_results}

To isolate classification performance from detection and label-generation artifacts, we evaluate the Stage-2 head on the \emph{detected-only} set: the subset of held-out \citet{Gatto2026} objects that \deepdisc{} both detects and matches one-to-one. This is a fair point of comparison with a catalog-level classifier such as the \citet{Gatto2026} Random Forest, which is handed already-detected Object-catalog photometry and asked only to label it; restricting to detected+matched sources removes the detection- and segmentation-completeness penalty that would otherwise be charged against \deepdisc{} but not against a pure classifier. To avoid scoring \deepdisc{} on data it was trained on, we restrict the evaluation to objects in \deepdisc{}'s held-out test tiles. On this set (the 3{,}934 real held-out \citet{Gatto2026} objects in the geographic test split: 3{,}132 galaxies and 802 stars, of which the no-class-weight model detects and matches 3{,}712 one-to-one; $\tau = 0.22$), the headline (no class weight) configuration reaches an overall classification accuracy of \DDnormAcc\%, with galaxy recall \DDnormGal\% and star recall \DDnormStar\%; star-weighted training trades a little galaxy recall for star recall (\DDswGal\% / \DDswStar\%, accuracy \DDswAcc\%). The geographic split (Stage-1 training tiles vs.\ held-out test tiles; \S\ref{sec:staged_training}) guarantees that these metrics are computed only on tiles the classification head never saw. Full-field \dpone{} \ecdfs{} inference is used only for the field-wide concordance with the Random Forest (\S\ref{sec:cross_compare}; Appendix~\ref{app:rf_concordance}), with train/validation-tile objects excluded. The strictly one-to-one comparison with the benchmark Random Forest of \citet{Gatto2026}, on the subset of these objects that is held out on the Random Forest side as well (associated by exact 64-bit Rubin \texttt{objectId}), is given in Table~\ref{tab:gatto_compare} and discussed in \S\ref{sec:cross_compare}. As an external check, the same Stage-2 head (whose geographic split already holds out the HST-overlap test tiles), evaluated against HST-confirmed \citet{Gatto2026} truths recovers HST galaxies/stars at \HSTGalRec\%/\HSTStarRec\% recall (\S\ref{sec:hst_results}). The galaxy value is consistent with the recall here; the star value is lower, but it rests on only 63 HST stars (Wilson 95\% CI 59.3--81.1\%) and on HST \texttt{CLASS\_STAR}, which becomes unreliable at faint magnitudes.

\begin{figure*}
\centering
\includegraphics[width=0.95\textwidth]{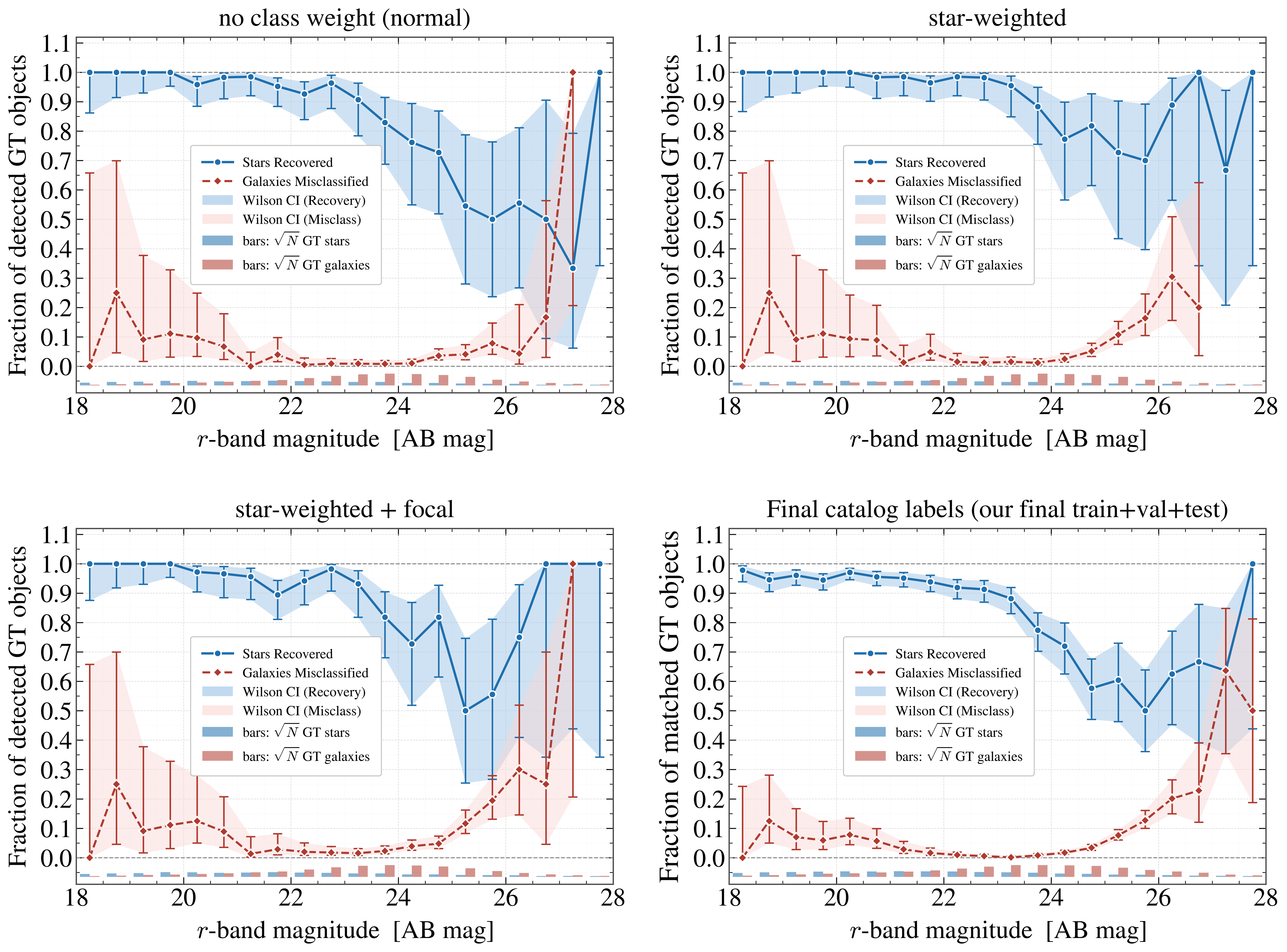}
\caption{Matched-set star/galaxy classification quality on \dpone{} \ecdfs{}+\edfs{} as a function of $r$-band magnitude, shown for the three Stage-2 loss configurations (no class weight, star-weighted, star-weighted$+$focal). The evaluation sample is the held-out \citet{Gatto2026} sources on the geographic test tiles, 3{,}934 objects (3{,}132 galaxies, 802 stars), classed by their \citet{Gatto2026} labels (\S\ref{sec:cls_results}); each configuration is scored on the subset of these it detects and matches one-to-one. Each panel plots the star recovery fraction and the galaxy$\to$star misclassification fraction with Wilson $95\%$ confidence intervals. The fourth panel applies the same metrics to the labels of our final training catalog itself (the \texttt{refExtendedness}-derived star/galaxy labels of the combined train/validation/test set, joined to the held-out \citet{Gatto2026} sample by \texttt{objectId}) and therefore shows the label quality the networks inherit from the \dpone{} catalog rather than the output of a trained model. The recovery curves degrade toward the faint end, where the catalog \texttt{refExtendedness} labels are known to become unreliable \citep{Bosch2018} and the held-out \citet{Gatto2026} sample is sparse.}
\label{fig:cls_quality}
\end{figure*}

The magnitude dependence is shown in Figure~\ref{fig:cls_quality}. As in the HSC analysis of \citet{Merz2023}, the faint end is limited partly by the labels: the Stage-1 features were learned from \texttt{refExtendedness} classes, which become unreliable for stars at $i \gtrsim 24$~mag \citep{LSSTDP1}, and the faint \citet{Gatto2026} stellar labels are photometric with $\sim 85\%$ purity (\S\ref{sec:staged_training}), so the apparent drop in stellar recall at faint magnitudes reflects a combination of model error and label noise. We caution that the held-out \citet{Gatto2026} sample is dominated by Gaia DR3 and spectroscopic stars at $r \lesssim 22$~mag, so the present comparison probes mainly the bright-to-intermediate regime, where multi-band colors and \texttt{refExtendedness} already perform well; the faint, blended regime where scene-level pixel information is expected to add the most value is sparsely labeled here and is the subject of the magnitude-resolved follow-up in \S\ref{sec:future}.

\subsection{HST/CANDELS true-purity validation}
\label{sec:hst_results}

\begin{figure*}
\centering
\includegraphics[width=\textwidth]{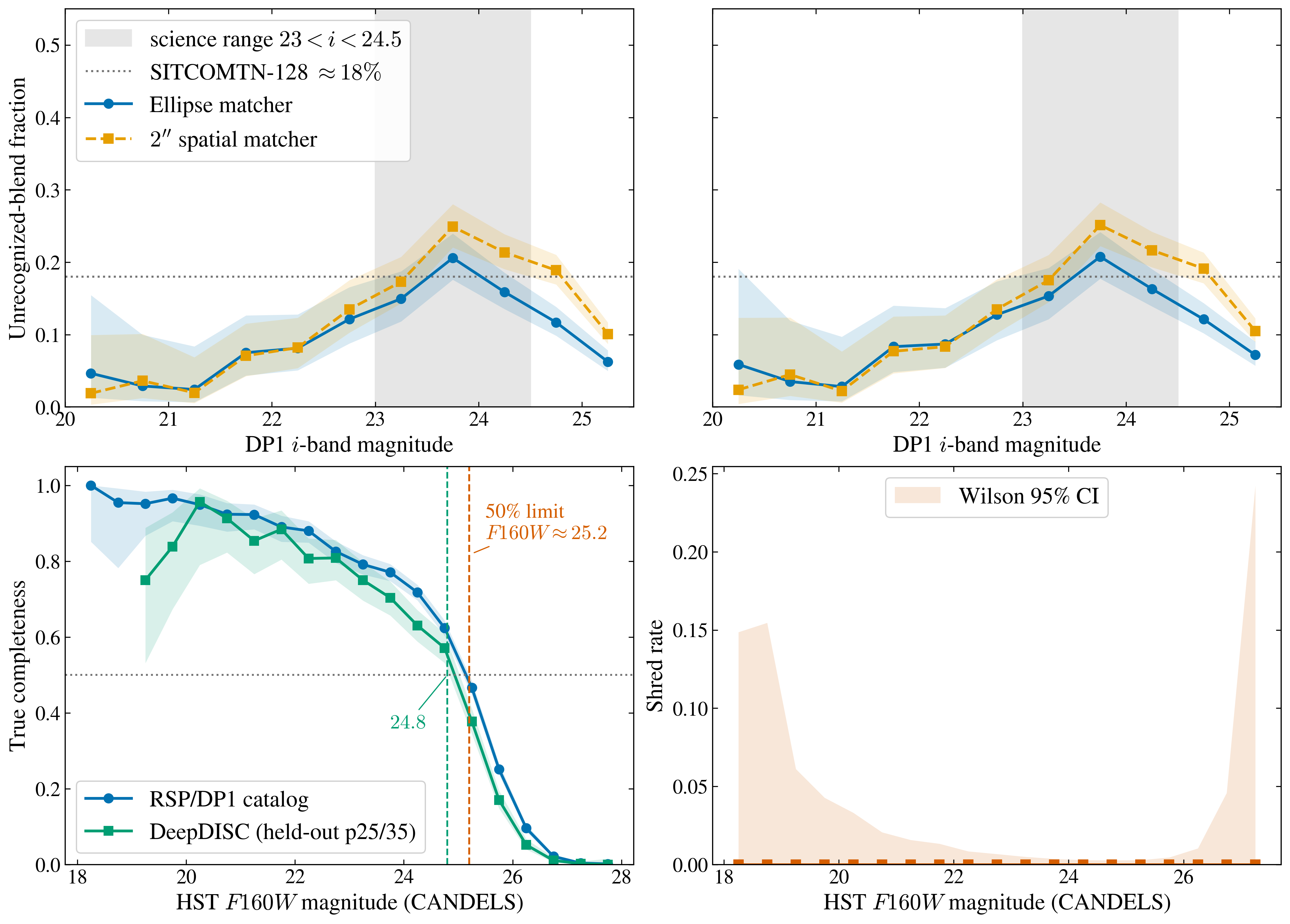}
\caption{HST/CANDELS-anchored validation of the full-field \dpone{} production catalog (cf.\ Table~\ref{tab:hst_validation}). \emph{Top row:} HST-anchored unrecognized-blend fraction versus \dpone{} $i$-band magnitude, reproducing Fig.~2 of \citet{Adari2025}, for all detections (left) and extended-only sources (\texttt{refExtendedness}$\,{=}\,1$; right); the ellipse matcher (primary, blue) and $2\arcsec$ spatial matcher (orange) are shown with Wilson $95\%$ bands, and the shaded $23<i<24.5$ science range gives the headline $\HSTRubinUB\%$ (the dotted line marks the SITCOMTN-128 / \citealt{Adari2025} reference of $\approx18\%$). \emph{Bottom row:} detection completeness (left) and shred rate (right) versus HST $F160W$ magnitude, showing both the full-field \rsp{}/DP1 catalog ($50\%$ limit at $F160W\approx\HSTComplim$~mag) and the \deepdisc{} predictions on the leakage-free held-out \texttt{T5063\_P25/35} ($F160W\approx\HSTDDComplim$~mag), the two columns of Table~\ref{tab:hst_validation}; completeness bins require $\geq 20$ HST sources. All panels use HST/CANDELS GOODS-S as the external reference.}
\label{fig:hst_ub}
\end{figure*}

This subsection first measures the \dpone{} production catalog, which provides the \scarlet{}-independent reference frame against which our own metrics are read, and then \deepdisc{} on a held-out footprint. Applying the cross-match protocol of \S\ref{sec:hst_validation}, we obtain the HST/CANDELS-anchored metrics summarized in Table~\ref{tab:hst_validation}. Cross-matching the DP1 production Object catalog against the CANDELS GOODS-S F160W catalog within the joint footprint (\HSTNdpone{} DP1 detections entering the match against \HSTNcandels{} CANDELS sources, $0.5\arcsec$ radius) yields a true completeness of $\HSTRubinComp\%$ ($\HSTNmatchcomp{}/\HSTNcandels{}$), a true purity of $\HSTRubinPur\%$ ($\HSTNmatchpur{}/\HSTNdpone{}$), and an HST-anchored UB fraction of $\HSTRubinUB\%$ (Wilson 95\% CI $[\HSTRubinUBlo,\HSTRubinUBhi]\%$; $\HSTRubinUBext\%$ for extended-only sources). This independently reproduces the $\approx 18\%$ measured by \citet{Adari2025} on the same field, and the full UB-versus-magnitude relation reproduces their Fig.~2 (Figure~\ref{fig:hst_ub}); a looser $2\arcsec$ spatial matcher gives a higher $\HSTRubinUBspatial\%$, showing the sensitivity of the UB fraction to the matcher. The shred rate is negligible ($\sim 0\%$: 2 of \HSTNmatchcomp{} HST sources split into multiple DP1 detections, 4 of \HSTNdpone{} DP1 detections are shred fragments), consistent with Rubin's $0.7\arcsec$ PSF merging rather than over-splitting sources. Detection completeness falls to $50\%$ at $F160W \approx \HSTComplim$~mag; the low overall completeness is expected because CANDELS reaches $F160W \approx 28$ well below the \dpone{} $5\sigma$ depth. This HST-anchored completeness ($\HSTRubinComp\%$, referenced to the much deeper CANDELS catalog) is a different quantity from the \scarlet{}-anchored matched-set completeness of \OverallComp\% in \S\ref{sec:blend_results}, which is referenced to the same-depth \scarlet{} truth; the two should not be compared directly. As a star/galaxy sanity check, DP1 \texttt{refExtendedness} recovers $\HSTRubinGalRec\%$ of HST galaxies and $\HSTRubinStarRec\%$ of HST stars (HST \texttt{CLASS\_STAR}$\,\geq 0.8$ as truth), the modest stellar recall reflecting the known \texttt{refExtendedness} ceiling that motivates our Stage-2 head (\S\ref{sec:cls_results}).

Restricting the same cross-match to the \dpone{} \texttt{objectId}s that fall in our model's train/validation/test tiles leaves the UB fraction stable ($\HSTSetUB\%$, within the \citet{Adari2025} confidence interval) while the true purity rises to $\HSTSetPur\%$ and the completeness falls to $\HSTSetComp\%$ (a subset effect); galaxy/star \texttt{refExtendedness} recall rise to $\HSTSetGalRec\%/\HSTSetStarRec\%$. The purity increase indicates that our tile-curation pipeline (\S\S\ref{sec:tiles},~\ref{sec:gt}) preferentially retains the cleaner, HST-confirmable detections. Any detector evaluated only on our curated tiles should therefore be compared with this $\HSTSetPur\%$ rather than with the full-field $\HSTRubinPur\%$.

To obtain a leakage-free \deepdisc{}$\times$HST measurement, we reuse the Stage-2
predictions already saved by the upstream pipeline, with no retraining or re-inference:
because the geographic train/test split fine-tunes the classification head on Stage-1
\emph{training} tiles only, the 50 cutouts of \texttt{T5063\_P25/35}, which
fall entirely within the Stage-1 test split, are already unseen by both the detector and
the classifier. On this held-out footprint ($7{,}426$ CANDELS sources, $2{,}552$
\deepdisc{} detections, $0.5\arcsec$ radius) \deepdisc{} reaches a true completeness of $\HSTComp\%$, a true purity of $\HSTPur\%$, and an HST-anchored UB fraction of $\HSTUB\%$ (Wilson 95\% CI
$[\HSTUBlo,\HSTUBhi]\%$; Table~\ref{tab:hst_validation}), with a $50\%$ completeness
limit at $F160W\approx\HSTDDComplim$. The low integrated completeness is largely HST-depth-limited:
CANDELS reaches $F160W\approx28$, roughly two magnitudes
below \dpone{}, so most reference sources lie below the \dpone{} detection floor;
restricted to the bright end, \deepdisc{} recovers $\sim$84--95\% of CANDELS sources
brighter than $F160W=23$. \deepdisc{} is nonetheless less complete than the production
\rsp{} catalog on the same patches ($\HSTRubinCompHeldout\%$), with a 50\% limit
$\sim 0.4$~mag brighter. Its purity is higher than that of the production catalog on
those patches ($\HSTPur\%$ vs $\HSTRubinPurHeldout\%$), but the production catalog
restricted to the objects on our curated tiles reaches $\HSTSetPur\%$ (above), so this
difference reflects at least partly which sky is evaluated rather than the detector, and
we do not claim a purity advantage over the production pipeline. Its predicted-class
recall against HST \texttt{CLASS\_STAR}$\,\geq0.8$ is $\HSTGalRec\%$ for galaxies and
$\HSTStarRec\%$ for stars (45 of 63 stars; Wilson 95\% CI 59.3--81.1\%), compared with
production \texttt{refExtendedness} recall of $\HSTRubinGalRec\%/\HSTRubinStarRec\%$ over
the full field and $\HSTSetGalRec\%/\HSTSetStarRec\%$ on our tiles. Because this true
purity counts \deepdisc{} detections that have no \scarlet{} counterpart but a CANDELS
counterpart as true sources, it breaks the \scarlet{}-based ceiling of the matched-set
metrics in \S\ref{sec:matchedset}. Finally, in the same $23<i<24.5$ range, the HST-anchored UB fraction of \deepdisc{}
($\HSTUB\%$, 95\% CI $[\HSTUBlo,\HSTUBhi]\%$) is lower than that of the production
catalog ($\HSTRubinUB\%$, $[\HSTRubinUBlo,\HSTRubinUBhi]\%$, full field; $\HSTSetUB\%$
restricted to our tiles; cf.\ $\approx 18\%$ from \citealt{Adari2025}). The intervals
overlap and the \deepdisc{} value comes from a smaller footprint, so these data are
consistent with a modest reduction in unrecognized blends rather than demonstrating one.

\begin{table*}
\raggedright
\caption{HST/CANDELS GOODS-S--anchored metrics. The \deepdisc{} column is the leakage-free
held-out measurement (\texttt{T5063\_P25/35}); \deepdisc{} detections
($n{=}2{,}552$) vs CANDELS ($n{=}7{,}426$), $0.5\arcsec$ radius. The \rsp{} column is the
directly-measured full-field DP1 production catalog $\times$ HST/CANDELS, whose UB of
\HSTRubinUB\% (Wilson 95\% CI $[\HSTRubinUBlo,\HSTRubinUBhi]\%$) reproduces \citet{Adari2025}.
\deepdisc{} UB Wilson 95\% CI $[\HSTUBlo,\HSTUBhi]\%$. \emph{Footprint note:} the two columns cover different sky. On the same
held-out patches the production catalog gives \HSTRubinCompHeldout\%/\HSTRubinPurHeldout\%
completeness/purity; restricted to the objects on our curated tiles it gives
\HSTSetPur\% purity, \HSTSetUB\% UB, and \HSTSetGalRec\%/\HSTSetStarRec\% galaxy/star recall
(\S\ref{sec:hst_results}), so purity and recall differences between the columns partly
reflect footprint. Recall is the \deepdisc{} predicted class
(held-out) and DP1 \texttt{refExtendedness} respectively, both against HST
\texttt{CLASS\_STAR}$\,\geq 0.8$; the \deepdisc{} star recall rests on 63 HST stars. The integrated completeness is HST-depth-limited
(CANDELS reaches $F160W\approx28$); see the 50\% limit row and \S\ref{sec:hst_results}.}
\label{tab:hst_validation}
\renewcommand{\arraystretch}{2}
\setlength{\tabcolsep}{24pt}

\begin{tabular}{lcc}
\hline\hline
Metric & This work, \deepdisc{} & \rsp{}\,/\,DP1 cat. \\
       & (held-out p25/35)      & (full field) \\
\hline
True completeness vs.\ CANDELS   & \HSTComp\%     & \HSTRubinComp\%    \\
True purity vs.\ CANDELS         & \HSTPur\%      & \HSTRubinPur\%     \\
HST-anchored UB fraction         & \HSTUB\%       & \HSTRubinUB\%      \\
Galaxy recall                    & \HSTGalRec\%   & \HSTRubinGalRec\%  \\
Star recall                      & \HSTStarRec\%  & \HSTRubinStarRec\% \\
Shred rate                       & $\sim 0$\%     & $\sim 0$\%         \\
50\% completeness limit (F160W)  & \HSTDDComplim  & \HSTComplim        \\
\hline
Clean HST/CANDELS sources        & \multicolumn{2}{c}{\HSTNclean{} clean; \HSTNcandels{} in cross-match} \\
\hline
\end{tabular}%

\end{table*}

\section{Discussion}
\label{sec:discussion}

\subsection{Limitations and Caveats}

Four limitations frame what follows. First, the training set is small in absolute terms (\Ntilestrain{} tiles), set by the \dpone{} area and our curation cuts. Second, the Stage-1 labels are pipeline-derived: the over-deblending of \S\ref{sec:overdeblend} is only partly mitigated by the two-step filter, and the \texttt{refExtendedness} classes are unreliable beyond $i \approx 24$~mag \citep{Bosch2018,Merz2023}. Third, and consequently, what the network learns in Stage 1 is a point-like versus extended distinction rather than a physical one; the physical star/galaxy distinction enters only through the Stage-2 classification step, trained on the curated sample of \citet{Gatto2026}, and every classification result below refers to that step. Neither label source defines a quasar class, so quasars are assigned to one of the two classes rather than separated out (\S\ref{sec:future}, item~3). Fourth, star completeness depends strongly on the matching protocol (\S\ref{sec:headline}), so no single number summarizes detection quality for point-like sources at \dpone{} depth.

\subsection{Cross-comparison: \dpone{} source injection, Gatto et al.\ (2026), and Jiang et al.\ (2026)}
\label{sec:cross_compare}

Three external reference points place these results in context.

\paragraph{(i) \dpone{} source-injection tests \citep[\S5.6]{LSSTDP1}.}
The official \dpone{} validation campaign injects synthetic stars and galaxies into \ecdfs{} coadds and reruns the production \rsp{}. At the 80\% completeness threshold, $\sim 20\%$ of injected sources, mostly galaxies, are misclassified as stars by \texttt{refExtendedness}, and the fraction of correctly classified injected stars falls to $\sim 50\%$ at $i = 23.8$~mag. Our cross-match to HST measures the same ceiling directly on this field: \texttt{refExtendedness} recovers $\HSTRubinGalRec\%$ of HST galaxies but only $\HSTRubinStarRec\%$ of HST stars. This is the morphology-only baseline that both the image-based and the color-based classifiers are meant to improve on. The magnitude-resolved comparison at the $i = 23.8$~mag benchmark needs the public injection cubes and is left to future work (\S\ref{sec:future}, item~4).

\begin{figure*}[t]
\centering
\includegraphics[width=0.49\textwidth]{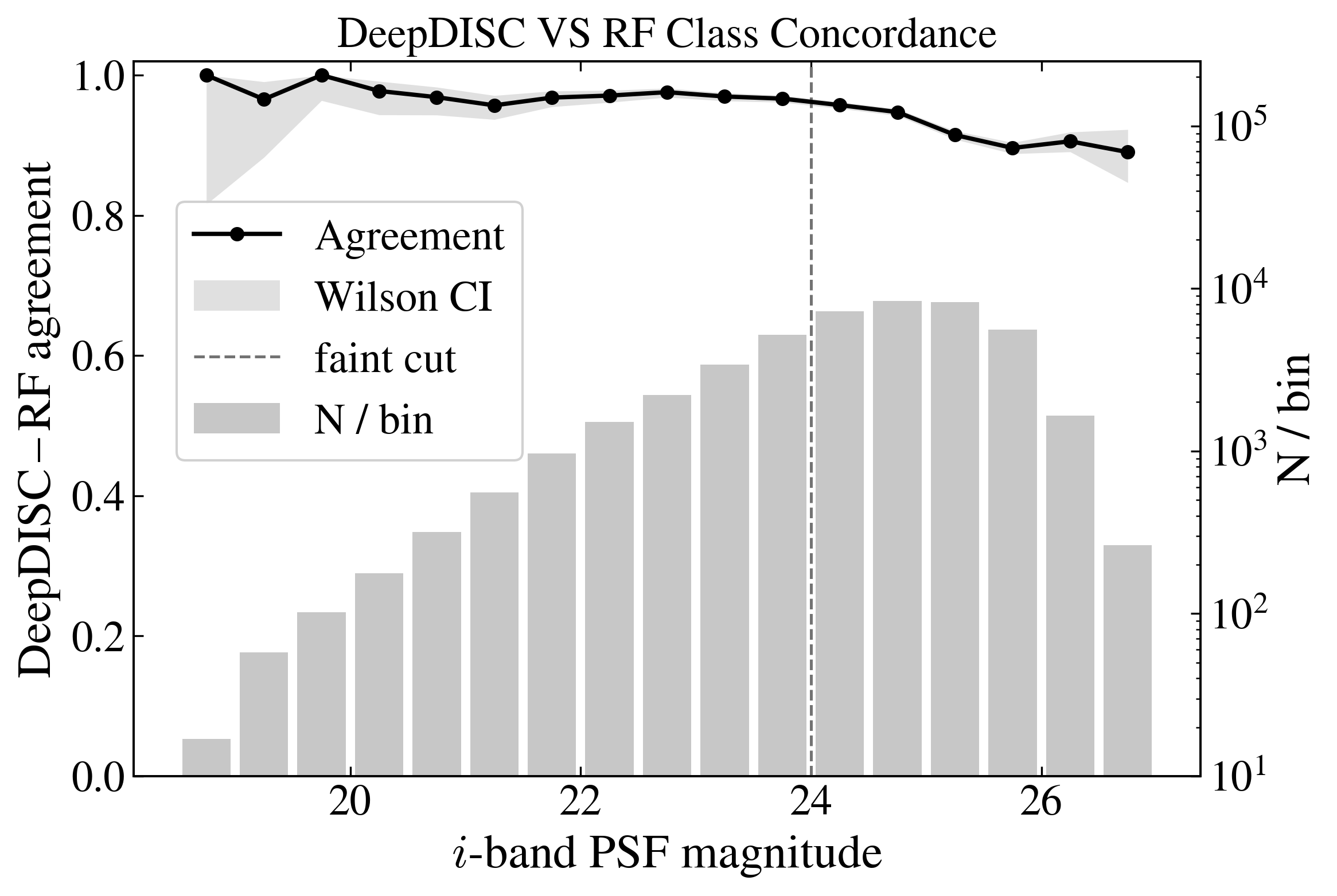}\hfill
\includegraphics[width=0.49\textwidth]{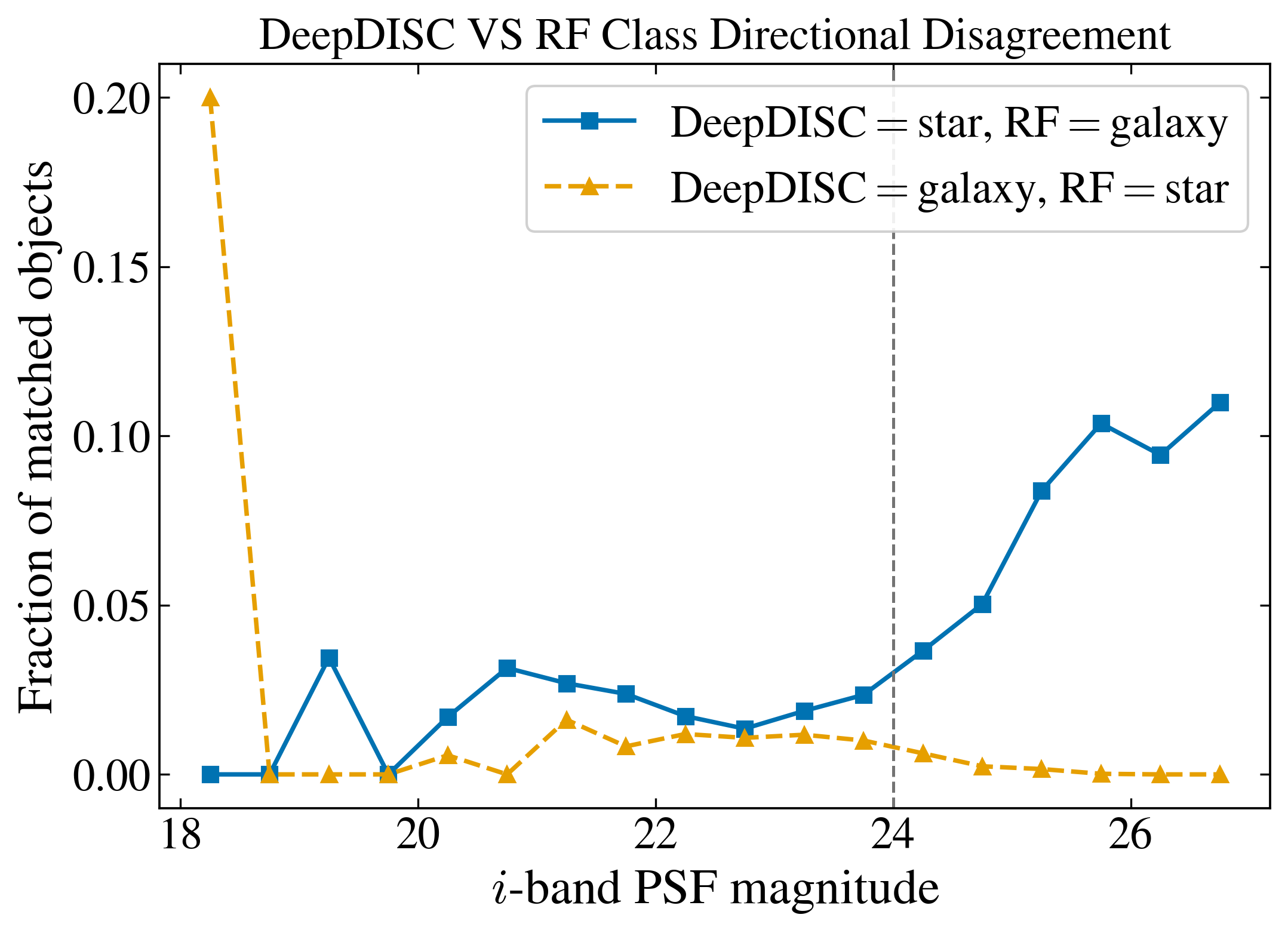}
\caption{Class agreement between \deepdisc{} and the full-field Random Forest
(\S\ref{sec:cross_compare}) on the \Nrfcommon{} \ecdfs{} objects held out by
both methods. The two assign the same class to \Nrfagree{} of them (\RFagree\%),
\Nrfagreegal{} galaxies and \Nrfagreestar{} stars, and differ on the remaining
\Nrfdisagree{}. Unlike the sample of Table~\ref{tab:gatto_compare}, these objects
carry no vetted labels, so the galaxy/star census is method-dependent: \deepdisc{}
calls \Nrfddgal{} galaxies and \Nrfddstar{} stars, the Random Forest \Nrfrfgal{}
and \Nrfrfstar{}. \emph{(a)}~Agreement per magnitude bin with Wilson $95\%$
intervals; the bootstrap aggregate is \RFagree\%, high to $i\approx23$ and
declining into the faint regime. \emph{(b)}~Direction of the \Nrfdisagree{}
disagreements: \NrfdisagreeDDstar{} are sources \deepdisc{} calls stars and the
Random Forest calls galaxies, against \NrfdisagreeDDgal{} in the opposite
direction, the former dominating toward the faint end.}
\label{fig:rf_agreement}
\end{figure*}

\paragraph{(ii) \citet{Gatto2026} Random Forest on \dpone{} \ecdfs{}.}
Our Stage-2 classification step and the Random Forest of \citet{Gatto2026} are trained on the same curated catalog, so they can be compared directly. We do so like for like: both are scored on classification alone, \deepdisc{} on the sources it detects and matches one-to-one (\S\ref{sec:cls_results}) so that it is not charged for detection completeness, and both drawn from the same \NholdoutHO{} objects, which neither method saw in training and which carry curated labels, and scored on the \NcommonHO{} of them that all three \deepdisc{} variants detect and match. Appendix~\ref{app:rf_concordance} documents that construction and the two Random Forest realizations it involves. The two methods land within about one percentage point of each other in overall accuracy (\DDnormAccHO\% versus \RFAccHO\%; Table~\ref{tab:gatto_compare}), are statistically indistinguishable on star recall, and differ mainly in star precision. Field-wide the picture is less favorable to \deepdisc{}: across \ecdfs{}, with both methods' training objects excluded, they agree on \RFagree\% of \Nrfcommon{} objects, high to $i \approx 23$ and declining into the faint regime, with \deepdisc{} increasingly calling stars what the Random Forest calls galaxies (Figure~\ref{fig:rf_agreement}); overall, \deepdisc{} assigns \Nrfddstar{} stars to the Random Forest's \Nrfrfstar{}. Where vetted labels exist to adjudicate the disagreements, they favor the Random Forest in about two thirds of cases (Appendix~\ref{app:rf_onetoone}).

We read the one-to-one result as near-parity, not as evidence that pixel-level information beats engineered colors; \deepdisc{} reaches it from a vetted label sample of similar size to the Random Forest's (Appendix~\ref{app:rf_concordance}). The test cannot decide the pixels-versus-colors question either way, because the vetted sample sits at $r \lesssim 22$~mag, where colors and morphology already work, and thins out exactly at the faint, blended end where scene-level information should matter most. Both methods also improve on morphology alone, although the available numbers are not on a common sample: $\sim 96$--$98\%$ accuracy here versus \texttt{refExtendedness} recall of \HSTRubinGalRec\%/\HSTRubinStarRec\% for HST galaxies/stars in reference point~(i) (see also the fourth panel of Figure~\ref{fig:cls_quality}). Two cautions. This classification accuracy is not the Stage-1 star \emph{detection} $F_1$ ($0.479$; \S\ref{sec:inference}), which also folds in detection misses and strict class-aware matching. And the methods are complementary rather than competing: a downstream stack could take detection and segmentation from \deepdisc{} and the final star/galaxy call from either classifier.

\begin{table*}
\raggedright
\caption{Star/galaxy classification on an object set held out by both methods: the benchmark RF's held-out 30\% validation sample (4{,}308 sources) joined to \deepdisc{}'s held-out test tiles by exact \texttt{objectId}, giving \NholdoutHO{} objects (863 galaxies, 241 stars) that neither method saw in training (Appendix~\ref{app:rf_concordance}). On its own validation split the benchmark RF reports galaxy precision/recall/$F_1$ of \RFgalPrec\%/\RFgalRec\%/\RFgalFone\% and star \RFstarPrec\%/\RFstarRec\%/\RFstarFone\%, for an overall \RFacc\% \citep[Table~2]{Gatto2026}; the columns here are recomputed on the common subset. All columns are scored on the \NcommonHO{} objects (796 galaxies, 230 stars) that all three \deepdisc{} variants detect and match one-to-one ($0.5\arcsec$ Hungarian matching at $\tau=0.22$), so \deepdisc{} is not charged for detection completeness ($\sim$95\% on the \NholdoutHO{} objects), matching the setting of the catalog-level RF, which classifies already-detected sources. The \deepdisc{} columns are the Stage-2 classification steps of the three loss configurations of Tables~\ref{tab:performance}--\ref{tab:performance_hung}; the RF column uses its per-object validation predictions (shared by Dr.\ Gatto), scored against the same vetted labels. All values percent; brackets are Wilson 95\% intervals, widest for stars because the subset contains only 230. Accuracy is the pooled fraction of correctly classified objects.}
\label{tab:gatto_compare}
\renewcommand{\arraystretch}{1.7}
\setlength{\tabcolsep}{18pt}

\begin{center}
\begin{tabular}{lcccc}
\hline\hline
Metric          & \deepdisc{}              & \deepdisc{}     & \deepdisc{}             & benchmark \\
(one-to-one)    & (no class weight)        & (star-weighted) & (star-weighted + focal) & RF \\
\hline
Galaxy recall    & \DDnormGalHO      & \DDswGalHO      & \DDswfGalHO      & \RFGalHO    \\[-2.5ex]
                 & [96.2, 98.4]      & [94.1, 96.9]    & [94.1, 96.9]     & [98.4, 99.7] \\
Star recall      & \DDnormStarHO     & \DDswStarHO     & \DDswfStarHO     & \RFStarHO   \\[-2.5ex]
                 & [93.9, 98.5]      & [95.0, 99.1]    & [94.4, 98.8]     & [92.7, 97.9] \\
Galaxy precision & \DDnormGalPrecHO  & \DDswGalPrecHO  & \DDswfGalPrecHO  & \RFGalPrecHO \\[-2.5ex]
                 & [98.2, 99.6]      & [98.5, 99.7]    & [98.3, 99.6]     & [97.9, 99.4] \\
Star precision   & \DDnormStarPrecHO & \DDswStarPrecHO & \DDswfStarPrecHO & \RFStarPrecHO \\[-2.5ex]
                 & [87.6, 94.6]      & [82.2, 90.5]    & [82.1, 90.4]     & [94.4, 98.8] \\
Galaxy $F_1$     & \DDnormGalFoneHO  & \DDswGalFoneHO  & \DDswfGalFoneHO  & \RFGalFoneHO \\
Star $F_1$       & \DDnormStarFoneHO & \DDswStarFoneHO & \DDswfStarFoneHO & \RFStarFoneHO \\
Overall accuracy & \DDnormAccHO      & \DDswAccHO      & \DDswfAccHO      & \RFAccHO      \\
\hline
\end{tabular}%
\end{center}

\end{table*}

\paragraph{(iii) \citet{Jiang2026} \deepdisc{}-Euclid on EDF-N.}
\citet{Jiang2026} are the closest contemporaneous \deepdisc{} reference: same framework, same \swinb{} backbone, and the same division of label sources between pipeline products for detection and spectroscopy for classification, but on Euclid Q1 imaging and with detection and classification trained together rather than in our two steps (\S\ref{sec:staged_training}). They report 93\% detection completeness against the Euclid MER catalog, $\sim 90\%$ true purity against JWST/NEXUS \citep{Shen2024}, and above 99\% recovery of spectroscopically confirmed stars and galaxies. Their spectroscopic label set is several times larger than ours and includes quasars as a third class, which motivates items~2 and~3 of \S\ref{sec:future}. Their purity and our HST-anchored \HSTPur\% (\S\ref{sec:hst_results}) are not directly comparable, given the different depths and reference catalogs, but they address the same question: how much of the apparent purity is real, and how much is inherited from the upstream pipeline.

\subsection{Comparison with the LSST production pipeline}
\label{sec:rsp_compare}

A direct comparison against the production \rsp{} catalog is complicated by differences in matching conventions, depth limits, and class definitions, and by the absence of a complete reference catalog over \ecdfs{}+\edfs{}. Two anchors are available. On the simulated DC2 footprint, processed through the same pipeline, \citet{EjjagiriMerz2026} find \deepdisc{} slightly more complete and slightly less pure than the \rsp{} Object catalog at $i < 25.3$, with a $\sim 2$ percentage-point lower UB fraction. On the real \dpone{} \ecdfs{} field, our HST-anchored UB measurement of the production catalog, $\HSTRubinUB\%$ (\S\ref{sec:hst_results}), reproduces the $\approx 18\%$ of \citet{Adari2025} under consistent matching conventions; residual protocol differences are treated in a separate methods paper.

\subsection{Connection to weak-lensing and large-scale-structure systematics}

The UB fraction is not only a detection-quality metric; it is a known cosmology systematic. \citet{Dawson2016} estimate a $\sim 14\%$ increase in shear-measurement noise at LSST depth from unrecognized blends, \citet{Boucaud2020} find photometric biases above $0.75$~mag for $\sim 12\%$ of artificially blended CANDELS galaxies, and Stage-IV forecasts \citep{Nourbakhsh2022} show that without mitigation the resulting shifts on $\Omega_m$ and $S_8$ become statistically significant. A detector that lowers the UB fraction at fixed completeness would therefore buy a real systematics margin; establishing whether \deepdisc{} does so requires a same-footprint comparison at matched completeness, which the present \dpone{}$\times$CANDELS overlap is too small to provide. Propagating our catalog into shear or 3$\times$2pt analyses is left to a study in preparation with the DESC photo-$z$ and weak-lensing working groups.

\subsection{Identified improvements and future work}
\label{sec:future}

Four improvements to the pipeline follow from the comparisons above, in priority order, together with two extensions beyond \dpone{}.

\paragraph{1. \mvit{} ablation against the \swinb{} baseline.}
We chose \swinb{} for the \citet{Merz2023} HSC results, the \citet{Jiang2026} Euclid precedent, and the availability of an in-domain LSST checkpoint at this backbone, but \citet{Merz2023} found \mvit{} marginally ahead at fixed compute and \citet{Merz2025} adopted it for the DC2 photo-$z$ extension. The HSC comparison suggests at most a modest difference, but a controlled retraining is worth doing.

\paragraph{2. Self-supervised in-domain pretraining at \dptwo{} scale.}
The LSST checkpoint is itself supervised on simulations, and the \dpone{} pool of unlabeled tiles is far smaller than \dptwo{} will offer. Pretraining directly on unlabeled six-band tiles \citep{Caron2021,Oquab2024,Simeoni2025} reduces the dependence on the \scarlet{}-derived labels, which would then be needed only for fine-tuning, and should gain disproportionately from the larger pool. This is underway on an \mvit{}-B backbone using the unlabeled tiles of this dataset, with preliminary evidence that the learned features already separate sources from sky (Appendix~\ref{app:ssl_pilot}).

\paragraph{3. Adding a quasar class.}
Neither of our label sources defines a quasar class, so \S\ref{sec:cls_results} reports a two-class decision. \citet{Jiang2026} recover quasars at $\sim 85\%$ once DESI DR1 spectroscopy is folded in, and \ecdfs{} has substantial DESI overlap, so adding those classifications and redshifts would give a third class and, with them, simultaneous galaxy and quasar photometric redshifts.

\paragraph{4. Absolute faint-end accuracy via source injection.}
The classification comparison of \S\ref{sec:cross_compare} is anchored at $r \lesssim 22$~mag, where colors and morphology already work. We extend it across \ecdfs{} through Appendix~\ref{app:rf_concordance} and the \deepdisc{}$\times$CANDELS purity measurement of \S\ref{sec:hst_results}, but neither settles \emph{absolute} faint-end accuracy: agreement compares two imperfect methods, and CANDELS arbitrates only where it overlaps and resolves. The decisive test is the public \dpone{} injection cubes \citep{LSSTDP1} at $i = 23.8$~mag, where the injected truth answers the question directly.

\paragraph{5. Joint Rubin+Roman training.}
Higher-resolution near-infrared imaging offers a second route to lower UB fractions. On overlapping simulated Rubin and Roman scenes, \citet{EjjagiriMerz2026} find that adding Roman $YJH$ inputs cuts the simulated UB fraction from $\sim 8.3\%$ for a resolution-matched Rubin-only model to $\sim 6.6\%$ at $i < 25.3$ at effectively unchanged purity, and that work also explores distilling a Rubin-only model from the joint one, since the joint-training regime is bounded by the small survey overlap; so far distillation has not recovered the joint-survey gain. Confirmation on real data awaits the first LSST DR1 and Roman HLWAS overlap in the south.

\paragraph{6. \dptwo{} and DR1.}
\dptwo{} (Early \dptwo{} released July 2026, with the full release expected later in 2026\footnote{\url{https://dp2.lsst.io}}) brings more area, deeper coadds, more uniform band coverage, and, with larger contiguous fields, far fewer of the edge tiles that required masking in \S\ref{sec:filter}. The method transfers directly, but the data-ingestion code must be ported to the \dptwo{} data products, which use cell-based coadds, a new sky tessellation, and new \scarlet{} model outputs. Survey-wide, including low Galactic latitudes, the galaxy:star ratio is closer to $\sim 1{:}1$ (\S\ref{sec:intro}) than the $\sim 4{:}1$ of these extragalactic fields, so any class weighting should be re-derived for each training footprint rather than carried over from this one.

\section{Conclusions}
\label{sec:conclusion}

We have presented the first deployment of the \deepdisc{} scene-level deep-learning framework on real LSST data, using the \dpone{} \ecdfs{} and \edfs{} fields. Our principal findings are:

\begin{enumerate}
\item \deepdisc{} trains stably on \dpone{} six-band deep coadds with a \swinb{} backbone and produces a coherent detection, instance segmentation, and star/galaxy classification output at scene-scale resolution. The analysis pipeline is available to DESC members (see Data Availability).
\item A curation pipeline for \dpone{} coadds and \scarlet{} labels (border crop, bright-star shredding filters, and tile-level quality gates) removes the dominant label artifacts, notably shredded star annotations around bright stars, before training.
\item In-domain (simulated-LSST) pretraining gives a much stronger zero-shot starting point than ImageNet-21k pretraining (whose ROI heads must be reinitialized), but the advantage largely vanishes after \dpone{} fine-tuning, when the two initializations converge to near-parity under identical training (Appendix~\ref{app:init}). Simulated-data pretraining is therefore a useful warm start but not a prerequisite for strong \dpone{} performance.
\item The matching-protocol dependence of the star numbers ($\sim$\CstarA\% under 5-condition vs.\ $\sim$\CstarH\% under Hungarian for the same model) shows that most of the stars missing under the 5-condition protocol are detected but not labelled star at Stage 1. Star-weighted training shifts the trade-off in the expected direction ($\sim$\CgalCWA\% / $\sim$\CstarCWA\% under 5-condition; $\sim$\CgalCWH\% / $\sim$\CstarCWH\% under Hungarian) at a cost in Stage-2 star precision; the appropriate weighting depends on the target population (the $\sim 6{:}1$ galaxy:star ratio of this extragalactic training set versus $\sim 1{:}1$ survey-wide).
\item A hybrid label strategy (\scarlet{}-derived per-pixel masks for detection and instance-segmentation supervision, and the curated \citet{Gatto2026} catalog for the Stage-2 classification supervision), combined with a two-stage training protocol in which the detector is frozen while the classification part is fine-tuned, keeps each label source in the regime where it is most reliable. Stage-2 classification fine-tuning is computationally cheap and does not require redoing detection training.
\item All headline metrics are computed on tiles free of \dpone{} band-dropout regions, where \scarlet{} produces no pseudo-ground truth and matching is ill-defined. As an external, \scarlet{}-independent check within the joint \dpone{}~$\cap$~CANDELS \ecdfs{} footprint, the \dpone{} production catalog cross-matched against HST/CANDELS yields a true purity of $\HSTRubinPur\%$, an HST-depth-limited true completeness of $\HSTRubinComp\%$, and an HST-anchored UB fraction of $\HSTRubinUB\%$ that reproduces the $\approx 18\%$ measurement of \citet{Adari2025} on the same field (Table~\ref{tab:hst_validation}). On a leakage-free held-out HST/CANDELS footprint (\texttt{T5063\_P25/35}, unseen by both the detector and the classifier), \deepdisc{} reaches a true purity of $\HSTPur\%$ and recovers HST galaxies and stars at $\HSTGalRec\%/\HSTStarRec\%$ recall (the latter from 63 stars). It is less complete than the production catalog on the same patches ($\HSTComp\%$ vs.\ $\HSTRubinCompHeldout\%$); its purity exceeds the production catalog's on those patches ($\HSTRubinPurHeldout\%$) but matches the production catalog restricted to our curated tiles ($\HSTSetPur\%$), so we do not claim a purity advantage. Its HST-anchored UB fraction ($\HSTUB\%$, 95\% CI $[\HSTUBlo,\HSTUBhi]\%$) is consistent with or modestly below the production value ($\HSTRubinUB\%$).

\item Under the two-step FOF+Hungarian protocol of \citet{Mao2021,EjjagiriMerz2026}, the \scarlet{}-anchored unrecognized-blend fraction is $\sim$\UBfrac\% at $L=1\arcsec$ and $\sim$\UBfracTwo\% at $L=2\arcsec$; the HST-anchored UB fractions ($\HSTUB\%$ for \deepdisc{} and $\HSTRubinUB\%$ for the \dpone{} production catalog) are the values directly comparable to the $\approx 18\%$ HST cross-match measurement of \citet{Adari2025} for the same field and to simulation-based predictions \citep{Troxel2023,Nourbakhsh2022}.
\item Cross-comparison with the official \dpone{} source-injection tests \citep{LSSTDP1}, the photometry-only Random Forest classifier of \citet{Gatto2026} on the same field, and the contemporaneous Euclid Q1 \deepdisc{} application of \citet{Jiang2026} establishes \deepdisc{} as a viable image-based complement to catalog-level classifiers. On a like-for-like, classification-only comparison (both methods scored on an identical, doubly-held-out object set of \NcommonHO{} detected+matched sources, so \deepdisc{} is not penalized for detection or segmentation), \deepdisc{} reaches \DDnormAccHO\% accuracy against the benchmark Random Forest's \RFAccHO\%, within about one percentage point: near-parity rather than a clear win for pixel-level features. Field-wide, \deepdisc{} labels more faint sources as stars than the Random Forest (\Nrfddstar{} vs.\ \Nrfrfstar{} stars), and independent labels favor the Random Forest in most disagreements. Because the held-out labels sit mostly at $r \lesssim 22$~mag, this comparison does not yet test the faint, blended regime where scene-level information is expected to add the most value; that test is deferred to the magnitude-resolved follow-up (\S\ref{sec:future}).
\end{enumerate}

This work establishes \deepdisc{} as a viable real-data complement to the production \rsp{} for LSST deep-field analysis and lays the groundwork for a forthcoming self-supervised, semi-supervised, and joint-survey extension. The method established here carries over to \dptwo{} once the data-ingestion code is ported to the \dptwo{} data products, and provides the supervised baseline against which forthcoming self-supervised pretraining and joint Rubin+Roman work will be compared.

\section*{Data Availability}

The \deepdisc{} code is publicly available at \url{https://github.com/grantmerz/deepdisc}. The analysis pipeline for this paper (data retrieval, curation, training, inference, matching, and figure scripts) is available to DESC members at \url{https://github.com/LSSTDESC/DeepDISC_plus}. Trained model weights (Stage 1 and Stage 2), the curated \dpone{} annotation catalog with per-stage tile lists, and a value-added catalog of 70{,}815 \dpone{} \ecdfs{} objects outside the training and validation footprint, each carrying the Stage-2 \deepdisc{} star/galaxy class, classification score, and matched \dpone{} \texttt{objectId} ($0.5\arcsec$ match), will be available upon request to Rubin data rights holders. Furthermore, \dpone{} data is already available to Rubin data rights holders via the Rubin Science Platform.

\section*{Acknowledgments}

This paper has undergone internal review within the LSST Dark Energy Science Collaboration (DESC). We sincerely thank our internal reviewers, Johann Cohen-Tanugi and Arun Kannawadi, for their thoughtful comments and suggestions, which helped improve the paper.

We sincerely thank Dr.\ M.~Gatto for sharing their curated \ecdfs{} star/galaxy catalog, the full-field Random Forest prediction catalog over \ecdfs{}, and the per-object predictions of the benchmark Random Forest's held-out validation sample, which enabled the one-to-one classification comparison of Table~\ref{tab:gatto_compare}.

Author contributions: P.Y.\ led the project: the data retrieval and curation, the pseudo-ground-truth construction, the two-stage training, the inference and matching pipelines, the HST/CANDELS and Random Forest comparisons, all figures and tables, the paper writing, and the response to the DESC internal review. X.L.\ advised P.Y., planned the tests and analyses, and contributed to writing the paper. Y.S.E.\ helped set up \deepdisc{} and the first inference and advised on constructing the \scarlet{} pseudo-ground truth. G.M.\ developed the \deepdisc{} codebase and helped set up \deepdisc{} and the first inference.

We thank the Rubin Observatory and the LSST Data Management team for the production of \dpone{} and the Rubin Science Platform. We thank the LSST Dark Energy Science Collaboration deblending working group for useful discussions on \scarlet{} systematics. We thank Drs.\ V. Kindratenko, S. Luo, and B. Bode at NCSA for GPU computing support.

We acknowledge support from the Illinois Campus Research Board Award RB25035, NSF grant AST-2308174, and NASA grants 80NSSC24K0219 and 80NSSC26K0333. This work was supported in part by LSST Discovery Alliance and the Heising-Simons Foundation through Grant 2026-HSF-ES-02-Liu. The views expressed are those of the author(s) and do not necessarily reflect the views of LSST Discovery Alliance or its funding partners.

This work used Delta and DeltaAI at NCSA through allocations PHY240290 and PHY250386 from the Advanced Cyberinfrastructure Coordination Ecosystem: Services \& Support (ACCESS) program, which is supported by U.S. National Science Foundation grants \#2138259, \#2138286, \#2138307, \#2137603, and \#2138296. All model training and inference were run on Delta and DeltaAI. The \dpone{} data retrieval and the \scarlet{} pseudo-ground-truth mask generation, which are CPU-only steps, used DESC computing resources at the National Energy Research Scientific Computing Center (NERSC).

The DESC acknowledges ongoing support from the Institut National de Physique Nucl\'eaire et de Physique des Particules in France; the Science \& Technology Facilities Council in the United Kingdom; and the Department of Energy and the LSST Discovery Alliance in the United States. DESC uses resources of the IN2P3 Computing Center (CC-IN2P3--Lyon/Villeurbanne - France) funded by the Centre National de la Recherche Scientifique; the National Energy Research Scientific Computing Center, a DOE Office of Science User Facility supported by the Office of Science of the U.S.\ Department of Energy under Contract No.\ DE-AC02-05CH11231; STFC DiRAC HPC Facilities, funded by UK BEIS National E-infrastructure capital grants; and the UK particle physics grid, supported by the GridPP Collaboration. This work was performed in part under DOE Contract DE-AC02-76SF00515.

This material is based upon work supported in part by the National Science Foundation through Cooperative Agreements AST-1258333 and AST-2241526 and Cooperative Support Agreements AST-1202910 and 2211468 managed by the Association of Universities for Research in Astronomy (AURA), and the Department of Energy under Contract No.\ DE-AC02-76SF00515 with the SLAC National Accelerator Laboratory managed by Stanford University. Additional Rubin Observatory funding comes from private donations, grants to universities, and in-kind support from LSST-DA Institutional Members.

This publication is based in part on proprietary Rubin Observatory Legacy Survey of Space and Time (LSST) data, and was prepared in accordance with the Rubin Observatory data rights and access policies. All authors of this publication meet the requirements for co-authorship of proprietary LSST data.

The use of AI language tools in preparing this manuscript and its analysis pipeline is declared in the AI Assistance Statement that follows.

\section*{AI Assistance Statement}

The authors used Anthropic's Claude (Claude Opus 5 and Claude Fable 5.1, through Claude Code). For the manuscript, the tool was used to refine, correct, edit, and format the text and \LaTeX{} source to improve clarity of language and readability, and, at the authors' direction, to cross-check stated numbers and method descriptions against the pipeline outputs. For the pipeline, it assisted with code debugging, author-specified modifications to the plotting scripts, and documentation of the repository released with this paper. Additionally, ChatGPT (OpenAI; GPT-5.6) was used minimally for occasional assistance with code debugging and implementation. All code and scientific results were reviewed and validated by the authors. The authors reviewed every proposed change and accepted, revised, or rejected it. All scientific ideas, methods, experiments, analyses, results, interpretations, and conclusions are the authors' own and were independently verified by them, and the authors take full responsibility for the final manuscript.

\facilities{Rubin LSST}

\software{
Detectron2 \citep{wu2019detectron2},
DeepDISC \citep{Merz2023,Merz2025,Merz2026,Jiang2026,EjjagiriMerz2026}
}
\clearpage
\appendix

\section{Data-curation summary}
\label{app:curation}

Table~\ref{tab:curation} consolidates the per-step parameters and the
tile/object funnel of the data-curation pipeline described in
\S\ref{sec:data} and \S\ref{sec:gt}, provided here for reproducibility.
Counts are listed as train\,/\,val\,/\,test where they differ by split;
percentile cuts are evaluated on the combined multi-tract \scarlet{} catalog.
Note that the class integers differ between products: the \dpone{} Object catalog's
\texttt{refExtendedness} is $0$ for point-like (star-like) and $1$ for extended
(galaxy-like) sources, whereas the training tiles use the \detectron{} convention
$\{0{:}\text{galaxy},\,1{:}\text{star}\}$; the value is flipped at tile-creation time.

\begin{table}[htbp]
\centering
\caption{Summary of the \dpone{} \ecdfs{}+\edfs{} data-curation pipeline
(\S\ref{sec:data}, \S\ref{sec:gt}): per-step parameters and the tile/object
funnel. Counts are train\,/\,val\,/\,test where they differ by split.}
\label{tab:curation}
\footnotesize
\setlength{\tabcolsep}{8pt}
\resizebox{0.58\columnwidth}{!}{%
\begin{tabular}{@{}lr@{}}
\hline\hline
Quantity & Value \\
\hline
\multicolumn{2}{@{}l}{\textbf{A. Patch level} (full $3400\times3400$ patches, all tracts)}\\
\hline
\multicolumn{2}{@{}l}{\textit{1. \scarlet{} pseudo-ground truth}}\\
Lower mask threshold                & $2\sigma$ \\
Parallel workers                    & 121 \\
\multicolumn{2}{@{}l}{\textit{2. Star percentile filter (combined multi-tract catalog)}}\\
Width / height cut (p98)            & $45$ px \\
Area cut (p98)                      & $1{,}980$ px$^2$ \\
SNR cut (p99)                       & $1{,}409$ \\
Stars removed                       & \NstarStageOneRem{} (\StageOneStarRemFrac\%) \\
\multicolumn{2}{@{}l}{\textit{3. Patch screening and border crop}}\\
Patch size                          & $3400\times3400$ px \\
Patches rejected                    & any NaN in any band; $<70\%$ valid coverage \\
Tracts retained                     & 2234, 2394, 4848, 4849, 5063 \\
Border crop                         & 200 px ($\to 3000\times3000$) \\
Annotations retained after crop     & $99.47\%$ \\
\multicolumn{2}{@{}l}{\textit{4. Patch split}}\\
Train / val / test                  & $70/11/20$ of \Npatches{} patches \\
\hline
\multicolumn{2}{@{}l}{\textbf{B. Tile level} ($760\times760$ cutouts, per split; counts train\,/\,val\,/\,test)}\\
\hline
\multicolumn{2}{@{}l}{\textit{5. Tiling}}\\
Tile size / stride                  & $760$ / $560$ px \\
Raw tiles (after spread filter)     & \Ntilestrainraw{} / \Ntilesvalraw{} / \Ntilestestraw{} \\
\quad total before tile-level filtering & \Ntilesrawtot{} \\
\multicolumn{2}{@{}l}{\textit{6. Fill-value / artifact masking}}\\
Tiles removed                       & \Ntilesfillmask{} \\
\multicolumn{2}{@{}l}{\textit{7. Bright-star mask}}\\
Bright-star area threshold          & $1{,}600$ px$^2$ \\
Inner / outer exclusion zone        & $0.6$ / $2.0\,r_{\rm eff}$ \\
Outer-zone overlap (IoS)            & $>1\%$ \\
Bright-star centers / stars removed & \Nbrightstar{} / \Nremoved{} \\
Star annotations after tiling       & $\sim$\NstarTilingRaw{} \\
\multicolumn{2}{@{}l}{\textit{8. Minimum-annotation filter}}\\
Min.\ annotations per tile          & 50 \\
Tiles after fill/mask/density filters & \NtilestrainPre{} / \NtilesvalPre{} / \NtilestestPre{} \\
\multicolumn{2}{@{}l}{\textit{9. Final image-quality gates}}\\
Low dynamic range (mean Lupton $\sigma$)  & $<0.003$ \\
Partial band coverage (band $\sigma$)     & $<10\%$ of median \\
Tiles dropped (low-DR / partial)    & \Ntilesdrop{} (\Ntilesdropdr{}/\Ntilesdropcov{}; two fail both) \\
Partial-coverage rejections involving $u$ / $y$ & \NuAny{} / \NyAny{} \\
\quad rejected on $u$ alone                    & \NuOnly{} \\
Annotations / objects lost in this pass         & \Nanndroplast{} / \Nobjdroplast{} \\
Final tiles                         & \Ntilestrain{} / \Ntilesval{} / \Ntilestest{} \\
\quad total / discarded by tile-level filtering & \Ntilestot{} / \Ntilesqcdrop{} ($20\%$) \\
\hline
\multicolumn{2}{@{}l}{\textit{10. Objects (final tile set)}}\\
Unique objects                      & \Nobjtrain{} / \Nobjval{} / \Nobjtest{} \\
Grand total                         & \Nobjall{} (\Ngalall{} galaxies, \Nstarall{} stars) \\
\hline
\end{tabular}
}
\end{table}

\clearpage
\section{Tiling and quality-control examples}
\label{app:tiles}

\begin{figure*}[!ht]
\centering
\includegraphics[width=1\textwidth]{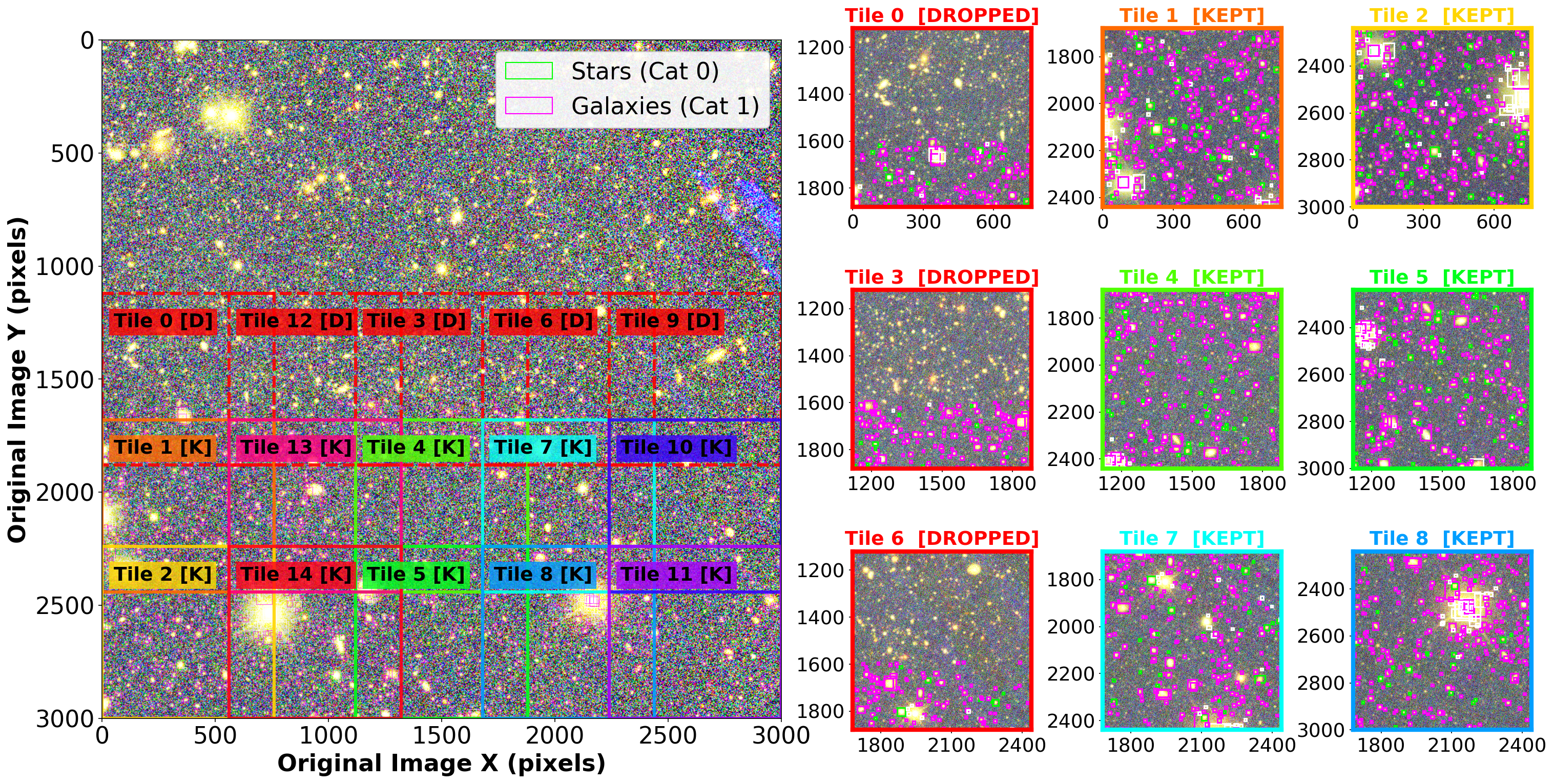}
\caption{Tiling of a \dpone{} patch into overlapping cutouts (\S\ref{sec:tiles}).
\emph{Left:} a cropped $3000\times3000$ patch with the $760\times760$ tile grid overlaid
(560-pixel stride, 200-pixel overlap); each tile is marked kept~[K] or dropped~[D], and
the \scarlet{} annotations are colour-coded (stars green, galaxies magenta).
\emph{Right:} the individual tile cutouts, labelled by index, kept/dropped status, and
pixel origin; tiles dropped by quality control are outlined in red. For this patch, 10
of the 15 tiles are kept and 5 are dropped ($2{,}072$ annotations total).}
\label{fig:cutout_tiling}
\end{figure*}

\begin{figure*}[!ht]
\centering
\includegraphics[width=0.9\textwidth]{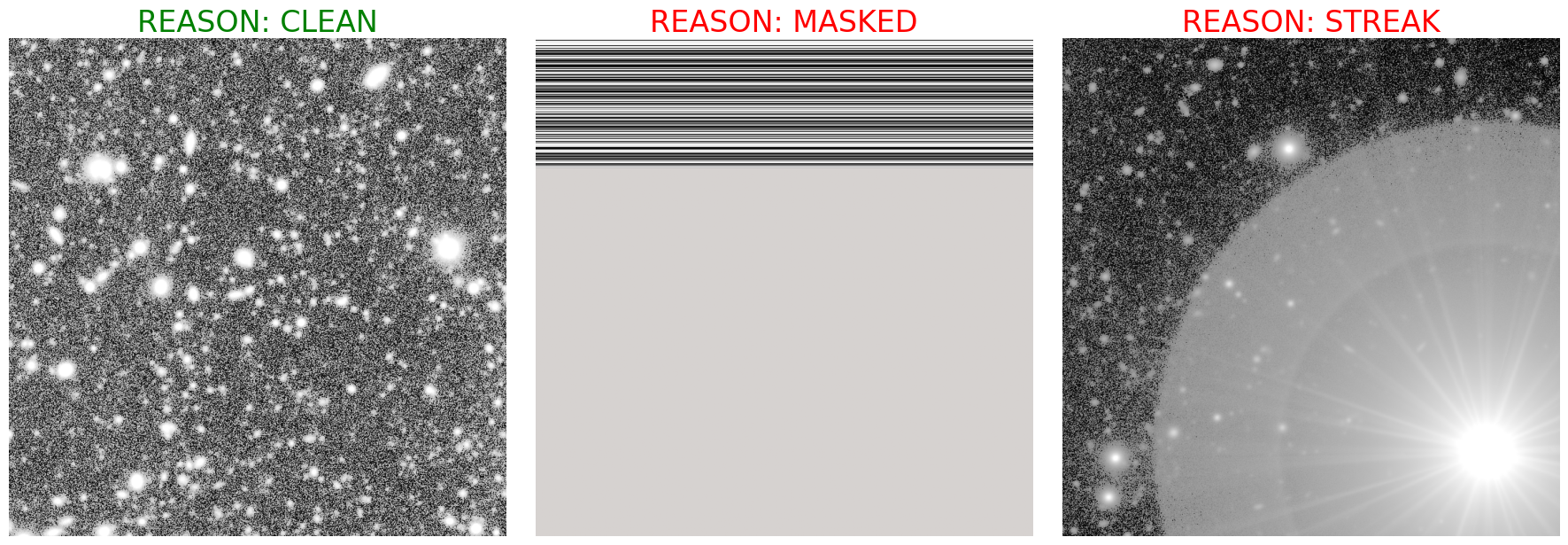}
\caption{Representative tile-level quality-control outcomes and their causes
(\S\ref{sec:filter}). \emph{Left (clean):} a tile with full
$ugrizy$ coverage and no artifacts, retained. \emph{Centre (masked):} a tile dominated
by no-data / partial-band regions (the uniform grey and striped bands, where one or
more $ugrizy$ bands are missing), rejected. \emph{Right (streak):} a tile contaminated
by a bright star with an extended halo and diffraction spikes, rejected. These
categories motivate the band-dropout/fill rejection and bright-star handling of the
curation pipeline (\S\ref{sec:gt}).}
\label{fig:qc_reasons}
\end{figure*}

\begin{figure*}[!ht]
\centering
\includegraphics[width=0.576\textwidth]{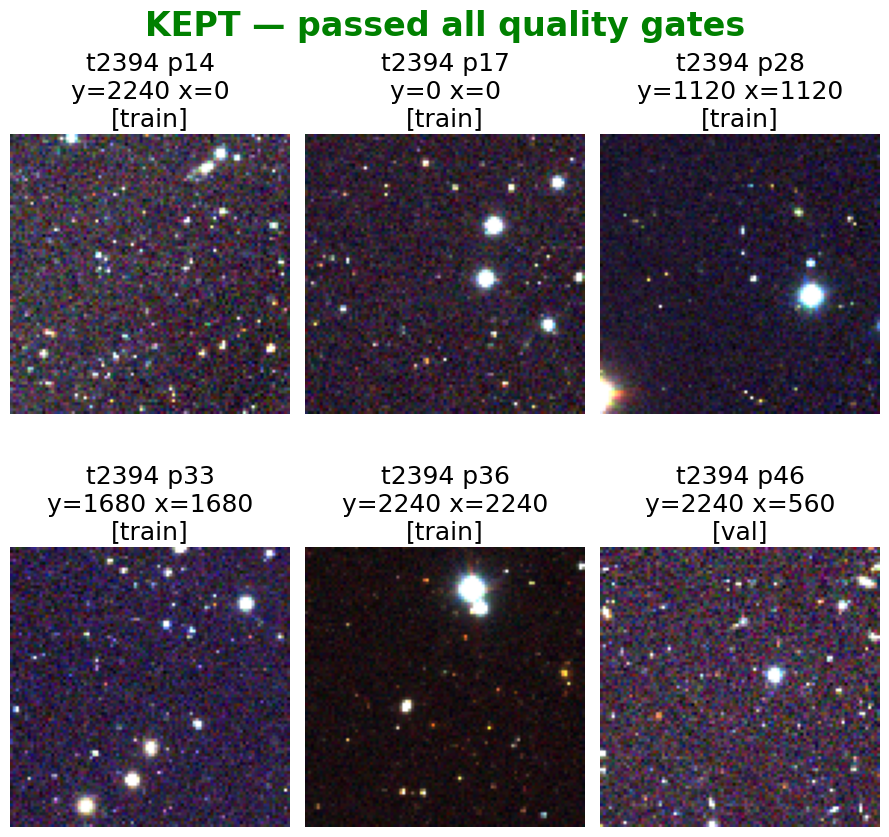}\\[3pt]
\includegraphics[width=0.576\textwidth]{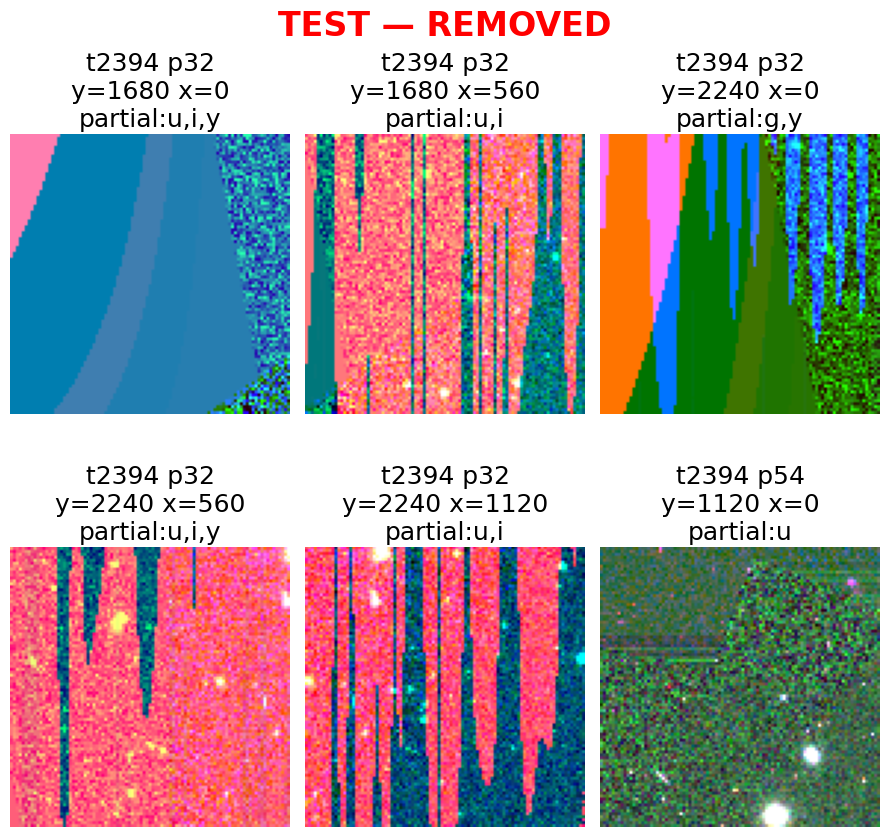}
\caption{Final outcome of the tile-level quality control (\S\ref{sec:filter}). \emph{Top:} a random sample of tiles that pass all quality gates and enter the $\Ntilestot{}$-tile training/validation/test set (each labelled by tract/patch, cutout origin, and split); these have full $ugrizy$ coverage, finite pixels, and adequate dynamic range. \emph{Bottom:} six of the test-split tiles removed by the final image-quality pass, each flagged for partial $ugrizy$ band coverage (the saturated false-colour stripes and bands, with the missing band(s) listed per tile). The contrast between clean multi-band tiles and band-dropout artifacts is precisely what the curation pipeline is built to separate (tiles shown at reduced resolution).}
\label{fig:final_kept_removed}
\end{figure*}

This appendix illustrates the tiling and quality-control steps of
\S\ref{sec:tiles}--\S\ref{sec:filter}. Figure~\ref{fig:cutout_tiling} shows how a
cropped patch is sliced into overlapping cutouts and which tiles survive quality
control, Figure~\ref{fig:qc_reasons} shows the principal rejection categories, and
Figure~\ref{fig:final_kept_removed} shows the final kept-versus-removed tile sets.

\clearpage
\section{Sample pseudo-ground-truth training tiles}
\label{app:gt_samples}

\begin{figure*}[!ht]
\centering
\includegraphics[width=0.49\textwidth]{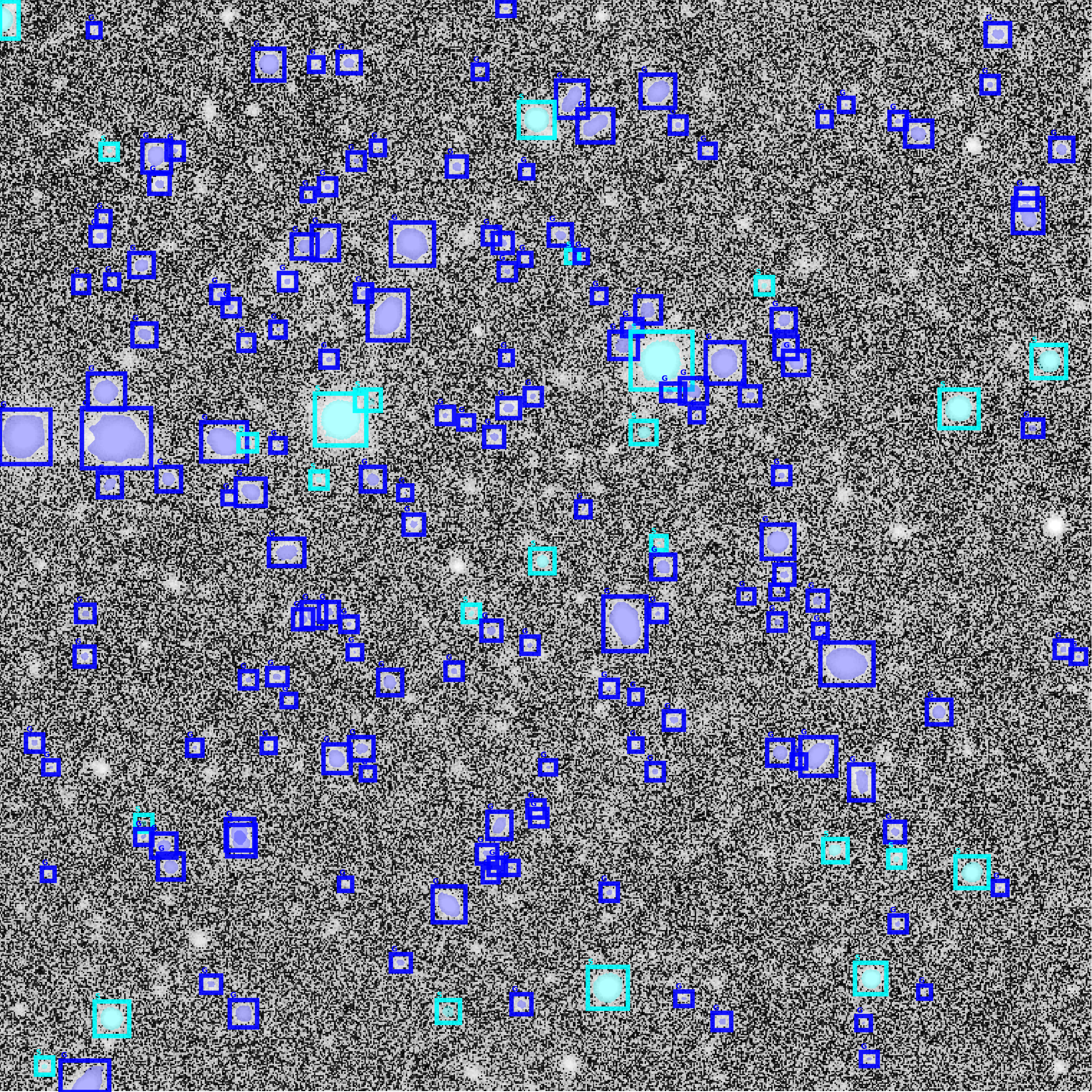}\hfill
\includegraphics[width=0.49\textwidth]{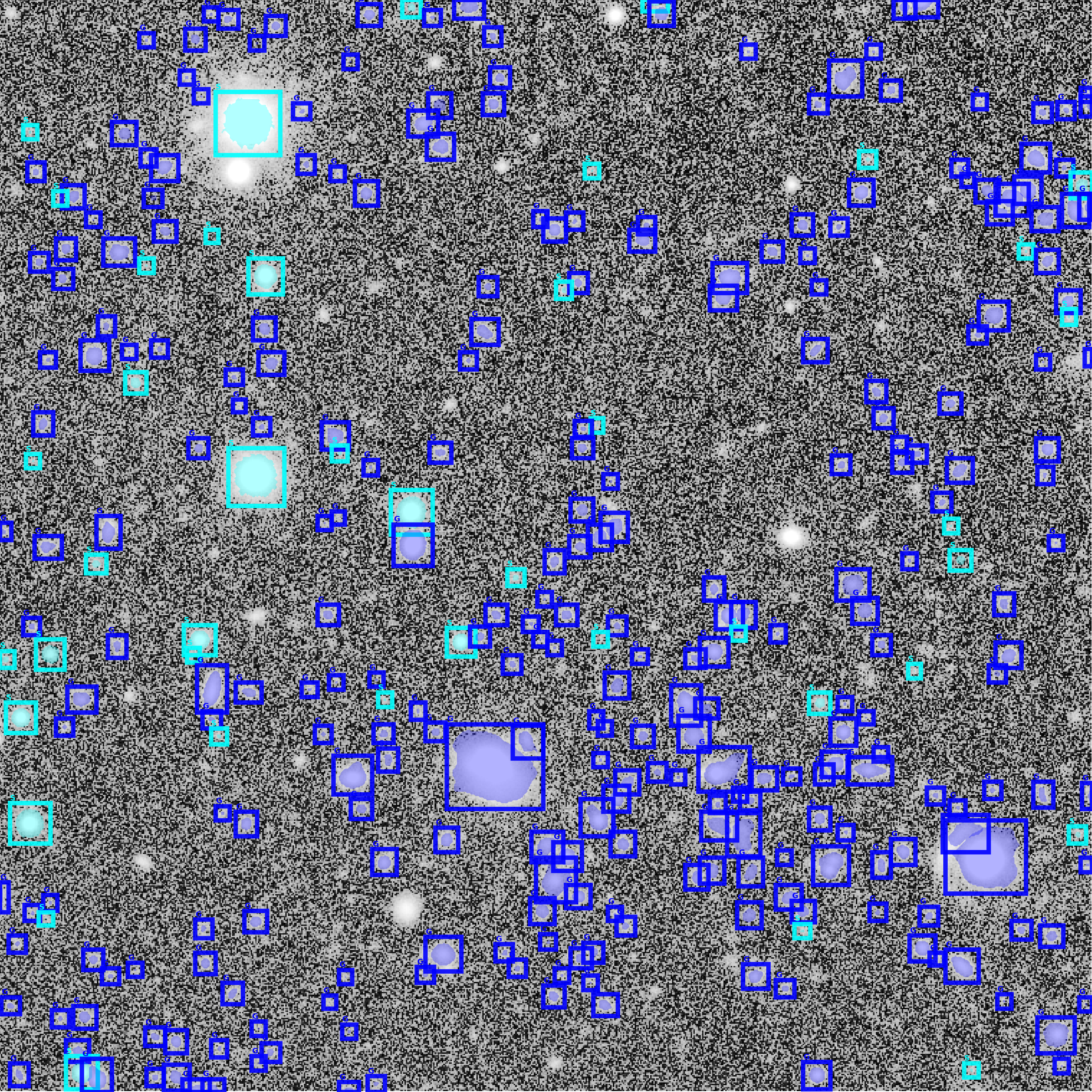}
\caption{Two representative pseudo-ground-truth tiles fed into \deepdisc{} Stage-1
(detection) training (\S\ref{sec:scarlet_gt}). The \scarlet{}-derived annotations are
overlaid on a grayscale rendering of each tile (shown for display only; the model ingests
the six-band stack, \S\ref{sec:multiband}), colour-coded by class (galaxies
blue, labelled ``G''; stars cyan, labelled ``S'').
\emph{Left:} tile \texttt{T2394\_P7\_C\_1120\_2240} (tract~2394, patch~7;
185 objects). \emph{Right:} tile \texttt{T4848\_P81\_C\_1120\_560} (tract~4848,
patch~81; 325 objects). These are the labels the model is trained to reproduce, and they inherit the
biases of the upstream \scarlet{} deblender discussed in \S\ref{sec:gt}.}
\label{fig:gt_samples}
\end{figure*}

Figure~\ref{fig:gt_samples} shows two representative pseudo-ground-truth tiles as they
enter \deepdisc{} Stage-1 (detection and instance-segmentation) training, with the
\scarlet{}-derived annotations overlaid (\S\ref{sec:scarlet_gt}). These are precisely the
COCO-format labels the network is trained to reproduce; the star/galaxy classification
labels used in Stage-2 come from the separate, curated
\citet{Gatto2026} sample and are not shown here. In each panel the \scarlet{} bounding
boxes are colour-coded by class (galaxies in blue, labelled ``G''; stars in cyan, labelled
``S''), with the tile identifier (encoding tract, patch, and cutout pixel origin) and the
total object count given in the caption. For display the boxes are overlaid on a
grayscale rendering of the tile; the model itself ingests the full six-band
stack (\S\ref{sec:multiband}). Note the high source density (185 and 325 objects,
respectively) and the mix of faint, blended, and bright saturated sources characteristic
of the \dpone{} \ecdfs{}+\edfs{} Deep Fields.

\clearpage
\section{Additional training details}
\label{app:training_detail}

\begin{figure*}[htbp]
\centering
\includegraphics[width=0.7\textwidth]{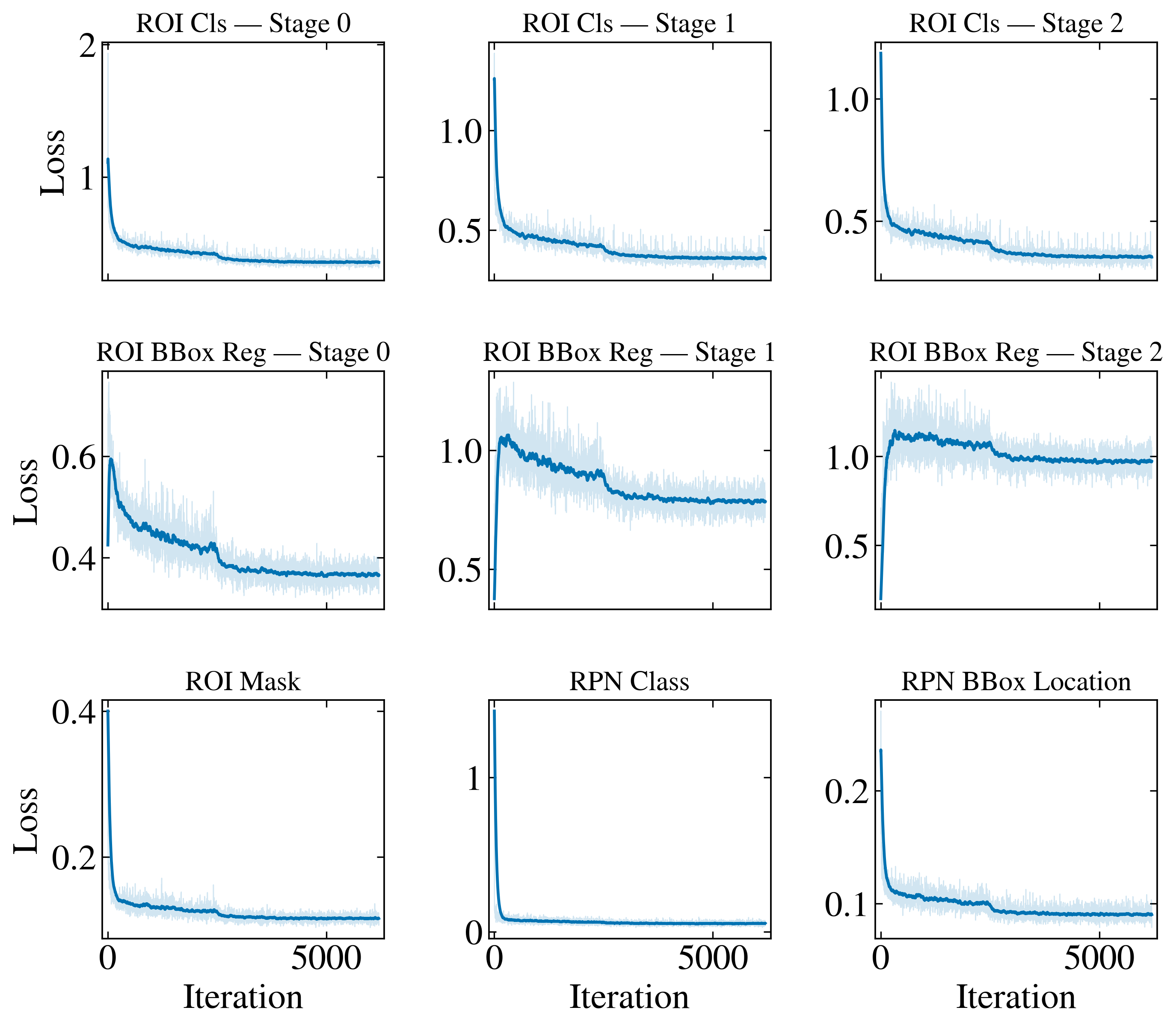}
\caption{Stage-1 training losses for the headline (no class weight) run initialized from the LSST checkpoint. \textit{Top row:} ROI-head classification losses for the three cascade stages. \textit{Middle row:} ROI-head bounding-box regression losses for the three cascade stages. \textit{Bottom row:} ROI-head mask loss (left) and the two RPN losses (class and box location, centre/right), which converge cleanly (see text).}
\label{fig:losses}
\end{figure*}

This appendix collects the training material that supports
\S\ref{sec:staged_training} and \S\ref{sec:training} but is not required to
follow the main results: the Stage-1 training-convergence diagnostics
(Appendix~\ref{app:loss_curves}), the Stage-2 classification-head
fine-tuning schematic (Appendix~\ref{app:stage2_schematic}), the magnitude
distribution of the Stage-2 classification labels (Appendix~\ref{app:label_mag_dist}), and the
comparison of in-domain and ImageNet-21k initialization (Appendix~\ref{app:init}).

\subsection{Stage-1 training convergence and Cascade R-CNN stage behavior}
\label{app:loss_curves}

Figure~\ref{fig:losses} shows the training loss curves for the headline run initialized from the LSST checkpoint. The RPN classification and bounding-box losses converge cleanly. The ROI cascade classification losses for the three stages converge to nearly the same value ($\approx 0.35$ averaged over the final 10\% of iterations). The bounding-box regression losses do not: Stages~1 and 2 plateau at higher values than Stage~0 ($\approx 0.79$ and $0.97$ versus $0.37$). This is expected under the cascade design \citep{Cai2018}: each later stage refines the previous stage's proposals at a stricter IoU threshold ($0.5$, $0.6$, $0.7$), and its regression targets are scaled by progressively larger weights (the \detectron{} cascade defaults), so the three regression losses are not on a common scale. The mask loss decreases smoothly to a stable floor.

\subsection{Stage-2 classification-head fine-tuning schematic}
\label{app:stage2_schematic}

\begin{figure*}[htbp]
\centering
\includegraphics[width=1\textwidth]{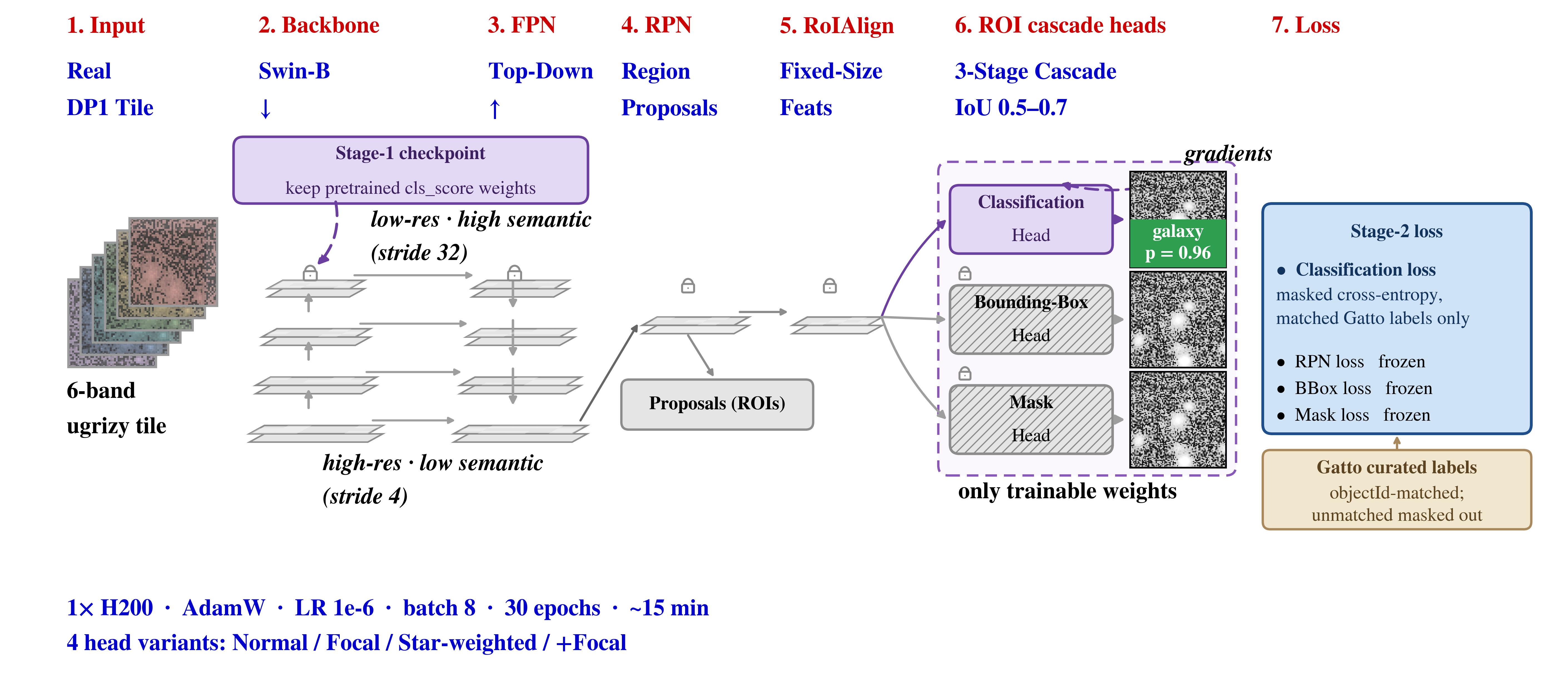}
\caption{\textbf{Stage-2 training: star/galaxy classification-head fine-tuning.}
Shown in the same seven-stage layout as Figure~\ref{fig:stage1_train} for direct
comparison. The model is initialized from the Stage-1 checkpoint with its pretrained
\texttt{cls\_score} weights \emph{kept}, and the entire detector (the \swinb{}
backbone, FPN, RPN, \texttt{RoIAlign}, and the ROI box and mask heads, all greyed
and padlocked here) is frozen (\texttt{requires\_grad=False}); \emph{only} the
classification head (highlighted) is updated. Its supervision is the
\citet{Gatto2026} curated star/galaxy sample, matched to the \scarlet{} annotations
by exact Rubin \texttt{objectId}; detections without a curated label are masked out
of the classification loss (\S\ref{sec:staged_training}). Consequently only the
classification term of the loss is active (a masked softmax cross-entropy
evaluated over the Gatto-labelled ROIs, while the RPN, bounding-box, and mask terms
are inactive and frozen), and gradients propagate back to the classification head
alone. This deliberately gentle fine-tune (learning rate $10^{-6}$, 30 epochs,
${\sim}15$~min on a single H200) replaces the \texttt{refExtendedness}-derived
labels, which become unreliable at $i\gtrsim24$~mag, with the curated
\citet{Gatto2026} labels while leaving the Stage-1 detection and
instance-segmentation performance untouched. Four head variants (normal, focal, star-weighted, and
star-weighted$+$focal) are trained (\S\ref{sec:training}).}
\label{fig:stage2_train}
\end{figure*}

Figure~\ref{fig:stage2_train} presents the Stage-2 fine-tuning step in the same
seven-stage layout as the Stage-1 schematic (Figure~\ref{fig:stage1_train}),
with every frozen component greyed out and padlocked: only the classification
head is updated, supervised by the \citet{Gatto2026} labels matched to the
\scarlet{} annotations by exact \texttt{objectId} (\S\ref{sec:staged_training}).

\clearpage
\subsection{Stage-2 label magnitude distribution}
\label{app:label_mag_dist}

\begin{figure*}[htbp]
\centering
\includegraphics[width=1\textwidth]{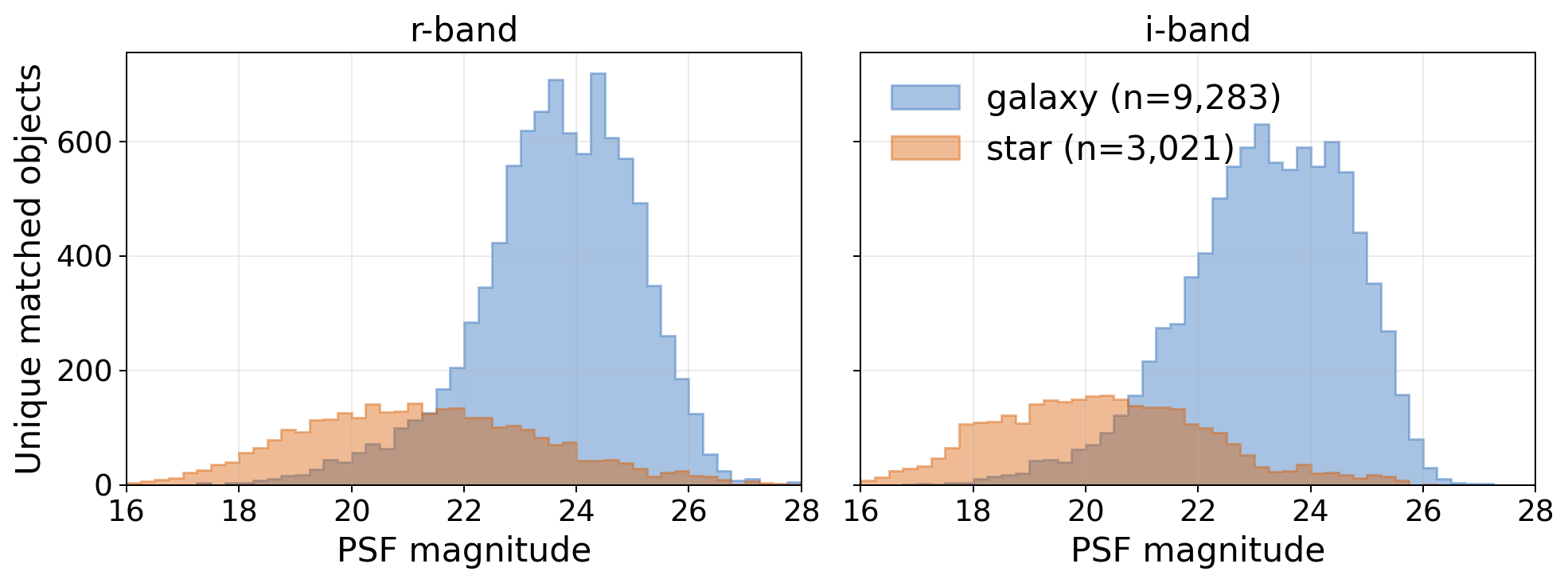}
\caption{$r$- and $i$-band PSF-magnitude distributions of the \citet{Gatto2026}-matched objects entering Stage-2 classification (training and test tiles combined: 8{,}370 training-tile plus 3{,}934 test-tile exact \texttt{objectId} matches; \S\ref{sec:staged_training}): 12{,}304 objects (9{,}283 galaxies, 3{,}021 stars), plotted with galaxies in blue and stars in orange. Each object is counted once, deduplicated by \texttt{objectId}, so sources recurring across overlapping tiles are not double-counted. The photometric stellar classifications (\S\ref{sec:staged_training}) extend the labelled stellar sample to $r>22$~mag, beyond the reach of the Gaia DR3 and spectroscopic identifications.}
\label{fig:label_mag_dist}
\end{figure*}

Figure~\ref{fig:label_mag_dist} shows the $r$- and $i$-band PSF-magnitude
distributions of the \citet{Gatto2026}-matched objects entering Stage-2
classification, split by class, illustrating the contribution of the faint-end photometric stellar classifications discussed in \S\ref{sec:staged_training}.

\subsection{Initialization: in-domain vs.\ ImageNet-21k pretraining}
\label{app:init}

\begin{table}[htbp]
\raggedright
\caption{Effect of initialization on \dpone{} detection completeness (no class weighting, strict 5-condition class-aware match, same test set and $\tau{=}0.22$ as Table~\ref{tab:performance}). In-domain (LSST checkpoint) pretraining dominates in the zero-shot regime but converges to near-parity with ImageNet-21k after \dpone{} fine-tuning. All values percent.}
\label{tab:init_ablation}
\renewcommand{\arraystretch}{1.25}
\setlength{\tabcolsep}{16pt}
\begin{tabular}{lcc}
\hline\hline
Initialization                          & Galaxies & Stars \\
\hline
\multicolumn{3}{l}{\emph{Zero-shot (no \dpone{} fine-tuning)}} \\
\quad ImageNet-21k                      & $\sim 0$      & $\sim 0$       \\
\quad LSST checkpoint (in-domain)       & \Cgalbase     & \Cstarbase     \\
\hline
\multicolumn{3}{l}{\emph{Fine-tuned on \dpone{}}} \\
\quad ImageNet-21k                      & \CgalIN       & \CstarIN       \\
\quad LSST checkpoint (in-domain)       & \CgalA        & \CstarA        \\
\hline
\end{tabular}%
\end{table}

We compare two initialization strategies:
\begin{enumerate}
\item \textbf{ImageNet-21k checkpoint} 

(``\texttt{cascade\_mask\_rcnn\_swin\_b\_in21k\_model.pkl}'')

This is the standard \detectron{} \swinb{} model pretrained on ImageNet-21k with Cascade Mask R-CNN heads on Microsoft COCO \citep{Lin2014}.

\item \textbf{LSST checkpoint (in-domain)} 

(``\texttt{lsst5\_30k\_4h200\_bs192\_ep50.pth}'')

This is a \swinb{} Cascade Mask R-CNN further pretrained for 50 epochs on $\sim 30{,}000$ simulated LSST DC2 scenes following \citet{Merz2025}, on a $4\times$H200 GPU node at batch size 192.
\end{enumerate}
The LSST checkpoint is already six-band: we load its backbone, FPN, RPN, and mask head and re-initialize only the cascade box heads (\S\ref{sec:multiband}). The ImageNet-21k checkpoint is three-band: its patch-embedding weights are duplicated across the two band triplets to form the six-channel convolution, its backbone, FPN, and RPN are loaded, and all COCO ROI heads (box heads, box predictors, and mask head) are discarded and re-initialized. The two initializations therefore differ not only in pretraining domain but also in whether the mask head starts from pretrained weights.

The in-domain checkpoint's advantage is confined to the zero-shot regime. Under identical training and evaluation (Table~\ref{tab:init_ablation}), an ImageNet-21k--initialized checkpoint, which recovers essentially $0\%$ of galaxies and stars \emph{before} fine-tuning, reaches $\sim$\CgalIN\%/$\sim$\CstarIN\% \emph{after} \dpone{} fine-tuning, close to the LSST-checkpoint result ($\sim$\CgalA\%/$\sim$\CstarA\%): the in-domain checkpoint holds a $+\Cgaldelta$~pp edge on galaxies while the ImageNet-21k checkpoint holds a $+\Cstardelta$~pp edge on stars. Fine-tuning on real \dpone{} data thus recovers essentially all of the domain-specific advantage that in-domain pretraining supplies at initialization, so simulated-data pretraining is a convenient warm start rather than a prerequisite for strong \dpone{} performance. This is consistent with the direct-transfer (pre-fine-tuning) advantage reported by \citet{Merz2026} on JWST, while showing that the gap does not persist once real-data fine-tuning is applied.

\section{Operating-point hyperparameter sweep}
\label{app:sweep}

\begin{figure}[!ht]
\centering
\includegraphics[width=0.6\columnwidth]{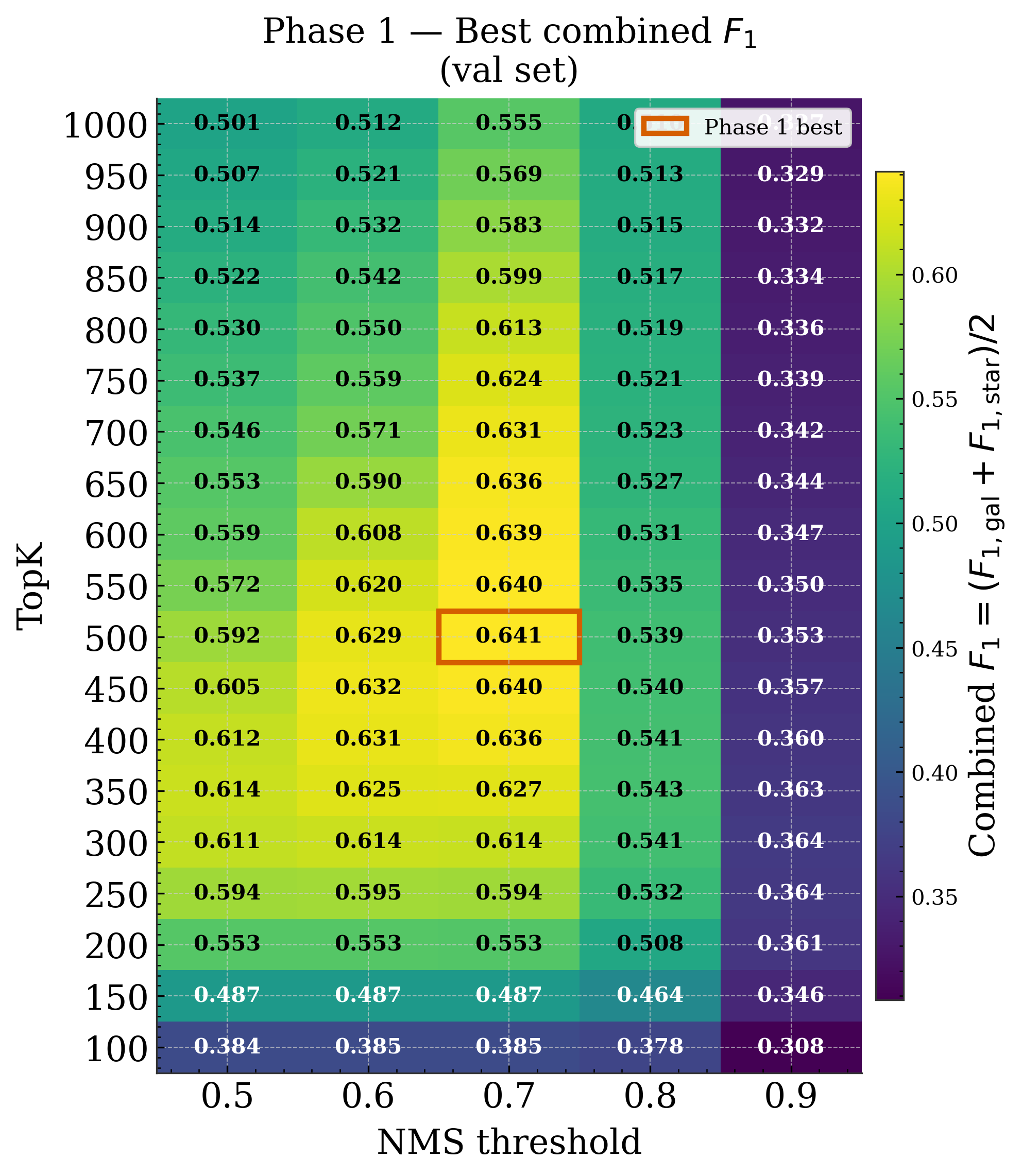}
\caption{Validation-set coarse (``Phase~1'') grid sweep used to fix the inference
hyperparameters (\S\ref{sec:inference}): the combined galaxy$+$star detection $F_1$
(mean of the per-class $F_1$) over the per-image detection cap (TopK) and the proposal
NMS threshold, on the held-out validation tiles. The orange box marks the coarse-grid
optimum ($\mathrm{TopK}{=}500$, $\mathrm{NMS}{=}0.7$); a subsequent fine sweep over TopK
and the score threshold $\tau$ refined this to the adopted operating point
$(\mathrm{TopK},\mathrm{NMS},\tau){=}(485,0.7,0.22)$ used for all test-set inference.}
\label{fig:sweep}
\end{figure}

The fixed inference operating point ($\mathrm{TopK}=485$, $\mathrm{NMS}=0.7$,
score threshold $\tau=0.22$; \S\ref{sec:inference}) was selected on the
validation set by a coarse-then-fine grid sweep over the per-image detection
cap (TopK), the proposal NMS threshold, and the score threshold, maximizing the
combined galaxy$+$star detection $F_1$. Figure~\ref{fig:sweep} shows the coarse-grid
combined-$F_1$ surface over $\mathrm{TopK}\times\mathrm{NMS}$; a subsequent fine sweep
over TopK and the score threshold refined the coarse optimum to the adopted operating
point. The same operating point is used for all test-set inference reported in this
paper.

\clearpage
\section{DeepDISC versus Random Forest concordance}
\label{app:rf_concordance}

\subsection{The two Random Forests and the comparison samples}
\label{app:rf_fullfield}

Section~\ref{sec:cross_compare} compares \deepdisc{} against the Random Forest
of \citet{Gatto2026} on two samples, and two distinct Random Forest
realizations are involved. Both are trained on the 14{,}360 sources of the
\citet{Gatto2026} curated catalog (\S\ref{sec:staged_training}) that have all
six bands and a valid \texttt{refExtendedness}. The \emph{benchmark RF} is the
published classifier of \citet[Table~2]{Gatto2026}, trained on 70\% of that
sample (10{,}052 sources) and evaluated on the held-out 30\% (4{,}308 sources;
3{,}109 galaxies, 1{,}199 stars), where it reaches an overall accuracy of
\RFacc\%. Dr.\ Gatto shared its per-object predictions on that validation
sample, which is what makes the object-by-object comparison of
Table~\ref{tab:gatto_compare} possible. The \emph{full-field RF} is trained on
all 14{,}360 sources and applied across the entire \ecdfs{} field; it supplies
the field-wide agreement of Figure~\ref{fig:rf_agreement}.

\paragraph{The object-by-object sample (Table~\ref{tab:gatto_compare}).}
This is the set held out on both sides: the benchmark RF's 4{,}308-source
validation sample, inner-joined to \deepdisc{}'s held-out test tiles by exact
64-bit Rubin \texttt{objectId}, giving \NholdoutHO{} objects (863 galaxies, 241
stars) that neither method saw in training and that carry the same
curated labels. The comparison is confined to the
classification step on both sides: \deepdisc{} is scored on the sources it
detects and matches one-to-one (\S\ref{sec:cls_results}), so it is not charged
for detection or segmentation completeness, which is the setting the
catalog-level Random Forest already works in. Table~\ref{tab:gatto_compare}
scores all columns on the \NcommonHO{} objects that all three \deepdisc{}
variants match, so the columns remain comparable to each other.
\deepdisc{} reaches its accuracy from a vetted label sample of similar size to
the RF's: the Stage-2 classification step is fine-tuned on the 8{,}370 exact
\texttt{objectId} matches on our training tiles, 9{,}331 including synthetic
recoveries (\S\ref{sec:staged_training}), against the benchmark RF's 10{,}052
sources. That statement concerns the vetted class labels only; Stage-1
detection is supervised by the far larger \scarlet{}-derived annotation set.

\paragraph{The full-field sample (Figure~\ref{fig:rf_agreement}).}
We run the Stage-2 classifier over full-field cutouts, match every \deepdisc{}
detection to the \dpone{} Object catalog by exact Rubin \texttt{objectId}, and
inner-join to the full-field RF prediction table, giving \Nrfcommon{} common
objects ($n = 14{,}543$ bright, $n = 31{,}373$ faint) with \RFagree\% aggregate
agreement (bootstrap interval $[93.9, 94.4]$). The full-field RF predictions
were shared in two feature variants, reference features alone and reference
features plus photometric uncertainties; we quote the latter throughout, to
match the benchmark RF's reported configuration, and note that the former gives
a consistent $93.8\%$ agreement. \emph{Agreement} is the fraction of matched
objects to which both methods assign the same class; per-bin uncertainties are
Wilson $95\%$ intervals and aggregate uncertainties are bootstrap intervals.
This sample too is held out on both sides: the shared full-field RF catalogs
were delivered with all $15{,}611$ curated \texttt{objectId}s removed, a
superset of the full-field RF's $14{,}360$ training sources, so no RF training
object can enter the metrics; and we drop every cutout matching a Stage-1 train
or validation tile, which under the geographic split
(\S\ref{sec:staged_training}) is exactly the set of objects whose labels the
classification step saw.

\subsection{Held-out concordance and HST tie-break}
\label{app:rf_heldout}

\begin{figure*}[!ht]
\centering
\includegraphics[width=0.49\textwidth]{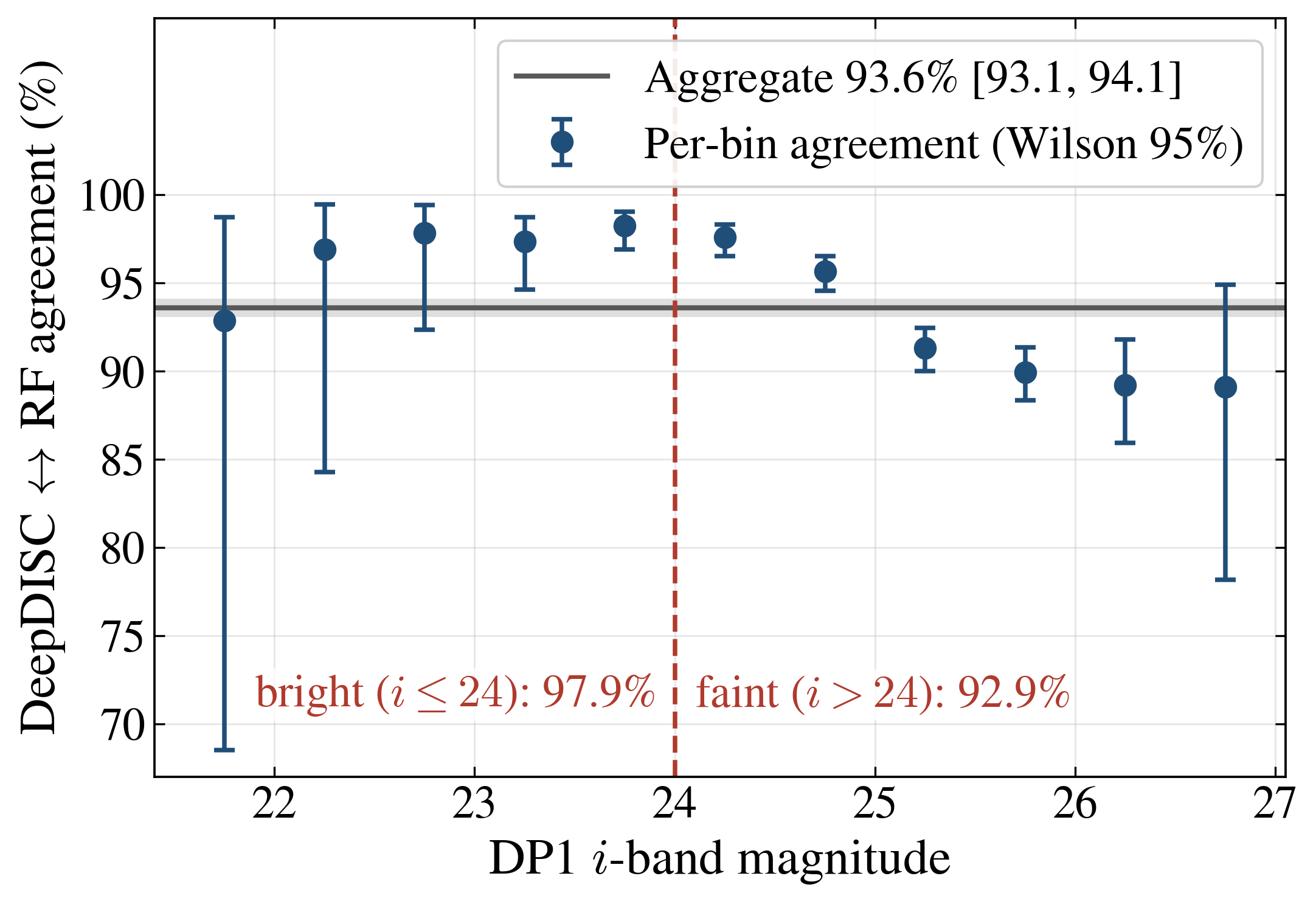}\hfill
\includegraphics[width=0.49\textwidth]{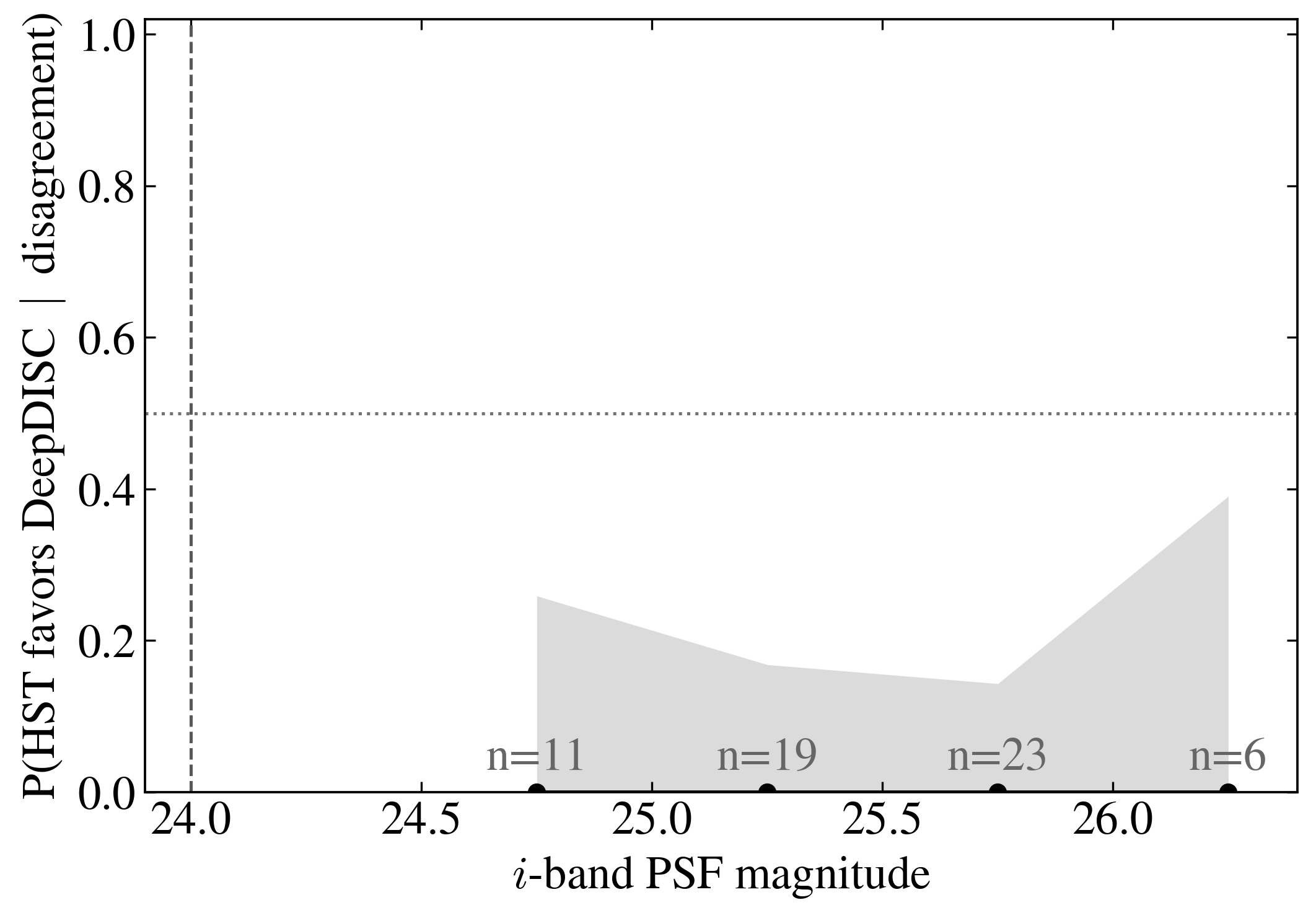}
\caption{Concordance and HST tie-break on the leakage-free held-out footprint
(\texttt{T5063\_P25/35}; $\sim8{,}000$ objects). \emph{(a)}~As panel~(a) of
Figure~\ref{fig:rf_agreement}, but on the held-out footprint; the aggregate agreement
\RFagreeHeldout\% closely matches the full-field value, indicating the concordance is
not driven by train/validation leakage. \emph{(b)}~HST/CANDELS tie-break of the
\deepdisc{}--RF disagreements versus $i$-band magnitude: HST \texttt{CLASS\_STAR}
morphology arbitrates each disagreement, a model-independent (if small-sample)
indication of which classifier is favored toward the faint end. Bootstrap
aggregates over the adjudicated disagreements: HST favors \deepdisc{} in
$1/62 = 0.016$ $[0.000, 0.048]$ of all cases and $0.000$ in the faint ($i>24$) bins.}
\label{fig:rf_agreement_heldout}
\end{figure*}

To remove any residual dependence on the full-field RF footprint, we repeat the
concordance on the leakage-free held-out footprint of \S\ref{sec:hst_results}
(\texttt{T5063\_P25/35}, unseen by both the detector and the classifier). On
these $\sim8{,}000$ matched objects the aggregate agreement is \RFagreeHeldout\%, with
the same bright-to-faint decline (Figure~\ref{fig:rf_agreement_heldout}); its
closeness to the full-field value indicates the concordance is not an artifact of
train/validation leakage. Finally, where the two methods disagree we use the deeper
HST/CANDELS imaging as an external arbiter, resolving each disagreement against the
HST \texttt{CLASS\_STAR} morphology as a function of magnitude
(Figure~\ref{fig:rf_agreement_heldout}b). HST favors the \deepdisc{} call in only $\sim2\%$ of
these disagreements (and essentially none at the faint end), i.e.\ it predominantly
confirms the RF: \deepdisc{}'s excess faint ``stars'' are largely HST-confirmed
galaxies, consistent with the elevated faint-end false-star rate seen against CANDELS
(\S\ref{sec:hst_results}). The HST-covered disagreement sample is small (a few tens of
objects, and only a handful of stars), so this is a secondary, fully external
cross-check; it points the same way as the disagreement-direction trend and
motivates the absolute source-injection benchmark proposed in \S\ref{sec:future};
the better-populated adjudication against the held-out curated
labels themselves follows in \S\ref{app:rf_onetoone}.

\subsection{One-to-one held-out set: magnitude-resolved agreement and truth-adjudicated disagreements}
\label{app:rf_onetoone}

\begin{figure*}[!t]
\centering
\includegraphics[width=\textwidth]{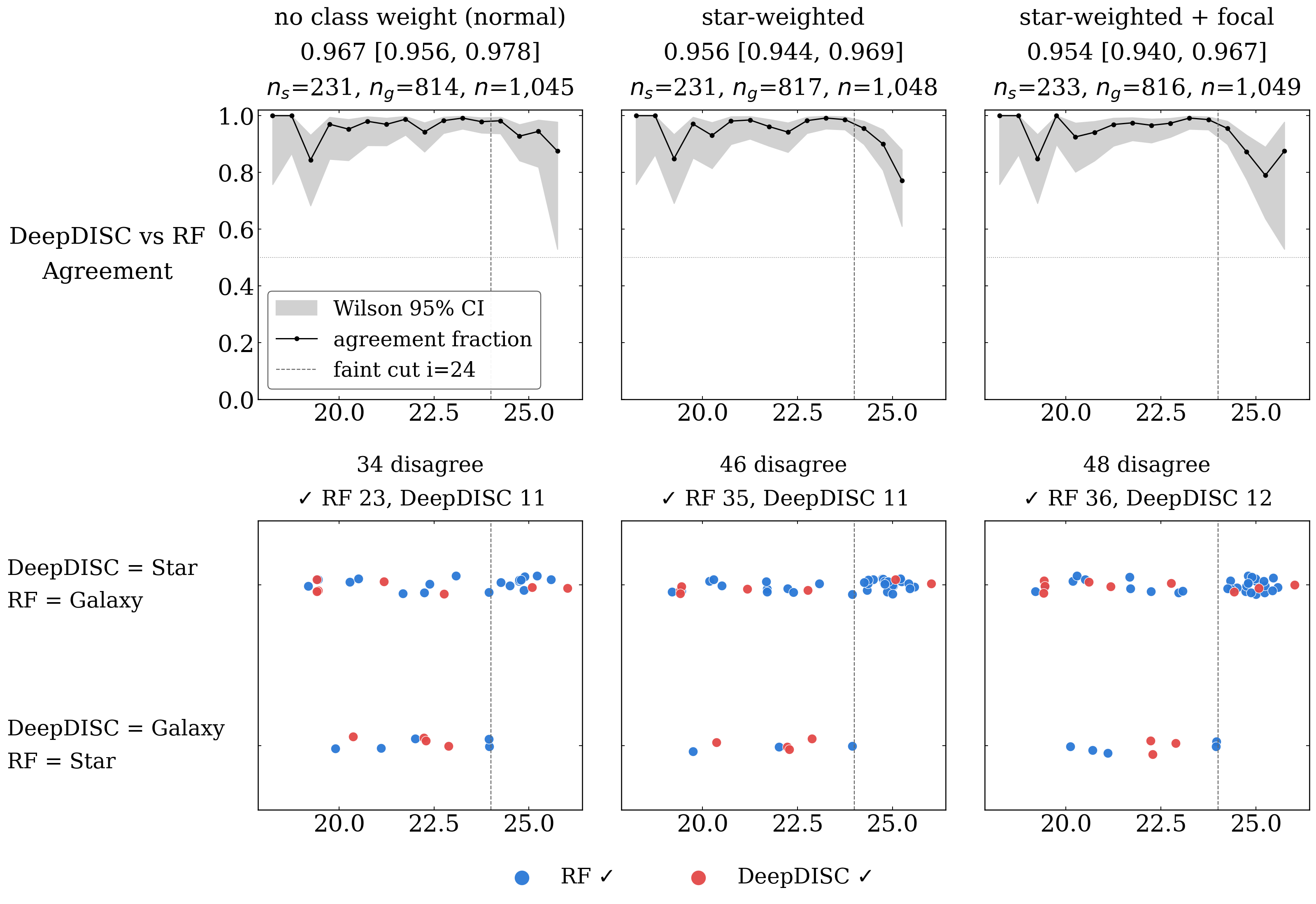}
\caption{Magnitude-resolved view of the comparison of Table~\ref{tab:gatto_compare}, with each of the three \deepdisc{} variants scored on the objects it detects and matches among the \NholdoutHO{} held out by both methods ($n = 1{,}045$, $1{,}048$, and $1{,}049$; galaxy and star counts in the panel headers), slightly more than the common subset of Table~\ref{tab:gatto_compare} since no column-to-column comparison is made here. The horizontal axis of every panel is $i$-band PSF magnitude. \emph{Top:} agreement with the benchmark RF ($0.5$-mag bins with $\geq 8$ objects; shaded bands are Wilson 95\% intervals; headers give the bootstrap aggregate and its interval). \emph{Bottom:} every disagreement, placed by magnitude and direction and coloured by the method the vetted label vindicates (check mark; counts in the headers). The methods agree on $96.7\%$, $95.6\%$, and $95.4\%$ of objects; the $34$, $46$, and $48$ disagreements are predominantly \deepdisc{}$=$star and RF$=$galaxy (25 of 34 for the no-class-weight head), and the labels favor the RF in most of them, consistent with the external HST tie-break of Figure~\ref{fig:rf_agreement_heldout}(b).}
\label{fig:rf_onetoone}
\end{figure*}

The full-field concordance above measures agreement between two imperfect
classifiers without reference to truth. The doubly-held-out one-to-one set of
Table~\ref{tab:gatto_compare} (\S\ref{sec:cross_compare}) permits a stronger
statement: every one of its \NholdoutHO{} objects carries a held-out
curated label: these objects are drawn from the benchmark
RF's 30\% validation sample and lie on \deepdisc{}'s held-out test tiles, so
neither classifier saw them in training and adjudicating by their labels is
fair to both models. Each \deepdisc{}--RF
disagreement can therefore be resolved directly rather than through an HST
proxy, and this is the primary disagreement analysis of the paper; the
HST/CANDELS tie-break above is the secondary, fully external cross-check.
Figure~\ref{fig:rf_onetoone} shows both views for the three \deepdisc{} models,
each evaluated on its detected+matched subset; this subsection records the
per-variant counts behind it.
Agreement with the benchmark RF is $96.7\%$ $[95.6, 97.8]$ for the headline
(no class weight) head ($n=1{,}045$; 34 disagreements; $97.2\%$ at $i\leq24$
versus $95.1\%$ at $i>24$) and $95.4$--$95.6\%$ for the two class-weighted
models (46--48 disagreements), higher throughout than the full-field
\RFagree\%, as expected for this brighter, curated sample. The disagreements
repeat the full-field pattern: they are predominantly objects \deepdisc{} calls
stars and the RF calls galaxies (25 of 34 for the no-class-weight head), and the
held-out curated labels favor the RF in most of them: $23/34$
($\approx 68\%$) with no class weighting, and $35/46$ and $36/48$
($\approx 75$--$76\%$) for star-weighted and star-weighted$+$focal
respectively. The HST tie-break of Figure~\ref{fig:rf_agreement_heldout}(b),
though based on far fewer objects, points the same way.

\clearpage
\section{Self-supervised pretraining pilot (in progress)}
\label{app:ssl_pilot}

\begin{figure*}[!ht]
\centering
\includegraphics[width=1\textwidth]{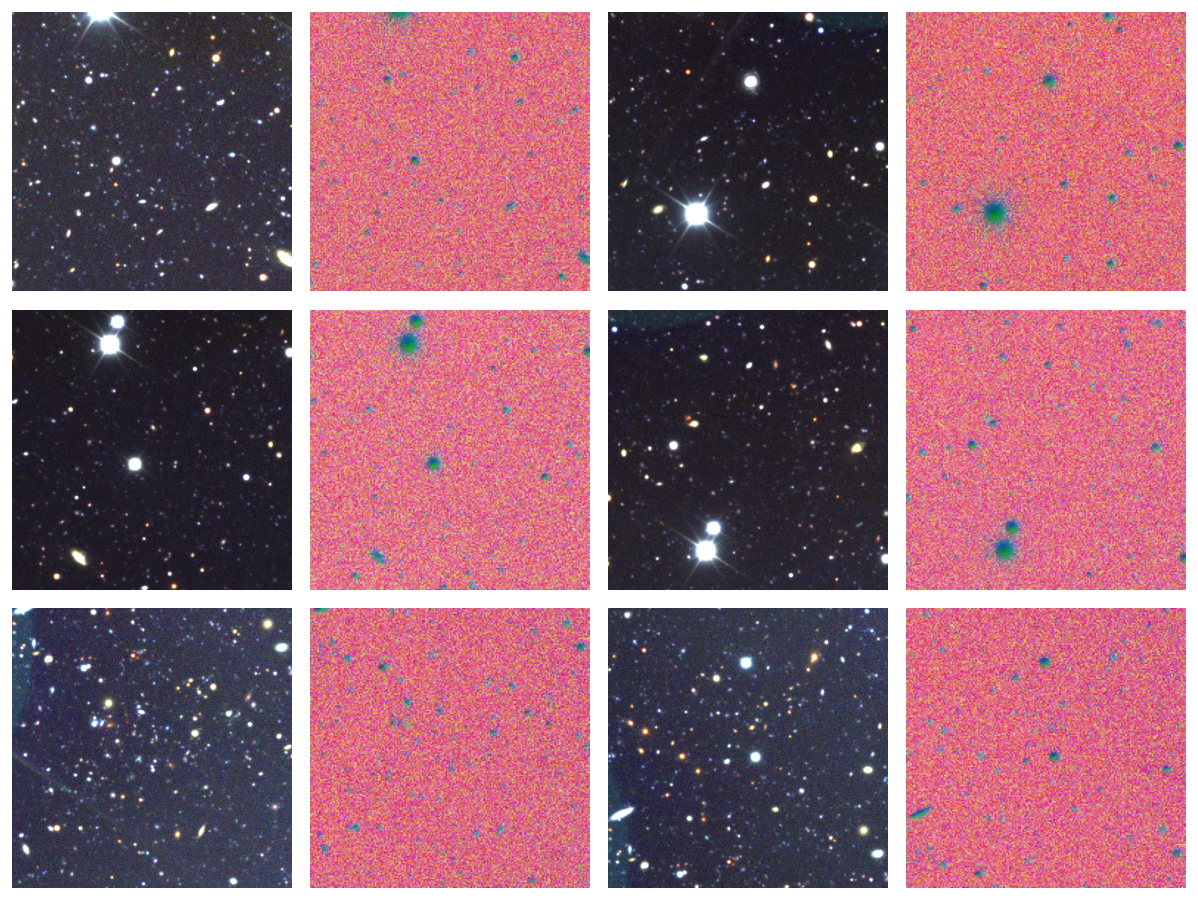}
\caption{Patch-token PCA visualization from the in-progress self-supervised pretraining pilot (Appendix~\ref{app:ssl_pilot}; \S\ref{sec:future}, item~2). For each of six example tiles (odd columns), the adjacent panel (even columns) maps the pretrained \mvit{}-B backbone's per-patch token features through a principal-component decomposition into RGB. Spatially coherent, source-associated color structure, rather than a uniform field, suggests that the backbone's features distinguish sources from background sky using only unlabeled six-band pixels, with no \scarlet{} or classification supervision.}
\label{fig:ssl_patch_pca}
\end{figure*}

As noted in \S\ref{sec:future} (item~2), we are already running an in-domain self-supervised pretraining effort as a precursor to the full \dptwo{}-scale program: a DINOv3-recipe (DINO~$+$~iBOT~$+$~Gram) objective applied to an \mvit{}-B backbone, trained on the unlabeled six-band tiles of the same \dpone{} dataset (\S\ref{sec:data}). Figure~\ref{fig:ssl_patch_pca} shows one qualitative diagnostic from this ongoing effort: a patch-token PCA projection of the pretrained backbone's feature map, rendered as an RGB image (channels = top three principal components). Distinct, spatially coherent colors around stars and galaxies suggest that the backbone's features already distinguish sources from background sky, learned from unlabeled pixels alone, without any \scarlet{} annotation or class label. We do not yet report downstream detection or classification numbers from this pretraining effort; a full evaluation is deferred to the \dptwo{}-scale extension of \S\ref{sec:future}.

\clearpage
\bibliography{dp1_refs}{}
\bibliographystyle{aasjournal}

\end{document}